\documentclass[aps,pra,reprint,nofootinbib,onecolumn,superscriptaddress,showpacs,showkeys,longbibliography]{revtex4-1}
\usepackage{eurosym}
\usepackage{amsmath,amssymb,amstext}
\usepackage[usenames,dvipsnames]{color}
\usepackage{graphicx}
\usepackage{braket}
\usepackage{natbib}
\usepackage{comment}
\usepackage{dcolumn}
\usepackage[english]{babel}
\usepackage{wasysym}
\usepackage[colorlinks,bookmarks=false,citecolor=blue,linkcolor=red,urlcolor=blue]{hyperref}

\begin{document}

\title{Effects of Gapless Bosonic Fluctuations on Majorana Fermions in
Atomic Wire Coupled to a Molecular Reservoir}
\date{\today}

\author{Ying Hu}
\affiliation{Institute for Quantum Optics and Quantum Information of the Austrian Academy of Sciences, A-6020 Innsbruck, Austria}

\author{Mikhail A. Baranov}

\affiliation{Institute for Quantum Optics and Quantum Information of the Austrian Academy of Sciences, A-6020 Innsbruck, Austria}
\affiliation{NRC Kurchatov Institute, Kurchatov Square 1, 123182 Moscow, Russia}

\begin{abstract}
We discuss the effects of quantum and thermal fluctuations on the Majorana
edge states in a topological atomic wire coupled to a superfluid molecular
gas with gapless excitations. We find that the coupling between the Majorana
edge states remains exponentially decaying with the length of the wire, even
at finite temperatures smaller than the energy gap for bulk excitations in
the wire. This exponential dependence is controlled solely by the
localization length of the Majorana states. The fluctuations, on the other
hand, provide the dominant contribution to the preexponential factor, which
increases with temperature and the length of the wire. More important is
that thermal fluctuations give rise to a decay of an initial correlation
between Majorana edge states to its stationary value after some
thermalization time. This stationary value is sensitive to the temperature
and to the length of the wire, and, although vanishing in the thermodynamic
limit, can still be feasible in a mesoscopic system at sufficiently low
temperatures. The thermalization time, on the other hand, is found to be
much larger than the typical time scales in the wire, and is sufficient for
quantum operations with Majorana fermions before the temperature-induced
decoherence sets in.
\end{abstract}

\maketitle

\section{Introduction}

Majorana fermions \cite{Wilczek2009} (or Ising anyons) are probably the
simplest example of non-Abelian anyons - quantum objects with exchange
operations resulting in non-commuting unitary transformations on the space
of degenerate ground states (see, for example \cite
{AnyonReview0,AnyonReview1,AnyonReview2} and references therein). The
emerging non-Abelian statistics has not only fundamental importance as an
alternative to the canonical bosonic and fermionic ones, but also provides
tools for topological quantum computation \cite
{AnyonReview0,TQC1,TQC2,TQC3,TQC4}. In many-body systems, non-Abelian anyons
can emerge as quasi-particles in topological ordered states \cite
{PfQHE,Read2000,Anyon1}. One of the simplest systems exhibiting Majorana
fermions, is a one-dimensional (1D) topological superconductor -- a system
of 1D spinless fermions with a nearest-neighbor (in a lattice realization 
\cite{Kitaev}) or $p$-wave (in a continuous one \cite{Cheng}) pairing
amplitude, in which Majorana fermions appear as edge states. A variety of
physical setups have been proposed for the realization of the corresponding
Hamiltonians both in solid-state structures \cite
{Majreview1,Majreview2,Majreview3,MajCMP0,MajCMP1,MajCMP2,MajCMP12,MajCMP3}
and in systems of ultracold atoms and molecules \cite{LiangJiang,Nascimbene,MajAtom2,MajAtom3,MajAtom10,MajAtom11}. Based on these proposals, recent experiments \cite
{MajEx1,MajEx2,MajEx30,MajEx3,Churchill2013,Finck2013,MajEx4} provide
strong evidences for the existence of Majorana states and make an important
step toward an experimental demonstration of the existence of objects with
non-Abelian statistics.

A key element of most of the considered setups for the realization of
Majorana states is a coupling of the one-dimensional fermions to a reservoir
which serves a source of pairs to generate an effective $p$-wave (or
nearest-neighbor) pairing amplitude. In the realizations with solid-state
systems \cite{MajCMP0,MajCMP1,MajCMP2,MajCMP12,MajCMP3}, the
reservoir is a bulk superconductor and the coupling is due to the proximity
effect. In the atom-molecule realizations \cite{LiangJiang,Nascimbene}, the reservoir is a cloud of molecular BEC and the coupling involves some
molecular dissociation mechanism. The two reservoirs, being absolutely
similar on a mean-field level in providing the $p$-wave pairing amplitude
for fermions, have very different low-energy excitations and, therefore,
their quantum and thermal fluctuations behave differently. In a solid state
superconducting reservoir, one has gapped single-particle excitations,
whereas the excitations in a superfluid molecular reservoir are gapless
collective modes -- Bogoliubov sound. As a result, the correlations between
fluctuations in a solid-state superconducting reservoir are short-range, and
their account do not change the mean-field result -- the coupling between
Majorana edge states remains exponentially decaying with the distance
between them \cite{Dloss}. On the other hand, the decay of correlations
between fluctuations in a molecular superfluid reservoir follows a power
law, raising the question of their effects on the mean-field results.

In this paper we discuss the effects of quantum and thermal fluctuations in
a molecular superfluid reservoir on the properties of Majorana fermion edge
states in a finite one-dimension system of fermionic atoms in a lattice. Our
consideration is based on a generic microscopic Hamiltonian describing a
coupled system of atoms in the lattice and a surrounded superfluid molecular
cloud.

The paper is organized as follows. In Sec. \ref{Model} we describe our
microscopic model and show the emergence of the Kitaev Hamiltonian for
fermions in the lattice. The properties of the Majorana edge states, as well
as fermionic excitations in the wire and bosonic ones in the reservoir are
discussed in Sec. \ref{Kitaev and quasiparticles}. The interactions between
the excitations are the topic of Sec. \ref{Interactions}. The analysis of
their effects on the properties of the Majorana fermions at zero temperature
is presented in Secs. \ref{Effects of InteractionsT0} and \ref
{EnergyCorrection}, and at finite temperatures in Sec. \ref{Temperature}.
The consequences and the proposals for optimal experimental conditions are
briefly discussed in Sec. \ref{Conclusion}. Technical details are given in
two Appendices: In Appendix \ref{AppendixA} we present a scenario leading to
our microscopic Hamiltonian. This can be viewed as a new proposal for
experimental realization of Majorana edge states, as well as an example
demonstrating the capability to control the microscopic Hamiltonian of the
topological wire under experimental conditions. Appendix \ref{AppendixB}
contains analytical solution of the Bogoliubov-de Gennes equations for the
wave function and the energy of the Majorana fermions in a finite Kitaev
wire with open boundary conditions, and the calculations of several
correlation functions in the bulk of the wire.

\section{Microscopic model}

\label{Model}

We consider a system of single-component fermionic atoms in a
one-dimensional (1D) optical lattice (wire) coupled to a Bose-condensed gas
of homonuclear molecules (reservoir) made of two fermionic atoms in
different internal states \cite{BlochDalibardZwerger}. The most essential
for our purposes part of this coupling is a process converting a molecule
from the reservoir into two atoms in the wire (and vice versa). An
underlying physical mechanism of this conversion could be, for example,
radio-frequency assisted dissociation \cite{LiangJiang} or tunneling \cite
{Nascimbene}. In Appendix \ref{AppendixA}, we present another possible
mechanism involving Raman transitions between different internal states of
atoms. To be more specific, we consider the Hamiltonian 
\begin{equation}
H={H}_{\mathrm{BEC}}+{H}_{\mathrm{L}}+{H}_{\mathrm{conv}}+{H}_{\mathrm{int}},
\label{Htotal0}
\end{equation}
where ${H}_{\mathrm{BEC}}$ is the Hamiltonian for the molecular reservoir, 
\begin{equation}
{H}_{\mathrm{BEC}}=\int d\mathbf{r}\hat{\phi}^{\dag }\left( -\frac{\hbar
^{2} }{2m}\nabla ^{2}-\mu _{\mathrm{M}}+\frac{g_{M}}{2}\hat{\phi}^{\dag }
\hat{\phi }\right) \hat{\phi},  \label{Hbec}
\end{equation}
with $\hat{\phi}(\mathbf{r})$ being the field operator of diatomic molecules
with the mass $m=2m_{a}$ and the binding energy $E_{b}=\hbar
^{2}/m_{a}a_{s}^{2}$, where $a_{s}$ is the scattering length between the 
\textbf{\ }atoms forming the molecule, $g_{\mathrm{M}}=4\pi \hbar ^{2}a_{ 
\mathrm{M}}/m$ is the molecular coupling constant with $a_{\mathrm{M}}$
being the molecule-molecule scattering length ($a_{\mathrm{M}}\approx
0.6a_{s}$, see \cite{PetrovSolomonSchlyapnikov,PetrovSolomonSchlyapnikov1}),
and $\mu _{\mathrm{M}}$ is the molecular chemical potential. In the
following, we will consider the regime of weak interaction $n_{\mathrm{M}
}a_{ \mathrm{M}}^{3}<1$, where $n_{\mathrm{M}}$ is the density of molecules.

The second term in Hamiltonian (\ref{Htotal0}) 
\begin{equation}
H_{\mathrm{L}}=\sum_{j}\left[ -J\left( \hat{a}_{j}^{\dag}\hat{a}_{j+1} +\hat{
a} _{j+1}^{\dagger}\hat{a}_{j}\right) -\mu_{0}\hat{a}_{j}^{\dag}\hat {a}_{j} 
\right],  \label{HL}
\end{equation}
describes fermionic atoms in the wire. Here $\hat{a}_{j}$ and $\hat{a}
_{j}^{\dag}$ are fermionic annihilation and creation operators on a site $j$
, respectively, $J$ is the hopping amplitude, and $\mu_{0}$ is the fermionic
chemical potential.

The conversion of a molecule from the reservoir into two atoms in the wire
is described by the third term in Hamiltonian (\ref{Htotal0}) 
\begin{equation}
{H}_{\mathrm{conv}}=\sum_{j}\int d\mathbf{r}\left[ K_{j}(\mathbf{r})\hat {a}
_{j}^{\dag}\hat{a}_{j+1}^{\dag}\hat{\phi}\left( \mathbf{r}\right) +\text{h.c}.
\right].  \label{Hconv}
\end{equation}
Here, the explicit form of the amplitude $K_{j}(\mathbf{r})$ relies on the
specific realization of the conversion mechanism (see, for example, Ref. 
\cite{Nascimbene} or Appendix \ref{AppendixA}).

Finally, the last term in Hamiltonian (\ref{Htotal0}) 
\begin{equation}
H_{\mathrm{int}}=\sum_{j}\int d\mathbf{r}g_{j}(\mathbf{r})\hat{a}_{j}^{\dag
} \hat{a}_{j}\hat{\phi}^{\dag }(\mathbf{r})\hat{\phi}(\mathbf{r})
\label{Hint}
\end{equation}
describes a short-range interaction between atoms and molecules (assuming
their spatial overlap) with $g_{j}(\mathbf{r})=g_{a\mathrm{M}}w^{2}(\mathbf{
r }-\mathbf{r}_{j})$, where $g_{aM}$ is the atom-molecule interaction and $
w( \mathbf{r}-\mathbf{r}_{j})$ is the Wannier function centered on the site $
j$ in the wire.

Note that in writing the Hamiltonians ${H}_{\mathrm{conv}}$ and ${H} _{
\mathrm{int}}$, we take into account only the nearest-neighbour and on-site
terms, respectively, assuming the condition $a_{s}<a$ that the size of the
molecule $a_{s}$ is smaller than the lattice spacing $a$. Intuitively, this
condition arises naturally in optimizing the conversion, because too small
or too large molecules will lead to smaller overlap of their wave function
with Wannier functions on \emph{different} sites of the wire, and therefore,
results in a smaller conversion amplitude $K$ (see, for example, Appendix 
\ref{AppendixA}).

Assuming zero temperature at the moment, we will treat the Hamiltonian ({\ref
{Htotal0}}) within the Bogoliubov framework by decomposing the molecular
field operator $\hat{\phi}(\mathbf{r})$ into a mean-field part and quantum
fluctuations, $\hat{\phi}(\mathbf{r})=\phi _{0}(\mathbf{r})+\delta \hat{\phi}
(\mathbf{r})$, with $\phi _{0}(\mathbf{r})=\langle \hat{\phi}(\mathbf{r}
)\rangle $ being the mean-field condensate function and $\delta \hat{\phi}( 
\mathbf{r})$ representing the quantum fluctuations respectively. With this
decomposition, Hamiltonian (\ref{Htotal0}) can be recast into a sum of three
components 
\begin{equation}
H=H_{\mathrm{BMF}}(\phi _{0})+{H}_{\mathrm{K}}(\phi _{0})+H_{\mathrm{f}},
\label{Heff0}
\end{equation}
where 
\begin{equation}
H_{\mathrm{BMF}}=\int d\mathbf{r}\phi _{0}^{\ast }\left( -\frac{\hbar ^{2}}{
2m}\nabla ^{2}-\mu _{\mathrm{M}}+\frac{g_{M}}{2}|\phi _{0}|^{2}\right) \phi
_{0}
\end{equation}
is the mean-field BEC Hamiltonian, 
\begin{equation}
H_{\mathrm{K}}=\sum_{j=1}^{L-1}(\!-\!J\hat{a}_{j}^{\dag }\hat{a}_{j+1}+{\
\Delta _{\phi _{0}}}\hat{a}_{j}^{\dag }\hat{a}_{j+1}^{\dag }+\text{h.c.}
)-\sum_{j=1}^{L}\mu _{f}\hat{a}_{j}^{\dag }\hat{a}_{j}  \label{HK}
\end{equation}
is the Kitaev Hamiltonian \cite{Kitaev} for fermionic atoms with the pairing
amplitude 
\begin{equation}
{\Delta _{\phi _{0}}=|\Delta |}e^{i\theta }=\int d\mathbf{r}K_{j}\left( 
\mathbf{r})\phi _{0}(\mathbf{r}\right)  \label{Delta}
\end{equation}
and the renormalized chemical potential for fermions 
\begin{equation}
\mu _{f}=\mu _{0}+\int d\mathbf{r}g_{j}(\mathbf{r})\left\vert \phi
_{0}\left( \mathbf{r}\right) \right\vert ^{2}.  \label{muf}
\end{equation}
The third component in Eq. (\ref{Heff0}), Hamiltonian $H_{\mathrm{f}}$,
contains the effects of bosonic fluctuations $\delta \hat{\phi}$.

In Eq. (\ref{HK}), we have shown the emergence of the Kitaev Hamiltonian $
H_{K} $, which has a gap parameter defined in terms of the mean-field
condensate function $\phi_{0}$. The Gross-Pitaevskii (GP) equation for the
condensate wave function $\phi_{0}$ can be found by demanding that the term
in $H_{f}$, which is linear in the fluctuations of the molecular field $
\delta\hat{\phi}$ only, vanishes. The resulting GP equation reads 
\begin{align}
\left( -\frac{\hbar^{2}\nabla^{2}}{2m}+g_{\text{M}}\left\vert \phi_{0}(
\mathbf{r} )\right\vert^{2}\right) \phi_{0}(\mathbf{r})+\sum_{j}\left[
K_{j}^{\ast} ( \mathbf{r})\left\langle \hat{a}_{j+1}\hat{a}_{j}\right\rangle
_{H_{\mathrm{K} }} +g_{j}(\mathbf{r})\phi_{0}(\mathbf{r})\left\langle \hat{a}
_{j}^{\dag} \hat{a}_{j}\right\rangle _{H_{\mathrm{K}} }\right] =\mu_{\text{M}
}\phi _{0}(\mathbf{r}),  \label{GPequation}
\end{align}
where $\langle\ldots\rangle_{H_{\mathrm{K}}}$ denotes the expectation value
with respect to the \emph{ground state} of the Hamiltonian ${H}_{\mathrm{K}
}(\phi_{0})$ in Eq. (\ref{HK}). Equation (\ref{GPequation}) thus determines
the condensate wave function $\phi_{0}$ self-consistently.

With the condensate wave function $\phi_{0}(\mathbf{r})$ satisfying Eq. (\ref
{GPequation}), the Hamiltonian $H_{\mathrm{f}}$ reduces to the sum, 
\begin{equation}
H_{\mathrm{f}}=H_{\mathrm{ph}}+H_{\mathrm{c}}+H_{\mathrm{ph-ph}},
\end{equation}
which consist of the Bogoliubov Hamiltonian for phonons $H_{\mathrm{ph}}$
(the part quadratic in $\delta\hat{\phi}$), the interaction of phonons with 
\emph{fermionic excitations} $H_{\mathrm{c}}$, and the phonon-phonon
interactions $H_{\mathrm{ph-ph}}$. More explicitly,

\begin{equation}
H_{\mathrm{ph}}=\int d\mathbf{r}\left\{ \delta \hat{\phi}^{\dag }\left[-
\frac{\hbar ^{2}\nabla ^{2}}{2m}-\mu_{\mathrm{M}}+2g\left\vert
\phi_{0}\right\vert ^{2}+\sum_{j}g_{j}\left\langle \hat{a}_{j}^{\dag }\hat{a}
_{j}\right\rangle _{{H}_{\mathrm{K}}}\right] \delta \hat{\phi}+g_{M}\left(
\phi _{0}^{2}\delta \hat{\phi}^{\dag }\delta \hat{\phi}^{\dag }+\phi
_{0}^{\ast 2}\delta \hat{\phi}\delta \hat{\phi}\right) \right\} ,
\label{Hph}
\end{equation}
and 
\begin{equation}
H_{\mathrm{c}}=H_{\mathrm{c}}^{(3)}+H_{\mathrm{c}}^{(4)}  \label{Hcoupling}
\end{equation}
with 
\begin{align}
H_{\mathrm{c}}^{(3)}& =\sum_{j}\int d\mathbf{r}\Big\{K_{j}(\mathbf{r})(\hat{a
}_{j}^{\dag }\hat{a}_{j+1}^{\dag }-\langle \hat{a}_{j}^{\dag }\hat{a}
_{j+1}^{\dag }\rangle _{H_{\mathrm{K}}})\delta \hat{\phi}(\mathbf{r})+\text{
h.c.}  \notag \\
\!\!& +g_{j}(\mathbf{r})(\hat{a}_{j}^{\dag }\hat{a}_{j}-\langle \hat{a}
_{j}^{\dag }\hat{a}_{j}\rangle _{H_{\mathrm{K}}})(\delta \hat{\phi}^{\dag
}\phi _{0}\!+\!\text{h.c.})\Big\},  \label{Hc3} \\
H_{\mathrm{c}}^{(4)}& =\sum_{j}\int d^{3}\mathbf{r}g_{j}(\mathbf{r})(\hat{a}
_{j}^{\dag }\hat{a}_{j}-\langle \hat{a}_{j}^{\dag }\hat{a}_{j}\rangle _{H_{ 
\mathrm{K}}})\delta \hat{\phi}^{\dag }(\mathbf{r})\delta \hat{\phi}(\mathbf{
r }),  \label{Hc4}
\end{align}
where $\hat{a}_{j}^{\dag }\hat{a}_{j+1}^{\dag }-\langle \hat{a}_{j}^{\dag} 
\hat{a}_{j+1}^{\dag }\rangle $ and $\hat{a}_{j}^{\dag }\hat{a}_{j}-\langle 
\hat{a}_{j}^{\dag }\hat{a}_{j}\rangle $ represent fermionic fluctuations
(this form is equivalent to the normal ordering of the fermionic
quasiparticle operators). The phonon-phonon interaction Hamiltonian $H_{
\mathrm{ph-ph}}$ contains cubic and quartic in $\delta\hat{\phi}$
contributions which can be easily obtained from Eq. (\ref{Heff0}). Here we
do not write down $H_{\mathrm{ph-ph}}$ explicitly, because the corresponding
terms contain no coupling to the fermions and result only in the
renormalization of the bosonic excitations (phonon modes) defined by the
Bogoliubov Hamiltonian (\ref{Hph}). This renormalization is not important
for our purposes, and we neglect the Hamiltonian $H_{\mathrm{ph-ph}}$
assuming that the Hamiltonian $H_{\mathrm{ph}}$ already contains the
\textquotedblleft true \textquotedblright excitations in the molecular BEC.
As a result, in describing the system we limit ourselves to the effective
Hamiltonian 
\begin{equation}
{H}_{\mathrm{eff}}={H}_{\mathrm{K}}+H_{\mathrm{ph}}+H_{\mathrm{c}},
\label{Heffreduced}
\end{equation}
accompanied by the GP equation (\ref{GPequation}) for the self-consistent
determination of the molecular condensate wave function $\phi_{0}(\mathbf{r}
)$. In the Hamiltonian ${H}_{\mathrm{eff}}$, the first two (quadratic) terms 
$H_{\mathrm{K}}$ and ${H}_{\mathrm{ph}}$ describe fermionic and bosonic
quasiparticles, respectively, and the last term ${H}_{\mathrm{c}}$
corresponds to interactions between them.

\section{Fermionic and Bosonic quasiparticles}

\label{Kitaev and quasiparticles}

Let us first consider the properties of fermionic and bosonic quasiparticles
described by the quadratic Hamiltonians (\ref{HK}) and (\ref{Hph}),
respectively. The properties of the Kitaev Hamiltonian ${H}_{\mathrm{K}}$ in
Eq. (\ref{HK}) for $1D$ spinless fermions in the lattice are well-known \cite
{Kitaev}. We summarize them here to make the presentation self-contained,
and to create a `reference' point for future discussion of the effects of
quantum fluctuations.

Being quadratic in fermionic operators of the form 
\begin{equation}
{H}_{\mathrm{K}}=-\frac{1}{2}\sum\limits_{i,j}t_{ij}(\hat{a}_{i}^{\dag} \hat{
a}_{j}+\text{h.c.})+\frac{1}{2}\sum\limits_{i,j} (\Delta_{ij}\hat{a}
_{i}^{\dag}\hat{a}_{j}^{\dag}+\text{h.c.}),  \notag
\end{equation}
with obvious expressions for $t_{ij}$ and $\Delta_{ij}=-\Delta_{ji}$ written
down from Eq. (\ref{HK}), the Hamiltonian ${H}_{\mathrm{K}}$ can be
diagonalized by the Bogoliubov transformation 
\begin{equation}
\hat{a}_{j}=\sum_{m}\left( u_{j,m}\hat{\alpha}_{m}+\upsilon_{j,m}^{\ast}\hat{
\alpha}_{m}^{\dag}\right) ,  \label{aj}
\end{equation}
where the quasiparticle (excitation) \emph{fermionic} annihilation and
creation operators $\hat{\alpha}_{m}$ and $\hat{\alpha}_{m}^{\dag}$ obey
canonical anticommutation relations. The amplitudes $u_{j,m}$ and $
\upsilon_{j,m}$ satisfy the conditions $\sum_{j}(u_{j,m_{1}}^{\ast}
u_{j,m_{2}}+\upsilon_{j,m_{1}
}^{\ast}\upsilon_{j,m_{2}})=\delta_{m_{1}m_{2}} $ and $\sum_{j}(u_{j,m_{1}
}\upsilon_{j,m_{2}}+\upsilon_{j,m_{1}}u_{j,m_{2}} )=0$, and can be found
from the Bogoliubov-de-Gennes (BDG) equations 
\begin{align}
\sum_{j}\left( -t_{ij}u_{j,m}+\Delta_{ij}\upsilon_{j,m}\right) &
=E_{m}u_{i,m},  \notag \\
\sum_{j}\left( t_{ij}\upsilon_{j,m}+\Delta_{ji}^{\ast}u_{j,m}\right) &
=E_{m}\upsilon_{i,m}  \notag
\end{align}
with the quasiparticle energy $E_{m}\geq0$. The diagonal form of the
Hamiltonian reads 
\begin{equation}
{H}_{\mathrm{K}}=E_{0}+\sum_{m}E_{m}\hat{\alpha}_{m}^{\dag}\hat{\alpha}_{m},
\label{HKdiag}
\end{equation}
where $E_{0}=-\sum_{j,m}E_{m}\left\vert \upsilon_{j,m}\right\vert ^{2}$ is
the energy of the ground state $\left\vert 0\right\rangle $ defined by the
conditions $\hat{\alpha}_{j,m}\left\vert 0\right\rangle =0$ for all $\hat{
\alpha}_{j,m}$.

A remarkable feature of the Kitaev Hamiltonian ${H}_{\mathrm{K}}$ is the
existence of the topological phase for $|\mu_{f}|<2J$ \cite{Kitaev}, in
which a robust `zero-energy' fermionic edge mode ($m=M$, to be specific)
emerges with an energy $E_{M}$ vanishing exponentially with the system size $
L$, while other modes (with $m=\nu\neq M$) are gapped $E_{\nu}>\left\vert
\Delta \right\vert $. In the thermodynamic limit $L\rightarrow\infty$, the
presence of such edge modes results in the degeneracy of the ground state:
The states $\left\vert 0\right\rangle $ and $\hat{\alpha}_{M}^{\dag}|0
\rangle $ have the same energy. Moreover, although having different
fermionic parity, the two states cannot be distinguished by local
measurements in the bulk of the wire. This is because they differ by the
occupation of the fermionic edge mode and, therefore, have the same local
correlations in the bulk.

The edge character of the `zero-energy' mode and its connection to Majorana
fermions can be revealed by writing the corresponding annihilation operator
in the form $\hat{\alpha}_{M}=(\hat{\gamma}_{L}+i\hat{\gamma}_{R})/2$, where 
\begin{equation*}
\hat{\gamma}_{L}=\hat{\alpha}_{M}+\hat{\alpha}_{M}^{\dag}=\sum_{j}\left[
(u_{jM}^{\ast}+\upsilon_{jM})\hat{a}_{j}+(u_{jM}+\upsilon_{jM}^{\ast})\hat {a
}_{j}^{\dag}\right] ,
\end{equation*}
and 
\begin{equation*}
\hat{\gamma}_{R}=-i(\hat{\alpha}_{M}-\hat{\alpha}_{M}^{\dag})=-i\sum _{j}
\left[ (u_{jM}^{\ast}-\upsilon_{jM})\hat{a}_{j}+(\upsilon_{jM}^{\ast
}-u_{jM})\hat{a}_{j}^{\dag}\right] ,
\end{equation*}
are two Hermitian \emph{Majorana} operators satisfying the conditions $\hat{
\gamma}_{L(R)}=\hat{\gamma}_{L(R)}^{\dag}$, $\hat{\gamma}_{L(R)}^{2}=1$, and 
$\gamma_{L}\gamma_{R}=-\gamma_{R}\gamma_{L}$. It turns out that (see Ref. 
\cite{Kitaev} and Appendix \ref{AppendixB} for details) 
\begin{equation}
f_{Lj}\equiv u_{jM}+\upsilon_{jM}^{\ast}\approx2\left\vert A\right\vert
\rho^{j}\sin(j\theta)\sim e^{-ja/l_{M}}  \label{fLj}
\end{equation}
and 
\begin{equation}
f_{Rj}\equiv u_{jM}-\upsilon_{jM}^{\ast}\approx2\left\vert A\right\vert
\rho^{L+1-j}\sin[(L+1-j)\theta]\sim e^{-(L+1-j)a/l_{M}},  \label{fRj}
\end{equation}
where we assume $4(J^{2}-\left\vert \Delta\right\vert ^{2})-\mu^{2}>0$ such
that 
\begin{equation*}
\left\vert A\right\vert =\sqrt{\frac{\left\vert \Delta\right\vert
(4J^{2}-\mu^{2})}{J(4J^{2}-4\left\vert \Delta\right\vert ^{2}-\mu^{2})}}
,\;\rho =\sqrt{\frac{J-\left\vert \Delta\right\vert }{J+\left\vert
\Delta\right\vert }}<1,\;\cos\theta=\frac{-\mu}{2\sqrt{J^{2}-\left\vert
\Delta\right\vert ^{2}}}
\end{equation*}
and the Majorana localization length (for details and for the general case
see Appendix \ref{AppendixB}) 
\begin{equation}
l_{M}=\frac{a}{\ln\rho^{-1}}.  \label{lM1}
\end{equation}
The localization length $l_{M}$ also enters the expression for the energy of
the mode $\hat{\alpha}_{M}$, 
\begin{equation}
E_{M}\approx\left\vert \Delta\right\vert \frac{4J^{2}-\mu^{2}}{
J(J+\left\vert \Delta\right\vert )}\rho^{L}\left\vert \frac{\sin[(L+1)\theta]
}{\sin\theta }\right\vert \sim e^{-La/l_{M}},  \label{EMexact1}
\end{equation}
which becomes exponentially small for $L\gg l_{M}/a$ [see Eq. (\ref{EMfinal})]. The above expressions show that the fermionic `zero-energy' mode $\hat{
\alpha}_{M}$ represents a non-local fermion associated with two \emph{
spatially separated} Majorana operators $\hat{\gamma}_{L}$ and $\hat{\gamma}
_{R}$ localized at the opposite edges of the wire. The following form of the
Hamiltonian $H_{\mathrm{K}}$ 
\begin{align}
H_{\mathrm{K}} & =E_{0\mathrm{f}}+E_{M}\hat{\alpha}_{M}^{\dag}\hat{\alpha }
_{M}+\sum_{\nu}E_{\nu}\hat{\alpha}_{\nu}^{\dag}\hat{\alpha}_{\nu }
\label{KHsplitted} \\
& =E_{0\mathrm{f}}+\frac{1}{2}E_{M}+\frac{i}{2}E_{M}\hat{\gamma}_{L}\hat{
\gamma}_{R}+\sum_{\nu}E_{\nu}\hat{\alpha}_{\nu}^{\dag}\hat{\alpha}_{\nu }, 
\notag
\end{align}
emphasizes this special `zero-energy' edge mode $\hat{\alpha}_{M}$ and its
connection to the Majorana edge modes $\hat{\gamma}_{L}$ and $\hat{\gamma}
_{R}$, as compared to the gapped bulk excitations $\hat{\alpha}_{\nu}$ with
energies $E_{\nu}>\left\vert \Delta\right\vert $ (for details on the bulk
gapped modes see \ref{AppendixB}). Note that the energy $E_{M}$ of the
fermionic mode can also be viewed as the coupling between the corresponding
Majorana modes $\hat{\gamma}_{L}$ and $\hat{\gamma}_{R}$.

The properties of bosonic quasi-particles are described by the Hamiltonian $
H_{\mathrm{ph}}$, Eq. (\ref{Hph}), which can be diagonalized by using the
standard bosonic Bogoliubov transformation 
\begin{equation}
\delta \hat{\phi}(\mathbf{r})=\sum_{\gamma }[\tilde{u}_{\gamma }(\mathbf{r})
\hat{b}_{\gamma }-\tilde{\upsilon}_{\gamma }^{\star }(\mathbf{r})\hat{b}
_{\gamma }^{\dag }]  \label{BogPhonons}
\end{equation}
in terms of bosonic quasiparticle (phonon) operators $\hat{b}_{\gamma }$,
where $\tilde{u}_{\gamma }$ and $\tilde{\upsilon}_{\gamma }$ are the
solutions of the corresponding Bogoliubov-de-Gennes equations. The
diagonalized Hamiltonian reads 
\begin{equation}
H_{\mathrm{ph}}=E_{0\mathrm{ph}}+\sum_{\gamma }\epsilon _{\gamma }\hat{b}
_{\gamma }^{\dag }\hat{b}_{\gamma },  \label{Hphdiag}
\end{equation}
where $E_{0\mathrm{ph}}$ is the quasiparticle ground state energy, and $
\epsilon _{\gamma }$ is the quasiparticle spectrum. In general, the
interaction with fermions results in a spatially non-uniform condensate wave
function $\phi _{0}(\mathbf{r})\neq \mathrm{const}$, as well as in the
appearance of a position-dependent external potential in Eq. (\ref{Hph}) for 
$H_{\mathrm{ph}}$. As a consequence, bosonic excitations are not
characterized by the momentum, and their wave functions are not plane waves
anymore. The problem of finding the coefficients $\tilde{u}_{\gamma }(
\mathbf{r})$ and $\tilde{\upsilon}_{\gamma }(\mathbf{r})$ of the Bogoliubov
transformation (\ref{BogPhonons}) and the corresponding eigen-energies $
\epsilon _{\gamma }$ in this case can only be addressed numerically. In the
considered case of a large (compare to the wire) BEC and weak coupling, the
interaction with fermions in the wire generates quantitatively small effects
on bosonic excitations in the reservoir. We will therefore neglect them and
consider a spatially homogeneous condensate with $\phi _{0}(\mathbf{r})=
\sqrt{n_{\mathrm{M}}}$ and bosonic excitations characterized by the wave
vector $\mathbf{q}$. The corresponding wave functions are then plane waves, $
[\tilde{u}_{\mathbf{q}}(\mathbf{r}),\tilde{\upsilon}_{\mathbf{q}}(\mathbf{r}
)]=(\tilde{u}_{q},\tilde{\upsilon}_{q})V^{-1/2}\exp (i\mathbf{qr})$, such
that 
\begin{equation}
\delta \hat{\phi}(\mathbf{r})=\frac{1}{\sqrt{V}}\sum_{\mathbf{q}}(\tilde{u}
_{q}\hat{b}_{\mathbf{q}}-\tilde{\upsilon}_{q}\hat{b}_{-\mathbf{q}}^{\dag
})\exp (i\mathbf{qr}),  \label{BogoliubovBose}
\end{equation}%
where $\tilde{u}_{q}^{2}(\tilde{v}_{q}^{2})=\left[ \left( \epsilon
_{q}^{0}+g_{\mathrm{M}}n_{\mathrm{M}}\right) /\epsilon _{q}\pm 1\right] /2$
with $\epsilon _{q}=\sqrt{\epsilon _{q}^{0}\left( \epsilon _{q}^{0}+2g_{
\mathrm{M}}n_{\mathrm{M}}\right) }$ and $\epsilon _{q}^{0}=\hbar
^{2}q^{2}/2m $. As usual, for small wave vectors $q\lesssim \xi _{\text{BEC}
}^{-1}$, where $\xi _{\text{BEC}}=\hbar /\sqrt{mg_{\mathrm{M}}n_{\mathrm{M}}}
$ is the coherence length of the condensate, the excitations are phonons $
\epsilon _{q}=\hbar cq$ with the sound velocity $c=\sqrt{g_{\mathrm{M}}n_{
\mathrm{M}}/m}$.

\section{Interaction between quasiparticles}

\label{Interactions}

Let us now analyze the effects of the interaction $H_{\mathrm{c}}$ between
fermions excitations in the lattice and fluctuations in the reservoir
(phonons) on the properties of the \textquotedblleft
zero-energy\textquotedblright fermionic edge mode $\hat{\alpha}_{M}$. We
will be primarily interested in corrections to the energy $E_{M}$ of the
mode, see Eq. (\ref{KHsplitted}).

By using the Bogoliubov transformations (\ref{aj}) and (\ref{BogPhonons}),
the Hamiltonian (\ref{Heffreduced}) reads 
\begin{equation}
H_{\mathrm{eff}}=E_{0}+H_{0}+{H}_{\mathrm{c}}^{(3)}+{H}_{\mathrm{c}}^{(4)},
\label{Hc1}
\end{equation}
where $E_{0}$ is the ground state energy of the system, 
\begin{equation}
H_{0}=\sum_{m}E_{m}\hat{\alpha}_{m}^{\dag }\hat{\alpha}_{m}+\sum_{\mathbf{q}
}\epsilon _{q}\hat{b}_{\mathbf{q}}^{\dag }\hat{b}_{\mathbf{q}}  \label{H0}
\end{equation}
describes fermionic and bosonic excitations, and the terms 
\begin{equation}
{H}_{\mathrm{c}}^{(3)}=\sum_{\mathbf{q}}\sum_{m,n}\Big[O_{\mathbf{q}%
mn}^{(n)} \hat{\alpha}_{m}^{\dag }\hat{\alpha}_{n}\hat{b}_{\mathbf{q}}^{\dag
}+\text{ h.c.}+(O_{\mathbf{q}mn}^{(a1)}\hat{\alpha}_{m}\hat{\alpha}_{n}+O_{%
\mathbf{q} mn}^{(a2)}\hat{\alpha}_{m}^{\dag }\hat{\alpha}_{n}^{\dag })\hat{b}%
_{\mathbf{q }}^{\dag }+\text{h.c.}\Big],  \label{Hc3qp}
\end{equation}
and 
\begin{equation}
{H}_{\mathrm{c}}^{(4)}=\sum_{\mathbf{q}_{1},\mathbf{q}_{2}}\sum_{m,n}\Big[V_{%
\mathbf{q}_{1}\mathbf{q}_{2}mn}\hat{\alpha}_{m}^{\dag }\hat{\alpha}_{n}\hat{b%
}_{\mathbf{q}_{1}}^{\dag }\hat{b}_{\mathbf{q}_{2}}+\ldots \Big],
\label{Hc4qp}
\end{equation}
provide interactions between them, where the dots in ${H}_{\mathrm{c}}^{(4)}$
denote all other possible terms containing two fermionic and two bosonic
operators with the corresponding matrix elements. The Hamiltonians (\ref%
{Hc3qp}) and (\ref{Hc4qp}) describe interactions between fermionic and
bosonic quasiparticles: The first line in ${H}_{\mathrm{c}}^{(3)}$
corresponds to the emission (absorption) of a phonon by a fermionic
quasiparticle accompanied by a change of its quantum states, $n\rightarrow m$
, while the second line describes processes involving emission (absorption)
of a phonon and annihilation (creation) of a pair of fermionic
quasiparticles. The Hamiltonian ${H}_{\mathrm{c}}^{(4)}$ contains processes
with creation (annihilation) of two fermionic excitations and emission
(absorption) of two phonons.

We will consider the effects of the interaction Hamiltonians ${H}_{\mathrm{c}
}^{(3)}$ and ${H}_{\mathrm{c}}^{(4)}$ in the weak coupling case $n_{M}
a_{s}^{3}<1$ by using systematic perturbation expansion in this small
parameter. In what follows, we limit ourselves to the lowest order
contributions: the first order in ${H}_{\mathrm{c}}^{(4)}$ and the second
order in ${H}_{\mathrm{c}}^{(3)}$.

The interactions between fermionic and bosonic quasiparticles results in
renormalization of their properties. More specifically, the interactions
modify the energies $E_{m}$ and $\epsilon_{q}$ of quasiparticles (adding
also imaginary parts responsible for the decay of quasiparticles when it is
allowed by conservation laws). In the considered case of a weak coupling
between the wire and the reservoir, the renormalization of bosonic and \emph{%
gapped} fermionic bulk excitations ($m=\nu$) does not lead to any
qualitative change in their properties, and we will ignore it. On the other
hand, the properties of the \textquotedblleft\emph{zero-energy}
\textquotedblright edge mode ($m=M$), in particular, its exponentially small
energy $E_{M}$, can be modified substantially due to the coupling to the
gapless phonon modes. Below we will focus on the effects of the effects of
gapless bosonic excitations on the Majorana fermions.

We start our analysis with calculating the effects of ${H}_{\mathrm{c}
}^{(4)}$. The leading first-order contribution can be obtained by averaging
over the bosonic fields in Eq. (\ref{Hc4}), $\delta\hat{\phi}^{\dag}(\mathbf{%
\ r})\delta\hat{\phi}(\mathbf{r})\rightarrow\left\langle \delta\hat{\phi}
^{\dag }(\mathbf{r})\delta\hat{\phi}(\mathbf{r})\right\rangle _{H_{\mathrm{%
ph }}}=n_{\text{M}}^{\prime}$ with $n_{\text{M}}^{\prime}$ being the
condensate depletion, which yields (we omit an unimportant constant) 
\begin{equation*}
H_{\mathrm{c}}^{(4)}\rightarrow\left\langle H_{\mathrm{c}}^{(4)}\right
\rangle _{H_{\mathrm{ph}}}=\sum_{j}\hat{a}_{j}^{\dag}\hat{a}_{j}\int d^{3} 
\mathbf{r}g_{j}(\mathbf{r})n_{\text{M}}^{\prime}.
\end{equation*}
This term provides the renormalization of the fermionic chemical potential $%
\mu_{f}$ in the Kitaev Hamiltonian by replacing the condensate density $%
\left\vert \phi_{0}\right\vert ^{2}$ with the total molecular density $n_{%
\text{M}}=\left\vert\phi_{0}\right\vert ^{2}+n_{\text{M}}^{\prime}$ in Eq. (%
\ref{muf}) for $\mu_{f}$. The corresponding changes can be trivially taken
into account by staring with the renormalized $\mu_{f}$ in the initial
Kitaev Hamiltonian (\ref{HK}).

The interaction Hamiltonian ${H}_{\mathrm{c}}^{(3)}$ contributes in the
second order of the perturbation theory. To select the contributions in ${H}%
_{\mathrm{c}}^{(3)}$ that couple the \textquotedblleft{zero-energy}\textquotedblright edge mode $\hat{\alpha}_{M}$ to other modes, we write the fermionic operator $\hat{a}_{j}$ in the
form 
\begin{align*}
\hat{a}_{j} & =u_{jM}\hat{\alpha}_{M}+\upsilon_{jM}^{\ast}\hat{\alpha}
_{M}^{\dag}+\hat{a}_{j}^{\prime} \\
& =\frac{1}{2}[f_{Lj}+f_{Rj}]\hat{\alpha}_{M}+\frac{1}{2}[f_{Lj}-f_{Rj} ]%
\hat{\alpha}_{M}^{\dag}+\hat{a}_{j}^{\prime},
\end{align*}
where we use Eqs. (\ref{fLj}) and (\ref{fRj}) to express the amplitudes $%
u_{j,M}$ and $v_{j,M}$ in terms of the wave functions of the Majorana edge
modes $f_{Lj}$ and $f_{Rj}$ (we take real $f_{Lj}$ and $f_{Rj}$), and $\hat {%
a}_{j}^{\prime}$ contains the operators $\hat{\alpha}_{\nu}$ and $\hat {%
\alpha}_{\nu}^{\dag}$ of the gapped modes only. The Hamiltonian ${H}_{%
\mathrm{c}}^{(3)}$ then takes the form 
\begin{align}
{H}_{\mathrm{c}}^{(3)} & ={H}_{\mathrm{c1}}^{(3)}+{H}_{\mathrm{c2}} ^{(3)}+{H%
}_{\mathrm{c3}}^{(3)}  \notag \\
& =\sum_{\mathbf{q}}\left[ O_{\mathbf{q}MM}^{(n)}\hat{\alpha}_{M}^{\dag} 
\hat{\alpha}_{M}\hat{b}_{\mathbf{q}}^{\dag}+\text{h.c.}\right]  \notag \\
& +\sum_{\mathbf{q},\nu}\Big[(O_{\mathbf{q}\nu M}^{(n1)}\hat{\alpha}_{\nu
}^{\dag}\hat{\alpha}_{M}+O_{\mathbf{q}\nu M}^{(n2)}\hat{\alpha}_{\nu} \hat{
\alpha}_{M}^{\dag}+2O_{\mathbf{q}\nu M}^{(a1)}\hat{\alpha}_{\nu} \hat{\alpha}
_{M}+2O_{\mathbf{q}\nu M}^{(a2)}\hat{\alpha}_{\nu}^{\dag} \hat{\alpha}
_{M}^{\dag})\hat{b}_{\mathbf{q}}^{\dag}+\text{h.c.} \Big]  \notag \\
& +\sum_{\mathbf{q},\nu,\mu}\left[ (O_{\mathbf{q}\nu\mu}^{(n)}\hat{\alpha}
_{\nu}^{\dag}\hat{\alpha}_{\mu}\!+\!O_{\mathbf{q}\nu\mu}^{(a1)}\hat{\alpha}
_{\nu}\hat{\alpha}_{\mu}\!+\!O_{\mathbf{q}\nu\mu}^{(a2)}\hat{\alpha}_{\nu
}^{\dag}\hat{\alpha}_{\mu}^{\dag})\hat{b}_{\mathbf{q}}^{\dag}\!+\!\text{h.c.}
\right] ,  \label{Hc3expanded}
\end{align}
where the terms in the second line (the Hamiltonian ${H}_{c1}^{(3)}$) couple
phonons to the \textquotedblleft zero-energy\textquotedblright mode $\hat{%
\alpha}_{M}$, the terms in the third line (${H}_{\mathrm{c2}}^{(3)}$)
correspond to the interaction of phonons with the mode $\hat{\alpha}_{M}$
and the gapped modes $\hat{\alpha}_{\nu}$, and the terms in the last line (${%
H}_{\mathrm{c3}}^{(3)}$) describe coupling of phonon to the gapped modes.

With the use of Eqs. (\ref{Hc3}), (\ref{fLj}) and (\ref{fRj}), it is easy to
see that the matrix element $O_{\mathbf{q}MM}^{(n)}$ contains the products
of the Majorana wave functions belonging to different edges, 
\begin{equation*}
O_{\mathbf{q}MM}^{(n)}=\frac{1}{\sqrt{V}}\int d\mathbf{re}^{-i\mathbf{qr}}
\sum_{j}\left\{ \frac{1}{2}[K_{j}(\mathbf{r})\tilde{\upsilon}_{q}-K_{j}
^{\ast}(\mathbf{r})\tilde{u}_{q}](f_{Lj}f_{Rj+1}-f_{Rj}f_{Lj+1})+g_{j} (%
\mathbf{r})[\tilde{u}_{q}\phi_{0}-\tilde{\upsilon}_{q}\phi_{0}^{\ast}
]f_{Lj}f_{Rj}\right\} ,
\end{equation*}
is exponentially small with the system size, $O_{\mathbf{q}MM}^{(n)}\sim
\exp(-L/l_{M})\sim E_{M}$. As a result, the leading (second order)
contribution of ${H}_{\mathrm{c1}}^{(3)}$ to $\delta E_{M}$ is proportional
to $E_{M}^{2}$ and can be neglected. We therefore have to consider only the
Hamiltonians ${H}_{\mathrm{c2}}^{(3)}$ and ${H}_{\mathrm{c3}}^{(3)}$.

The Hamiltonian ${H}_{\mathrm{c2}}^{(3)}$ can be conveniently written in the
form 
\begin{equation}
{H}_{\mathrm{c2}}^{(3)}=(H_{L}^{(-)}+H_{R}^{(-)})\hat{\alpha}_{M}+(H_{L}
^{(+)}+H_{R}^{(+)})\hat{\alpha}_{M}^{\dag},  \label{H3c2}
\end{equation}
where 
\begin{align}
H_{L}^{(+)} & =H_{L}^{(-)}=-H_{L}^{(+)\dag}  \notag \\
& =\frac{1}{2}\sum_{jj^{\prime}}f_{Lj}\int d\mathbf{r}[-K_{jj^{\prime} }( 
\mathbf{r})\delta\hat{\phi}(\mathbf{r})\hat{a}_{j^{\prime}}^{\prime\dag
}+K_{jj^{\prime}}^{\ast}(\mathbf{r})\delta\hat{\phi}^{\dag}(\mathbf{r})\hat {
a}_{j^{\prime}}^{\prime}]+\frac{1}{2}\sum_{j}f_{Lj}\int d\mathbf{r} g_{j}( 
\mathbf{r})(\phi_{0}\delta\hat{\phi}^{\dag}+\phi_{0}^{\ast}\delta \hat{\phi}
)(\hat{a}_{j}^{\prime\dag}-\hat{a}_{j}^{\prime})  \label{HLplus}
\end{align}
and 
\begin{align}
H_{R}^{(+)} & =-H_{R}^{(-)}=H_{R}^{(+)\dag}  \notag \\
& =\frac{1}{2}\sum_{jj^{\prime}}f_{Rj}\int d\mathbf{r}[-K_{jj^{\prime} }(\mathbf{r})\delta\hat{\phi}(\mathbf{r})\hat{a}_{j^{\prime}}^{\prime\dag
}-K_{jj^{\prime}}^{\ast}(\mathbf{r})\delta\hat{\phi}^{\dag}(\mathbf{r})\hat {
a}_{j^{\prime}}^{\prime}]-\frac{1}{2}\sum_{j}f_{Rj}\int d\mathbf{r} g_{j}(
\mathbf{r})(\phi_{0}\delta\hat{\phi}^{\dag}+\phi_{0}^{\ast}\delta \hat{\phi}
)(\hat{a}_{j}^{\prime\dag}+\hat{a}_{j}^{\prime})  \label{HRplus}
\end{align}
[here $K_{j_{1}j_{2}}(\mathbf{r})=-K_{j_{2}j_{1}}(\mathbf{r})\equiv K_{j_{1}
}(\mathbf{r})\delta_{j_{2},j_{1}+1}$] contain the wave function of the left $
f_{Lj}$ and of the right $f_{Rj}$ Majorana modes, respectively, and are
linear in both bosonic operators of the reservoir $\delta\hat{\phi} (\mathbf{
r})$ and $\delta\hat{\phi}^{\dag}(\mathbf{r})$, and in fermionic operators
of the gapped modes $\hat{a}_{j}^{\prime}$ and $\hat{a}_{j} ^{\prime\dag}$.
The relations between the $+$ and $-$ operators suggest another form for ${H}
_{\mathrm{c2}}^{(3)}$, 
\begin{equation}
{H}_{\mathrm{c2}}^{(3)}=\left( H_{L}^{(+)}-H_{R}^{(+)}\right) \hat{\alpha }
_{M}+\left( H_{L}^{(+)}+H_{R}^{(+)}\right) \hat{\alpha}_{M}^{\dag},
\label{H3relevant}
\end{equation}
which will be used below for the analysis of different contributions to $\delta E_{M}$.

\section{Effects of interactions between quasiparticles. Zero temperature}

\label{Effects of InteractionsT0}

In order to calculate the energy correction $\delta E_{M}$ to the energy $
E_{M}$ resulting from the second line of Eq. (\ref{Hc3expanded}), one has to
compare the corrections to the energies of the ground state $\left\vert
0\right\rangle $ and of the state $|M\rangle=\hat{\alpha}_{M}^{\dag}|0
\rangle $ in which only the edge mode is populated, given by 
\begin{equation}
\delta E_{M}=\delta E_{|M\rangle}-\delta E_{|0\rangle}.  \notag
\end{equation}

The corrections to the ground state energy originates from the processes
with simultaneous creation and then annihilation of two fermionic
excitations (the edge mode and a bulk one) and one phonon, described by the $
O_{\mathbf{q}\nu M}^{(a2)}$-term, while the correction to the energy of the
state $|M\rangle$ involves simultaneous annihilation of the edge-mode
excitation and creation of a bulk fermionic excitation and a phonon, $O_{
\mathbf{q}\nu M}^{(n1)}$-term, followed by the reverse process. Direct
application of the perturbation theory yields 
\begin{align}
& \delta E_{M}=\sum_{\mathbf{q},\nu}\frac{\left\vert O_{\mathbf{q}\nu
M}^{(n1)}\right\vert ^{2}}{E_{M}-(E_{\nu}+\epsilon_{q})}-\sum_{\mathbf{q}
,\nu }\frac{4\left\vert O_{\mathbf{q}\nu M}^{(a2)}\right\vert ^{2}}{
-(E_{M}+E_{\nu }+\epsilon_{q})}  \notag \\
& \approx\sum_{\mathbf{q},\nu}\frac{4\left\vert O_{\mathbf{q}\nu M}
^{(a2)}\right\vert ^{2}\!-\!\left\vert O_{\mathbf{q}\nu M}^{(n1)}\right\vert
^{2}}{E_{\nu}+\epsilon_{q}}\!-\!E_{M}\frac{4\left\vert O_{\mathbf{q}\nu
M}^{(a2)}\right\vert ^{2}\!+\!\left\vert O_{\mathbf{q}\nu
M}^{(n1)}\right\vert ^{2}}{(E_{\nu}\!+\!\epsilon_{q})^{2}}\notag \\
& =\delta E_{M}^{(1)}+\delta E_{M}^{(2)},  \label{deltaEM}
\end{align}
where in the second line we have neglected terms $\sim
E_{M}^{2}/\Delta_{m}\ll E_{M}$. It should be mentioned that the Hamiltonian $
{H}_{\mathrm{c3}}^{(3)}$ contributes equally to the energies of the two
states and, hence, the corresponding contributions cancel each other. Note
that the relevant intermediate states contain a phonon and a gapped bulk
excitation such that $E_{\nu}+\epsilon_{q}>\left\vert \Delta\right\vert $ in
the denominators in Eq. (\ref{deltaEM}). Having also in mind that the matrix
elements in the numerators involve the wave functions of the edge modes, we
therefore could expect an exponential decay of $\delta E_{M}$ with the
system size $L$.

Another form of the expression for $\delta E_{M}$ can be obtained by writing
the matrix elements in the form [see Eqs. (\ref{Hc3expanded}) and (\ref
{H3relevant})] 
\begin{align}
O_{\mathbf{q}\nu M}^{(n1)} & =\left\langle \mathbf{q}\nu\left\vert
H_{L}^{(+)}-H_{R}^{(+)}\right\vert 0\right\rangle ,  \label{On1} \\
2O_{\mathbf{q}\nu M}^{(a2)} & =\left\langle \mathbf{q}\nu\left\vert
H_{L}^{(+)}+H_{R}^{(+)}\right\vert 0\right\rangle.  \label{Oa2}
\end{align}
After straightforward calculations we then obtain the following expressions
for $\delta E_{M}^{(1)}$ and $\delta E_{M}^{(2)}$ 
\begin{align}
\delta E_{M}^{(1)}  &  \approx2\sum_{\mathbf{q},\nu}\frac{\left\langle
\mathbf{0}\left\vert H_{R}^{(+)}{}^{\dag}\right\vert \mathbf{q}\nu
\right\rangle \left\langle \mathbf{q}\nu\left\vert H_{L}^{(+)}\right\vert
0\right\rangle +\left\langle \mathbf{0}\left\vert H_{L}^{(+)}{}^{\dag
}\right\vert \mathbf{q}\nu\right\rangle \left\langle \mathbf{q}\nu\left\vert
H_{R}^{(+)}\right\vert 0\right\rangle }{E_{\nu}+\epsilon_{q}}=4\sum
_{\mathbf{q},\nu}\frac{\operatorname{Re}\left\langle \mathbf{0}\left\vert
H_{R}^{(+)}{}^{\dag}\right\vert \mathbf{q}\nu\right\rangle \left\langle
\mathbf{q}\nu\left\vert H_{L}^{(+)}\right\vert 0\right\rangle}{E_{\nu
}+\epsilon_{q}},\label{DE1}\\
\delta E_{M}^{(2)}  &  \approx-2E_{M}\sum_{\mathbf{q},\nu}\frac{\left\vert
\left\langle \mathbf{q}\nu\left\vert H_{L}^{(+)}\right\vert 0\right\rangle
\right\vert ^{2}+\left\vert \left\langle \mathbf{q}\nu\left\vert H_{R}
^{(+)}\right\vert 0\right\rangle \right\vert ^{2}}{(E_{\nu}+\epsilon_{q})^{2}
}, \label{deltaEMnew}
\end{align}which can also be obtained by direct application of the perturbation theory
with the interaction Hamiltonian ${H}_{\mathrm{c2}}^{(3)}$ given by Eq. (\ref
{H3relevant}).

The expressions (\ref{deltaEM}) for $\delta E_{M}$ can be recast into a more
transparent form in terms of correlation functions as

\begin{equation}
\delta E_{M}=-\frac{i}{\hbar}\int_{0}^{\infty}d\tau e^{-\delta\tau}\left\{
\left\langle M\left\vert H_{\mathrm{c2I}}^{(3)}(\tau)H_{\mathrm{c2I}}
^{(3)}(0)\right\vert M\right\rangle -\left\langle 0\left\vert H_{\mathrm{c2I}
}^{(3)}(\tau)H_{\mathrm{c2I}}^{(3)}(0)\right\vert 0\right\rangle \right\} ,
\label{DeltaEMcorrelators}
\end{equation}
where $H_{\mathrm{c2I}}^{(3)}(\tau)$ is the interaction Hamiltonian ${H}_{
\mathrm{c2}}^{(3)}$ in the interaction representation, $H_{\mathrm{c2I}
}^{(3)}(\tau)=e^{iH_{0}\tau/\hbar}{H}_{\mathrm{c2}}^{(3)}e^{-iH_{0}
\tau/\hbar}$, $\,$and $\delta\rightarrow+0$. [Eq. (\ref{deltaEM}) is
recovered after inserting the complete set of intermediate state $|\mathbf{q}
\nu\rangle$ with one bosonic and one gapped fermionic excitation.] This
expression shows that the energy change $\delta E_{M}$ results entirely from
the edges of the wire because, as it was mentioned above, all correlations
in the bulk for the two states $|M\rangle$ and $|0\rangle$ are equal. After
using the form of ${H}_{\mathrm{c2}}^{(3)}$ given by Eq. (\ref{H3relevant}),
the expression (\ref{DeltaEMcorrelators}) can be rewritten as 
\begin{align}
\delta E_{M} & =-\frac{i}{\hbar}\int_{0}^{\infty}d\tau e^{-\delta\tau} 2\cos(
\frac{E_{M}\tau}{\hbar})\left[ \left\langle 0\left\vert H_{L}^{(+)}
{}^{\dag}e^{-\frac{i}{\hbar}H_{0}\tau}H_{R}^{(+)}\right\vert 0\right\rangle
+(L\leftrightarrow R)\right]  \notag \\
& +\frac{i}{\hbar}\int_{0}^{\infty}d\tau e^{-\delta\tau}2i\sin(\frac {
E_{M}\tau}{\hbar})\left[ \left\langle 0\left\vert H_{L}^{(+)}{}^{\dag }e^{-
\frac{i}{\hbar}H_{0}\tau}H_{L}^{(+)}\right\vert 0\right\rangle
+(L\rightarrow R)\right] ,  \label{deltaEintegral}
\end{align}
which provides another form of Eqs. (\ref{DE1}), and (\ref{deltaEMnew}) for $
\delta E_{M}^{(1)}$ and $\delta E_{M}^{(2)}$.

\begin{figure}[th]
\includegraphics[width=0.472\textwidth]{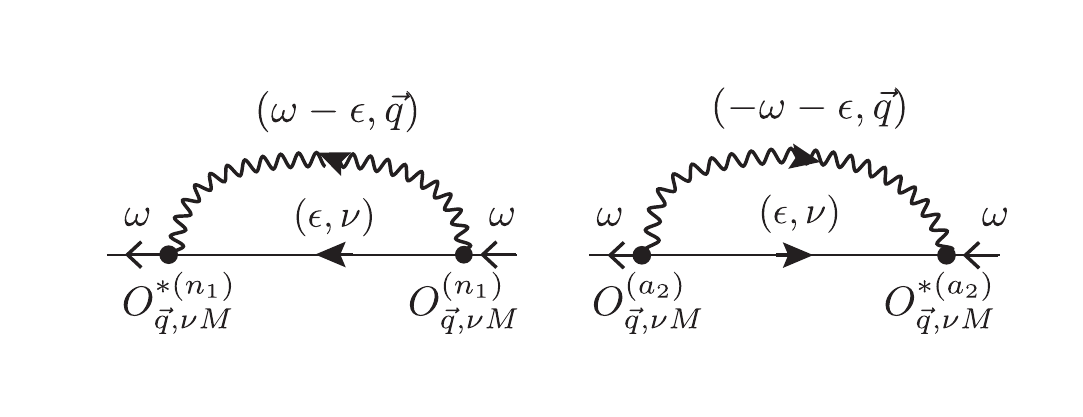}
\caption{Feynman diagrams for normal contributions to the self-energy of the 
$\protect\alpha$-mode at zero temperature. Solid lines correspond to gapped
fermionic excitations in the wire and wavy lines to bosinic excitations in
the molecular condensate.}
\label{Sigma0N}
\end{figure}

The above results show that the correction to the energy of the $\alpha_{M}$
mode involve two different types of correlations: The contribution $\delta
E_{M}^{(1)}$ involves long-range correlations between different edges, while
the contribution $\delta E_{M}^{(2)}$ contains local correlations at the
edges. As a result, the system-size dependence of $\delta E_{M}^{(2)}$
originates from the $L$-dependence of the energy $E_{M}$ of the mode, $
\delta E_{M}^{(2)}\sim E_{M}\sim\exp(-La/l_{M})$, while the $L$-dependence
of $\delta E_{M}^{(1)}$ results from the combined effect of the localization
of the edge modes described by $l_{M}$, and of the short-range correlations
in the bulk of the wire described by the coherence length $\xi_{\text{BCS}
}=l_{M}$ (see Appendix \ref{AppendixB}). We therefore also expect that the
leading dependence of $\delta E_{M}^{(1)}$ on the system size is
exponential, $\delta E_{M}^{(1)}\sim\exp(-La/l_{M})$.

An alternative derivation of the energy splitting $\delta E_{M}$ is based on
the Green's function technique (see, for example, Ref. \cite{Fetter}), which
applies to both the zero temperature and finite temperature regime which we
will discuss later. In the Green's function approach, the energies of
excitation correspond to the poles of the Green's function considered as a
function of the frequency $\omega $. The Green's function for the
\textquotedblleft zero-energy\textquotedblright edge mode $\alpha _{M}$ is
defined as 
\begin{eqnarray}
G_{M}(\tau ) &=&-i\left\langle \mathrm{T}\{\alpha _{M}(\tau )\alpha
_{M}^{\dag }(0)\}\right\rangle  \notag \\
&=&\int \frac{d\varepsilon }{2\pi \hbar }G_{M}(\varepsilon )\exp (-\frac{i}{
\hbar }\varepsilon \tau ).  \notag
\end{eqnarray}%
Here, $\mathrm{T}\{\alpha _{M}(\tau )\alpha _{M}^{\dag }(0)\}$ is the
time-ordered product of Heisenberg operators $\alpha _{M}(\tau )$ and $
\alpha _{M}(0)=\alpha _{M}$, where the evolution is defined by the
Hamiltonian ${H}^{(0)}+{H}_{\mathrm{c2}}^{(3)}+{H}_{\mathrm{c2}}^{(3)}$, and
the averaging is over the exact ground state of this Hamiltonian. The
Green's function $G_{M}(\varepsilon )$ can be found from the Dyson equation 
\begin{equation}
G_{M}^{-1}(\varepsilon )=G_{M}^{(0)-1}(\varepsilon )-\Sigma _{M}(\varepsilon
)=\varepsilon -E_{M}-\Sigma _{M}(\varepsilon ),  \notag
\end{equation}
where $G_{M}^{(0)}(\varepsilon )=(\varepsilon -E_{M}+i0)^{-1}$ is the bare
Green's function and $\Sigma _{M}(\varepsilon )$ is the self-energy of the $
\alpha $-mode, such that finding the renormalized energy of the $\alpha$
-mode reduces to solving the equation 
\begin{equation}
\varepsilon -E_{M}-\Sigma _{M}(\varepsilon )=0.  \label{pole0}
\end{equation}

At zero temperature and in the considered second-order of the perturbation
theory, the self-energy $\Sigma _{M}(\varepsilon )$ results from only two
normal (with one incoming and one outgoing lines of the $\alpha $-mode)
contributions as illustrated in Fig. \ref{Sigma0N}. There, the solid line
corresponds to the bare Green's function of a gapped fermionic excitation $
G_{\nu }^{(0)}(\varepsilon )=(\varepsilon -E_{\nu }+i0)^{-1}$ and the wavy
line to the bare bosonic excitation $D_{\mathbf{q}}^{(0)}(\varepsilon
)=(\varepsilon -\epsilon _{q}+i0)^{-1}$. The corresponding analytic
expression of $\Sigma _{M}(\varepsilon )=\Sigma _{M}^{(n)}(\varepsilon )$
reads 
\begin{equation}
\Sigma _{M}(\varepsilon )=\sum_{\mathbf{q},\nu }\frac{\left\vert O_{\mathbf{q
}\nu M}^{(n1)}\right\vert ^{2}}{\varepsilon -(E_{\nu }+\epsilon _{q})}+\sum_{
\mathbf{q},\nu }\frac{4\left\vert O_{\mathbf{q}\nu M}^{(a2)}\right\vert ^{2}
}{\varepsilon +E_{\nu }+\epsilon _{q}}.  \label{SigmaM0}
\end{equation}
After solving Eq. (\ref{pole0}) to the lowest order in the perturbation, 
\begin{equation}
\varepsilon \approx E_{M}+\Sigma _{M}^{(n)}(E_{M}),  \label{omega_solution}
\end{equation}
we recover the expression (\ref{deltaEM}) for $\delta E_{M}$.

\begin{figure}[th]
\includegraphics[width=0.472\textwidth]{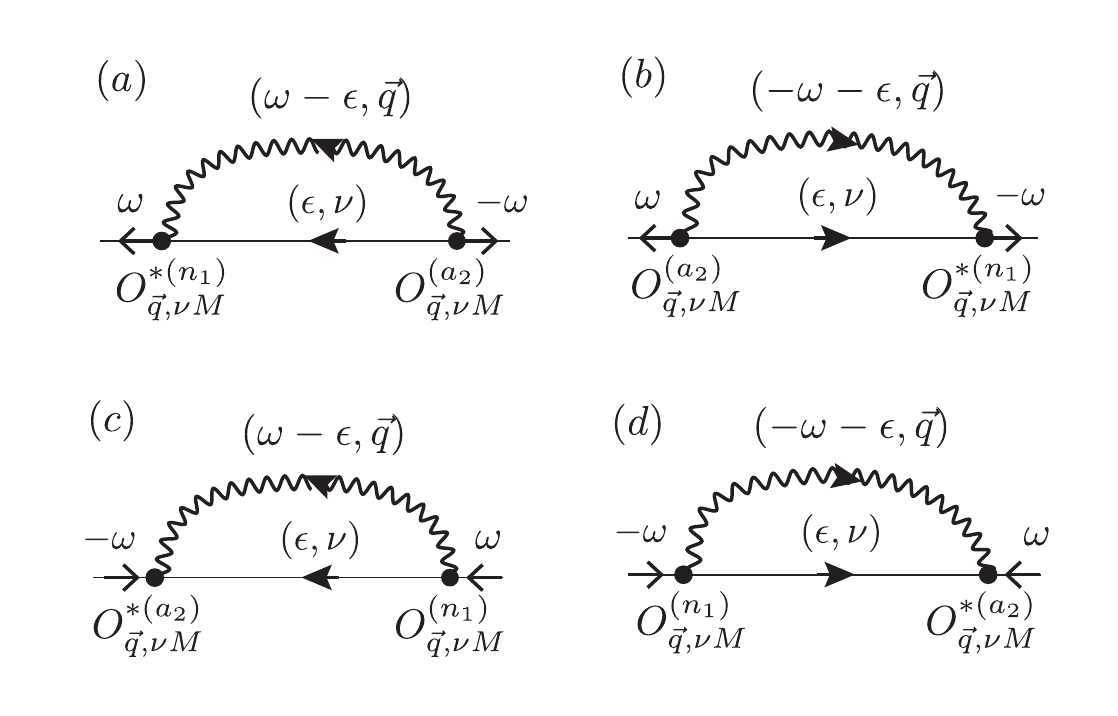}
\caption{Feynman diagrams for the anomalous terms $\Delta_{M}$ [(a) and (b)]
and $\Delta_{M}^{\ast}$ [(c) and (d)]\ for the $\protect\alpha$-mode. Solid
lines correspond to gapped fermionic excitations in the wire and wavy lines
to bosinic excitations in the molecular condensate.}
\label{Anomalous}
\end{figure}

It should be mentioned that the terms in $H_{\mathrm{c}}^{(3)}$ with matrix
elements $O_{\mathbf{q}\nu M}^{(a1)}$ and $O_{\mathbf{q}\nu M}^{(a2)}$ [see
Eq. (\ref{Hc3expanded})] generate in the second order also the \textquotedblleft{anomalous}\textquotedblright 
terms with two $\alpha _{M}$-lines going out [$\Delta _{M}(\varepsilon )$,
see Figs. \ref{Anomalous}(a) and \ref{Anomalous}(b)], or going in [$\Delta
_{M}^{\ast }(\varepsilon )$, see Figs. \ref{Anomalous}(c) and \ref{Anomalous}
(d)], which contribute to the \textquotedblleft{anomalous}\textquotedblright
part of the self-energy $\Sigma _{M}^{(a)}(\varepsilon )=\left\vert \Delta
_{M}(\varepsilon )\right\vert ^{2}[\varepsilon +E_{M}+\Sigma
_{M}^{(n)}(\varepsilon )]^{-1}$. These \textquotedblleft
anomalous\textquotedblright terms, however, are proportional to the
frequency, $\Delta _{M}(\varepsilon )\sim $ $\varepsilon $ for small $
\varepsilon $ (as a consequence of the Fermi-Dirac statistics) and,
therefore, do not affect the leading second-order solution (\ref
{omega_solution}) or (\ref{deltaEM}) of the equation (\ref{pole0}). In other
words, these frequency-proportional anomalous terms do not \textquotedblleft
open a gap\textquotedblright. This is in contrast to the standard pairing
case where frequency-independent anomalous terms open a finite gap $\sim
\left\vert \Delta \right\vert ^{2}$ which does not depend on the size of the
system.

\section{Evaluation of the energy correction}

\label{EnergyCorrection}

We now proceed with evaluation of the energy correction $\delta E_{M}$. To
be specific, we assume the following properties of the condensate: $
a_{s}\sim a$ and $n_{\text{M}}a_{s}^{3}\lesssim1$ (but not $\ll1$), which
correspond to an optimum atom-molecule conversion with a sufficiently large
amplitude $K$, see Appendix \ref{AppendixA}. In this case, $\xi_{\text{BEC}
}\sim a$ and, as a result, only phonon excitations with $q\lesssim\xi_{\text{
BEC}}^{-1}$ in the condensate are relevant for processes at distances $
\gtrsim a$. For the wire we assume that ${\alpha}\equiv\left\vert
\Delta\right\vert /J$ and $\beta\equiv\mu/2J$ satisfy the conditions $
\alpha\lesssim1$ and $\left\vert \beta\right\vert \leq1-\alpha ^{2}$, such
that the bulk quasiparticle spectrum $E_{\nu}\rightarrow E_{k}=2J\sqrt{(\cos
ka+\beta)^{2}+\alpha^{2}\sin^{2}ka}$ has two minima $\Delta_{m}=2\left\vert
\Delta\right\vert \sqrt{1-\beta^{2}/(1-\alpha^{2})}>0$ at $k=\pm k_{F}=\pm
a^{-1}\arccos[-\beta/(1-\alpha^{2})]$ inside the Brillouin zone $-\pi/a\leq
k\leq\pi/a$. (We neglect the effects of the boundary on the properties of
the extended wave functions in the bulk.) Under these conditions we can use
the local approximation for the fermionic-bosonic couplings in Eqs. (\ref
{HLplus}) and (\ref{HRplus}): $K_{j}(\mathbf{r})\rightarrow K_{0} \delta(
\mathbf{r-r}_{j})$ and $g_{j}(\mathbf{r})\rightarrow g_{0} \delta(\mathbf{r-r
}_{j})$ with real $K_{0}=\int d\mathbf{r}K_{j}(\mathbf{r})$ and $g_{0}=\int d
\mathbf{r}g_{j}(\mathbf{r})$, respectively, and the standard BCS expressions
for the wave functions (Bogoliubov amplitudes $u_{k}$ and $v_{k}$) of the
gapped fermionic modes.

In the following we evaluate the energy correction $\delta E_{M}^{(1)}$
which dominates for large $L$ [as we will see, $\delta E_{M}^{(1)}\sim L\exp
(-La/l_{M})$ as compared to $\delta E_{M}^{(2)}\sim
E_{M}\sim\exp(-La/l_{M}), $ see Eq.(\ref{deltaEMnew})]. With the usage of
Bogoliubov transformations (\ref{BogoliubovBose}) and (\ref{BogoliubovFermi}
), the matrix elements entering expression Eq. (\ref{DE1}) for $\delta
E_{M}^{(1)}$ can be written in the form (with $\nu=k$) 
\begin{eqnarray}
\left\langle\mathbf{q}k\left\vert H_{L}^{(+)}\right\vert 0\right\rangle&=& 
\frac{1}{2\sqrt{LV}}\sum_{j}f_{Lj}\int d\mathbf{r} e^{-i\mathbf{qr}}\left[
\sum_{j^{\prime}}K_{jj^{\prime}}(\mathbf{r} )e^{-ikj^{\prime}a}(\tilde{
\upsilon}_{q}u_{k}-\tilde{u}_{q}v_{k} )+g_{j}(\mathbf{r})\phi_{0}(\tilde{u}
_{q}-\tilde{\upsilon}_{q})(u_{k} +v_{k})e^{-ikja}\right],  \notag \\
\left\langle\mathbf{0}\left\vert H_{L}^{(+)\dag}\right\vert \mathbf{q}
k\right\rangle&=&-\left\langle \mathbf{q}k\left\vert H_{L}^{(+)}\right\vert
0\right\rangle^{*},  \notag \\
\left\langle \mathbf{q}k\left\vert H_{R}^{(+)}\right\vert 0\right\rangle&=& 
\frac{1}{2\sqrt{LV}}\sum_{j}f_{Rj}\int d\mathbf{r} e^{-i\mathbf{qr}}\left[
\sum_{j^{\prime}}K_{jj^{\prime}}(\mathbf{r} )e^{-ikj^{\prime}a}(\tilde{
\upsilon}_{q}u_{k}+\tilde{u}_{q}v_{k} )-g_{j}(\mathbf{r})\phi_{0}(\tilde{u}
_{q}-\tilde{\upsilon}_{q})(u_{k} -v_{k})e^{-ikja}\right],  \notag \\
\left\langle \mathbf{0}\left\vert H_{R}^{(+)\dag}\right\vert \mathbf{q}
k\right\rangle&=&\left\langle \mathbf{q}k\left\vert H_{R} ^{(+)}\right\vert
0\right\rangle^{*},  \notag
\end{eqnarray}
where we consider $K_{jj^{\prime}}(\mathbf{r})$ and $\phi_{0}$ to be real.
As we have mentioned before, the main contribution comes from the
phonon-part of the bosonic excitation spectrum with wave vectors $q\lesssim
a^{-1}$ ($\sim \xi_{\text{BEC}}^{-1}$), for which $\tilde{u}_{q}\approx
\tilde{\upsilon}_{q} \approx(1/2)\sqrt{\varepsilon_{q}/\varepsilon_{q}^{(0)}}
\equiv f_{q}\sim q^{-1/2}$. This, together with the local approximations for 
$K_{jj^{\prime} }(\mathbf{r})$ and $g_{j}(\mathbf{r})$, allows us to write
the matrix elements in a simpler form: 
\begin{eqnarray}
\left\langle \mathbf{q}k\left\vert H_{L}^{(+)}\right\vert 0\right\rangle&&
\approx\frac{-iK_{0}}{\sqrt{LV}}f_{q}(u_{k}-v_{k})\sin ka\sum_{j}
f_{Lj}e^{-i(q_{x}+k)ja},  \label{melementHL} \\
\left\langle \mathbf{q}k\left\vert H_{R}^{(+)}\right\vert 0\right\rangle&&
\approx\frac{-iK_{0}}{\sqrt{LV}}f_{q}(u_{k}+v_{k})\sin ka\sum_{j}
f_{Rj}e^{-i(q_{x}+k)ja}.  \label{melementHR}
\end{eqnarray}
The expression for $\delta E_{M}^{(1)}$ now reads 
\begin{eqnarray}
\delta E_{M}^{(1)}&\approx&4K_{0}^{2}\int\frac{d\mathbf{q}}{(2\pi)^{3}}
\int_{-\pi/a}^{\pi/a}\frac{adk}{2\pi}\frac{f_{q}^{2}(u_{k}+v_{k})^{2}\sin
^{2}ka}{E_{k}+\hbar cq}\sum_{j_{1},j_{2}}f_{Lj_{1}}f_{Rj_{2}}e^{i(q_{x}
+k)(j_{1}-j_{2})a}  \notag \\
&=&4K_{0}^{2}\int\frac{d\mathbf{q}}{(2\pi)^{3}}\int_{-\pi/a}^{\pi/a} \frac{
adk}{2\pi}\frac{f_{q}^{2}}{E_{k}+\hbar cq}\frac{\xi_{k}+2i\Delta\sin ka}{
E_{k}}\sin^{2}ka\sum_{j_{1},j_{2}}f_{Lj_{1}}f_{Rj_{2}}e^{i(q_{x}
+k)(j_{1}-j_{2})a}  \notag \\
&=&-4K_{0}^{2}\int\frac{d\mathbf{q}}{(2\pi)^{3}}\int_{-\pi/a}^{\pi/a} \frac{
adk}{2\pi}\frac{f_{q}^{2}\sin^{2}ka}{E_{k}+\hbar cq}\sqrt{\frac{\xi
_{k}+2i\Delta\sin ka}{\xi_{k}-2i\Delta\sin ka}}\sum_{j_{1},j_{2}}f_{Lj_{1}
}f_{Rj_{2}}e^{i(q_{x}+k)(j_{1}-j_{2})a},  \notag
\end{eqnarray}
where we use the expressions (\ref{bulk u and v}) for $u_{k}$ and $v_{k}$.

We start the evaluation of this expression with integration over $\mathbf{q}$
: 
\begin{eqnarray}
\int\frac{d\mathbf{q}}{(2\pi)^{3}}\frac{f_{q}^{2}}{E_{k}+\hbar cq}
e^{iq_{x}(j_{1}-j_{2})a}&=&\frac{m}{2\hbar^{2}}\int\frac{d\mathbf{q}} {
(2\pi)^{3}}\frac{1}{q}\frac{e^{iq_{x}(j_{1}-j_{2})a}}{q+\lambda E_{k}/2J} 
\notag \\
&=&\frac{m}{4\pi^{2}\hbar^{2}a}\frac{1}{\left\vert j_{1}-j_{2}\right\vert}
\int_{0}^{\infty}dq\frac{\sin(qa\left\vert j_{1}-j_{2}\right\vert )}{q+\lambda E_{k}/2J}  \notag \\
&\approx&\frac{m}{8\pi\hbar^{2}a}\frac{1}{\left\vert j_{1}-j_{2}\right\vert }
\frac{1}{1+\pi\lambda\left\vert j_{1}-j_{2}\right\vert E_{k}/4J},
\end{eqnarray}
where $\lambda=2Ja/\hbar c$ and the last line (an interpolation between the
limiting cases of small and large $\lambda\left\vert j_{1}-j_{2}\right\vert
E_{k}/J$) provides a very good approximation to the integral. Note that the
result of the integration diverges for $j_{1}=j_{2}$. This divergence is not
physical because it originates from the fact that the coupling $K_{0}$
between molecules in the reservoir and atoms in the lattice is $q$
-independent. In reality however, the coupling disappears for large $q$
because the molecular kinetic energy breaks the resonance condition for the
conversion of a molecule into a pair of atoms. This effectively limits the
integration to $q\lesssim a_{s}^{-1}\sim a$, and we can therefore estimate
the value of the integral for $j_{1}=j_{2}$ as $m/8\pi\hbar^{2}a$.

In performing the integration over $k$ we notice that the parameter $\lambda 
$ is a ratio between the Fermi velocity (when $\mu \approx 0$) and the sound
velocity, and under assumed conditions (see Appendix \ref{AppendixA}), we
have $\lambda \ll 1$. For this reason, the term $\pi \lambda \left\vert
j_{1}-j_{2}\right\vert E_{k}/4J$ becomes comparable with unity only for
large $\left\vert j_{1}-j_{2}\right\vert $, for which fermionic correlations
are already exponentially suppressed, see Appendix \ref{AppendixB}. We can
therefore neglect this term and write the expression for $\delta E_{M}^{(1)}$
in the form 
\begin{equation}
\delta E_{M}^{(1)}=-\frac{mK_{0}^{2}}{2\pi \hbar ^{2}a}\sum_{j_{1},j_{2}} 
\frac{f_{Lj_{1}}f_{Rj_{2}}}{\left\vert j_{1}-j_{2}\right\vert }\int_{-\pi
/a}^{\pi /a}\frac{adk}{2\pi }\sin ^{2}ka\sqrt{\frac{\xi _{k}+2i\Delta \sin
ka }{\xi _{k}-2i\Delta \sin ka}}e^{ik(j_{1}-j_{2})a},  \notag
\end{equation}
where $\left\vert j_{1}-j_{2}\right\vert $ has to be replaced with $1$ for $
j_{1}=j_{2}$. To perform the $k$-integration, we split the summation over $
j_{1}$ and $j_{2}$ into three parts: $j_{1}=j_{2}$, $j_{1}>j_{2}$, and $
j_{1}<j_{2}$, and denote the corresponding contributions to $\delta
E_{M}^{(1)}$ as $I_{0}$, $I_{+}$, and $I_{-}$, respectively, 
\begin{eqnarray}
\delta E_{M}^{(1)} &=&I_{0}+I_{+}+I_{-}  \notag \\
&=&\frac{mK_{0}^{2}}{2\pi \hbar ^{2}a}\left\{
\sum_{j}f_{Lj}f_{Rj}K_{0}+
\sum_{j_{1}>j_{2}}f_{Lj_{1}}f_{Rj_{2}}K_{+}(j_{1}-j_{2})+
\sum_{j_{1}<j_{2}}f_{Lj_{1}}f_{Rj_{2}}K_{-}(\left\vert
j_{1}-j_{2}\right\vert )\right\}  \notag
\end{eqnarray}
with 
\begin{eqnarray}
K_{0} &=&-\int_{-\pi /a}^{\pi /a}\frac{adk}{2\pi }\sin ^{2}ka\sqrt{\frac{\xi
_{k}+2i\Delta \sin ka}{\xi _{k}-2i\Delta \sin ka}},  \label{K0} \\
K_{\pm }(s) &=&-\frac{1}{s}\int_{-\pi /a}^{\pi /a}\frac{adk}{2\pi }\sin
^{2}ka\sqrt{\frac{\xi _{k}+2i\Delta \sin ka}{\xi _{k}-2i\Delta \sin ka}}
e^{\pm iksa}.  \label{Kplusminus}
\end{eqnarray}
The integrals over $k$ can be calculated in the same way as in Appendix \ref
{AppendixB}: After introducing the variable $z=\exp (-ia)$, the integrals
over $k$ are transformed into integrals over the unit circle $\mathrm{S}_{1}$
in the complex plane of $z$ (see Fig. \ref{Pole} in Appendix \ref{AppendixB}
), and for the kernels $K_{0}$ and $K_{\pm }$ we obtain 
\begin{eqnarray}
K_{0} &=&\oint_{\mathrm{S}_{1}}\frac{dz}{2\pi iz}\frac{(z-z^{-1})^{2}}{4}
\rho \sqrt{\frac{(z-x_{+}\rho ^{-2})(z-x_{-}\rho ^{-2})}{(z-x_{+})(z-x_{-})}}
,  \notag \\
K_{\pm }(s) &=&\frac{1}{s}\oint_{\mathrm{S}_{1}}\frac{dz}{2\pi iz}\frac{
(z-z^{-1})^{2}}{4}\rho ^{-1}\left[ \frac{(z-x_{+}\rho ^{-2})(z-x_{-}\rho
^{-2})}{(z-x_{+})(z-x_{-})}\right] ^{\pm 1/2}z^{s}.
\end{eqnarray}
The contour $\mathrm{S}_{1}$ is then deformed into the contour around the
cut \textrm{C}$_{1}$ (see Fig. \ref{Pole}), which connects the points $
x_{+}=\rho \exp (i\theta )$ and $x_{-}=\rho \exp (-i\theta )$ inside $
\mathrm{S}_{1}$ (during this deformation we also pick up the contributions
from the pole at $z=0$ in $K_{0}$ and in $K_{\pm }(s)$ for $s=1$ and $s=2$),
and for $\rho \ll 1$ we obtain 
\begin{align*}
K_{0}& \approx \frac{3}{4}\rho \cos \theta , \\
K_{+}(1)& \approx -\frac{1}{2},\ K_{+}(2)\approx -\frac{3}{8}\rho \cos
\theta ,\ K_{+}(s>2)\approx \frac{1}{4s}\oint_{\mathrm{C}_{1}}\frac{dz}{2\pi
i}\frac{z^{s-3}}{\sqrt{(z-x_{+})(z-x_{-})}}, \\
K_{-}(1)& \approx \frac{1}{4},\ K_{-}(2)\approx -\frac{1}{8}\rho \cos \theta
,\ K_{-}(s>2)\approx \frac{1}{4s}\oint_{\mathrm{C}_{1}}\frac{dz}{2\pi i} 
\sqrt{(z-x_{+})(z-x_{-})}z^{s-3},
\end{align*}
where we keep only the leading terms in $\rho \ll 1$ for $s=1$ and $s=2$,
and the leading powers in small $z$ in the integrals. The integrals can be
performed in the same way as in Appendix \ref{AppendixB} with the results 
\begin{align*}
\oint_{\mathrm{C}_{1}}\frac{dz}{2\pi i}\frac{1}{\sqrt{(z-x_{+})(z-x_{-})}}
z^{s-3}& =\rho ^{N-3}\frac{1}{\pi }\int_{0}^{\pi }d\phi (\cos \theta +i\sin
\theta \cos \phi )^{s-3}=\rho ^{s-3}\mathrm{P}_{s-3}^{0}(\cos \theta ) \\
& \overset{s\gg 1}{\rightarrow }\rho ^{s-3}\sqrt{\frac{2}{\pi s\sin \theta }}
\cos [(s-\frac{5}{2})\theta -\frac{\pi }{4}]
\end{align*}
and 
\begin{align*}
\oint_{\mathrm{C}_{1}}\frac{dz}{2\pi i}\sqrt{(z-x_{+})(z-x_{-})}z^{s-3}&
=\sin ^{2}\theta \rho ^{s-1}\frac{1}{\pi }\int_{0}^{\pi }d\phi \sin ^{2}\phi
(\cos \theta +i\sin \theta \cos \phi )^{s-3}=\rho ^{s-1}\sin \theta \mathrm{P%
}_{N-3}^{-1}(\cos \theta ) \\
& \overset{s\gg 1}{\rightarrow }\rho ^{s-1}\frac{\sin \theta }{s}\sqrt{\frac{%
2}{\pi s\sin \theta }}\sin [(s-\frac{5}{2})\theta -\frac{\pi }{4}]
\end{align*}
for integer $s>2$ with $\mathrm{P}_{\mu }^{\nu }(x)$ being the associate
Legendre function [$\mathrm{P}_{n}^{0}(x)=\mathrm{P}_{n}(x)$ is the Legendre
polynomial of degree $n$], which give an exponential decay of the kernels $%
K_{\pm }(s)$ for $s\gg 1$.

Using the expression 
\begin{align}
f_{Lj_{1}}f_{Rj_{2}} & =4\left\vert A\right\vert
^{2}\rho^{L+1+j_{1}-j_{2}}\sin(j_{1}\theta)\sin[(L+1-j_{2})\theta]
\label{fLfR} \\
& =2\left\vert A\right\vert
^{2}\rho^{L+1+j_{1}-j_{2}}[\cos(L+1-j_{1}-j_{2})\theta-\cos(L+1+j_{1}-j_{2})%
\theta]  \notag
\end{align}
for the product of the Majorana wave functions $f_{Lj_{1}}$ and $f_{Rj_{2}}$%
, we can easily calculate the leading contribution to $I_{0}$:%
\begin{equation}
I_{0}\approx-\frac{mK_{0}^{2}}{2\pi\hbar^{2}a}2\left\vert A\right\vert
^{2}L\rho^{L+1}\cos[(L+1)\theta]\frac{3}{4}\rho\cos\theta,  \label{I0final}
\end{equation}
where we neglect the sum over oscillating with $j$ terms. To calculate $%
I_{\pm}$ we first perform summation over $j_{2}$ for a fixed $j_{1}$ and
then over $j_{1}$: 
\begin{align*}
I_{+}+I_{-} & =\frac{mK_{0}^{2}}{2\pi\hbar^{2}a}\sum_{j_{1}=2}^{L}\left[
\sum_{1\leq j_{2}<j_{1}}f_{Lj_{1}}f_{Rj_{2}}K_{+}(j_{1}-j_{2})+\sum_{L\geq
j_{2}>j_{1}}f_{Lj_{1}}f_{Rj_{2}}K_{-}(\left\vert j_{1}-j_{2}\right\vert)%
\right] \\
& \approx\frac{mK_{0}^{2}}{2\pi\hbar^{2}a}\sum_{j_{1}=1}^{L}\left[
\sum_{s=1}^{\infty}f_{Lj_{1}}f_{Rj_{1}-s}K_{+}(s)+\sum_{s=1}^{
\infty}f_{Lj_{1}}f_{Rj_{1}+s}K_{-}(s)\right] ,
\end{align*}
where we extend the summation over $s=\left\vert j_{1}-j_{2}\right\vert $ to
infinity because of the fast convergency of the sums (as we will see below,
the main contribution comes from $j_{1}$ being in the bulk). Keeping in mind
the asymptotic behavior of the kernels $K_{\pm}(s)$ for $s\gg1$, 
\begin{equation*}
K_{\pm}(s)\sim s^{-\alpha_{\pm}}\rho^{s-2\mp1}\{\exp[i(s-\frac{5}{2})\theta-i \frac{\pi}{4}]\pm\exp[i(s-\frac{5}{2})\theta-i\frac{\pi}{4}]\},
\end{equation*}
where $\alpha_{+}=3/2$ and $\alpha_{-}=5/2$, the leading ($\sim L$)
contribution to $I_{+}+I_{-}$ reads 
\begin{align*}
I_{+}+I_{-} & =-\frac{mK_{0}^{2}}{2\pi\hbar^{2}a}2\left\vert A\right\vert
^{2}L\rho^{L+1} \\
& \times\sum_{s=1}^{\infty}\left\{ \rho^{s}\cos[(L+1+s)\theta]
K_{+}(s)+\rho^{-s}\cos[(L+1-s)\theta]K_{-}(s)\right\}
\end{align*}
(we neglect the sum over terms which oscillate with $j_{1}$). Note that the
convergency of the second sum is due to the factor $s^{-\alpha_{\pm}}=s^{-5/2}$ in the asymptotics of $K_{-}(s)$ (the factors depending on $s$ exponentially cancel each other). The leading for small $\rho$
contribution comes from the second term in the sum [with $K_{-}(s)$] and is 
\begin{align}
I_{+}+I_{-} & \approx-\frac{mK_{0}^{2}}{2\pi\hbar^{2}a}2\left\vert
A\right\vert ^{2}L\rho^{L+1}\frac{1}{\rho}\left\{ \cos(L\theta)\frac{1}{4}-\cos[(L+1)\theta]\frac{\cos\theta}{8}\right\}  \notag \\
& =-\frac{mK_{0}^{2}}{2\pi\hbar^{2}a}2\left\vert A\right\vert ^{2}L\rho ^{L}\frac{3\cos(L\theta)-\cos[(L+2)\theta]}{16},  \label{Ipmfinal}
\end{align}
where we keep only numerically dominant terms with $s=1$ and $s=2$. After
comparing Eqs. (\ref{I0final}) and (\ref{Ipmfinal}), we finally obtain for
the energy correction at zero temperature in the considered regime $4J^{2}-4\Delta^{2}-\mu^{2}>0$ and $0<J-\Delta\ll J+\Delta$ 
\begin{align}
\delta E_{M} & \approx\delta E_{M}^{(1)}\approx-\frac{mK_{0}^{2}}{16\pi
\hbar^{2}a}\frac{\Delta(4J^{2}-\mu^{2})}{J(4J^{2}-4\Delta^{2}-\mu^{2})}
\{3\cos(L\theta)-\cos[(L+2)\theta]\}Le^{-La/l_{M}}  \label{deltaEMT0} \\
& \sim\frac{ma^{2}}{\hbar^{2}}\Delta^{2}\frac{1}{n_{M}a^{3}}
Le^{-aL/l_{M}}\sim E_{M}\left( \frac{ma^{2}\Delta}{\hbar^{2}}\right) \frac{1 
}{n_{M}a^{3}}L\sim E_{M}\frac{\Delta}{E_{R}}L.  \label{deltaEMT0estimate}
\end{align}

This result shows that the energy correction due to quantum fluctuations
remains exponentially small with the length of the wire $L$, but contains an
extra linear dependence on $L$. It therefore dominates over $E_{M}$ for
sufficiently large $L$. For values of the ratio $\Delta/E_{R}$ of the order
of $10^{-2}$ (see, for example, Ref. \cite{Nascimbene} and Appendix \ref{AppendixA}), this happens for $L\gtrsim10^{2}$. For such values of $L$,
however, $E_{M}$ itself becomes practically zero provided the localization
length of the Majorana states $l_{M}$ is of the order of a few lattice
spacing. We can therefore conclude that in any practical discussion in which
the finite value of $E_{M}$ becomes an issue (for example, in determining
the lower bound for adiabatic operations with Majorana states), one can
ignore the correction due to quantum fluctuations and use the zero-order
value, Eq. (\ref{EMexact1}).

\section{Effects of interactions between quasiparticles. Finite temperatures}

\label{Temperature}

Let us now turn to the case of finite but small temperatures $T\ll\left\vert
\Delta\right\vert $. Note that because $\left\vert \Delta\right\vert \ll
E_{b}\sim E_{R}$, we can completely ignore the processes of molecular
dissociation and vortex formation in the condensate such that the only
relevant excitation in the reservoir are bosonic excitations described by
the operators $\hat{b}_{\mathbf{q}}$. This implies that the parity of the
wire is conserved.

The studies of temperature effects are most easily done using Matsubara
technique (see, for example \cite{Fetter}), in which one calculates the
Matsubara Green's function $G_{TM}(i\varepsilon_{n})$ of the mode $%
\alpha_{M} $ as a function of Matsubara frequencies $\varepsilon_{n}=\pi
T(1+2n)$. Being analytically continued in the upper half-plane of (complex)
frequency $i\varepsilon_{n}\rightarrow\varepsilon+i0$ from Matsubara $%
\varepsilon_{n}$ to real frequency $\varepsilon$, one obtains the retarded
Green's function $G_{M}^{R}(\varepsilon)$. The pole of this function is in
general at some complex frequency $\varepsilon_{\ast}=\varepsilon_{\ast}^{%
\prime} +i\varepsilon_{\ast}^{\prime\prime}$ with $\varepsilon_{\ast}^{%
\prime}$ determining the eigenenergy and $\varepsilon_{\ast}^{\prime%
\prime}=1/\tau$ the life-time $\tau$ of the mode.

\begin{figure}[th]
\includegraphics[width=0.472\textwidth]{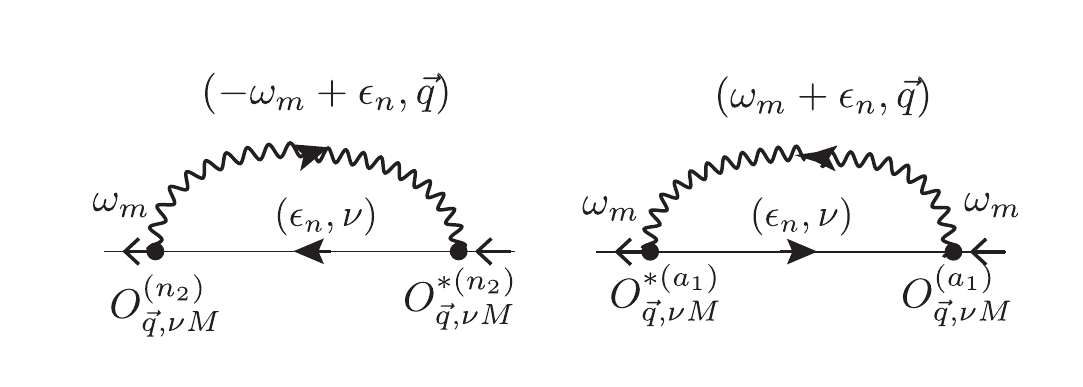}
\caption{Additions Feynman diagrams for normal contributions to the
self-energy of the $\protect\alpha$-mode at finite temperatures. Solid lines
correspond to gapped fermionic excitations in the wire and wavy lines to
bosinic excitations in the molecular condensate.}
\label{SigmaTN}
\end{figure}

The calculation of the Matsubara Green's function is very similar to that of
the Green's function at zero temperature and based on the Dyson equation 
\begin{equation*}
G_{TM}^{-1}(i\varepsilon_{n})=G_{TM}^{(0)-1}(i\varepsilon_{n})-\Sigma
_{TM}(i\varepsilon_{n})=i\varepsilon_{n}-E_{M}-\Sigma_{TM}(i\varepsilon_{n})
\end{equation*}
with the Matsubara self-energy $\Sigma_{TM}(i\varepsilon_{n})$ and $%
G_{TM}^{(0)}(i\varepsilon_{n})=(i\varepsilon_{n}-E_{M})^{-1}$. The lowest
(second-order) contribution to the self-energy are shown in Fig. \ref%
{Sigma0N} (with real frequencies replaced by Matsubara ones) and Fig. \ref%
{SigmaTN} where the solid and dashed lines corresponds to $%
G_{T\nu}^{(0)}(i\varepsilon _{n})=(i\varepsilon_{n}-E_{\nu})^{-1}$ and $D_{T%
\mathbf{q}}^{(0)}(i\omega _{m})=(i\omega_{m}-\epsilon_{q}+i0)^{-1}$,
respectively. (Note that, similar to the $T=0$ case, the "anomalous"
contributions can be ignored.) After performing the summation over the
(bosonic) Matsubara frequency $\omega _{m}=2\pi Tm$, we obtain 
\begin{equation*}
\Sigma_{TM}(i\varepsilon_{n})=\Sigma_{TM}^{(1)}(i\varepsilon_{n})+\Sigma
_{TM}^{(2)}(i\varepsilon_{n}),
\end{equation*}
where 
\begin{align*}
\Sigma_{TM}^{(1)}(i\varepsilon_{n}) & =\sum_{\mathbf{q},\nu}\left[ \frac{%
\left\vert O_{\mathbf{q}\nu M}^{(n1)}\right\vert ^{2}}{i\varepsilon
_{n}-(E_{\nu}+\epsilon_{q})}+\frac{4\left\vert O_{\mathbf{q}\nu
M}^{(a2)}\right\vert ^{2}}{i\varepsilon_{n}+E_{\nu}+\epsilon_{q}}\right] \\
& \times\left[ 1+n_{B}(\epsilon_{q})-n_{F}(E_{\nu})\right]
\end{align*}
and 
\begin{align*}
\Sigma_{TM}^{(2)}(i\varepsilon_{n}) & =\sum_{\mathbf{q},\nu}\left[ \frac{%
\left\vert O_{\mathbf{q}\nu M}^{(n2)}\right\vert ^{2}}{i\varepsilon
_{n}-(E_{\nu}-\epsilon_{q})}+\frac{4\left\vert O_{\mathbf{q}\nu
M}^{(a1)}\right\vert ^{2}}{i\varepsilon_{n}+E_{\nu}-\epsilon_{q}}\right] \\
& \times\left[ n_{B}(\epsilon_{q})+n_{F}(E_{\nu})\right]
\end{align*}
with $n_{F\nu}(T)$ and $n_{B\mathbf{q}}(T)$ being the fermionic and bosonic
occupation numbers of the gapped modes $\alpha_{v}$ and excitations $b_{%
\mathbf{q}}$ in the condensate, respectively. With the analytic continuation 
$i\varepsilon_{n}\rightarrow\varepsilon+i0$, an approximate solution of the
equation $G_{M}^{R}(\varepsilon)^{-1}=\varepsilon-E_{M}-\Sigma_{TM}(%
\varepsilon+i0)=0$ for the pole of the Green's function reads 
\begin{equation*}
\varepsilon_{M\ast}\approx
E_{M}+\Sigma_{TM}(E_{M}+i0)=E_{M}+\delta_{T}E_{M}-i\gamma_{M},
\end{equation*}
where 
\begin{align*}
\Sigma_{TM}(E_{M}+i0) &
=\Sigma_{TM}^{(1)}(E_{M}+i0)+\Sigma_{TM}^{(2)}(E_{M}+i0) \\
& =\sum_{\mathbf{q},\nu}\left[ \frac{4\left\vert O_{\mathbf{q}\nu
M}^{(a2)}\right\vert ^{2}}{E_{\nu}+\epsilon_{q}+E_{M}}-\frac{\left\vert O_{%
\mathbf{q}\nu M}^{(n1)}\right\vert ^{2}}{E_{\nu}+\epsilon_{q}-E_{M}}\right]
[1+n_{B}(\epsilon_{q})-n_{F}(E_{\nu})] \\
& +\sum_{\mathbf{q},\nu}\left[ \frac{4\left\vert O_{\mathbf{q}\nu
M}^{(a1)}\right\vert ^{2}}{E_{\nu}-\epsilon_{q}+E_{M}+i0}-\frac{\left\vert
O_{\mathbf{q}\nu M}^{(n2)}\right\vert ^{2}}{E_{\nu}-\epsilon_{q}-E_{M}-i0}%
\right] [n_{B}(\epsilon_{q})+n_{F}(E_{\nu})] \\
& =\delta_{T}E_{M}-i/\tau_{M}
\end{align*}
provides the correction $\delta_{T}E_{M}=\mathrm{\text{Re}}%
\Sigma_{TM}(E_{M}+i0)$ to the energy of the mode $\alpha_{M}$, as well as
its inverse life-time $\tau_{M}^{-1}=-\mathrm{\text{Im}}%
\Sigma_{TM}(E_{M}+i0) $. Note that the first term $\Sigma_{TM}^{(1)}(E_{M})$
which generalizes Eq. (\ref{SigmaM0}) to finite temperatures, contributes to 
$\delta_{T}E_{M}$ only because the energy denominators are never zero (for
this reason we skipped the $i0$ there), while the second term $%
\Sigma_{TM}^{(2)}(E_{M}+i0)$ which is non-zero only at finite temperatures,
contributes to both $\delta_{T}E_{M}$ and $\tau_{M}^{-1}$.

By using Eqs. (\ref{On1}) and (\ref{Oa2}) for the matrix elements $O_{%
\mathbf{q}\nu M}^{(n1)}$ and $O_{\mathbf{q}\nu M}^{(a2)}$, the terms $%
\Sigma_{TM}^{(1)}(E_{M})$ and can be written in the form 
\begin{equation}
\Sigma_{TM}^{(1)}(E_{M})\approx\sum_{\mathbf{q},\nu}\left[ \frac {4\text{Re}%
(\left\langle 0\left\vert H_{R}^{(+)}{}^{\dag}\right\vert \mathbf{q}%
\nu\right\rangle \left\langle \mathbf{q}\nu\left\vert H_{L}
^{(+)}\right\vert 0\right\rangle )}{E_{\nu}+\epsilon_{q}}-2E_{M} \frac{%
\left\vert \left\langle \mathbf{q}\nu\left\vert H_{L}^{(+)}\right\vert
0\right\rangle \right\vert ^{2}+\left\vert \left\langle \mathbf{q}
\nu\left\vert H_{R}^{(+)}\right\vert 0\right\rangle \right\vert ^{2}}{%
(E_{\nu }+\epsilon_{q})^{2}}\right] [1+n_{B}(\epsilon_{q})-n_{F}(E_{\nu})]
\label{SigmaT1}
\end{equation}
which recovers Eqs. (\ref{DE1}) and (\ref{deltaEMnew}) for $T=0$. The term $%
\Sigma_{TM}^{(2)}(E_{M})$ can also be written in the form involving matrix
elements of the operators $H_{L,R}^{(+)}$, if one notices [see Eqs. (\ref%
{H3relevant}) and (\ref{Hc3expanded})] that 
\begin{equation}
O_{\mathbf{q}\nu M}^{(n2)}=\left\langle \nu\left\vert H_{L}^{(+)}+H_{R}
^{(+)}\right\vert \mathbf{q}\right\rangle  \notag
\end{equation}
and 
\begin{equation}
2O_{\mathbf{q}\nu M}^{(a1)}=\left\langle \nu\left\vert H_{L}^{(+)}-H_{R}
^{(+)}\right\vert \mathbf{q}\right\rangle.  \notag
\end{equation}
The corresponding expression reads 
\begin{align*}
\Sigma_{TM}^{(2)}(E_{M}+i0) & =\sum_{\mathbf{q},\nu}\left\{ 2\text{Re}%
(\left\langle \mathbf{q}\left\vert H_{R}^{(+)}{}^{\dag}\right\vert
\nu\right\rangle \left\langle \nu\left\vert H_{L}^{(+)}\right\vert \mathbf{q}%
\right\rangle )\left[ \frac{1}{E_{\nu}-\epsilon_{q}+E_{M}+i0}+\frac{1}{%
E_{\nu}-\epsilon_{q}-E_{M}-i0}\right] \right. \\
& \left. +(\left\vert \left\langle \nu\left\vert H_{L}^{(+)}\right\vert 
\mathbf{q}\right\rangle \right\vert ^{2}+\left\vert \left\langle
\nu\left\vert H_{R}^{(+)}\right\vert \mathbf{q}\right\rangle \right\vert
^{2})\left[ \frac{1}{E_{\nu}-\epsilon_{q}+E_{M}+i0}-\frac{1}{%
E_{\nu}-\epsilon_{q}-E_{M}-i0}\right] \right\} \\
& \times\lbrack n_{B}(\epsilon_{q})+n_{F}(E_{\nu})]
\end{align*}
and contains again two different types of correlations: long-range
correlations between the edges (the first line) and short-range correlations
at the edges (the second line). The real part of $%
\Sigma_{TM}^{(2)}(E_{M}+i0) $ contains terms with the correlations of the
both types: 
\begin{align}
\mathrm{\text{Re}}\Sigma_{TM}^{(2)}(E_{M}+i0) & \approx\mathrm{p.V.}\sum_{ 
\mathbf{q},\nu}\frac{4\text{Re}(\left\langle \mathbf{q}\left\vert
H_{R}^{(+)}{}^{\dag}\right\vert \nu\right\rangle \left\langle \nu\left\vert
H_{L}^{(+)}\right\vert \mathbf{q}\right\rangle )}{E_{\nu}-\epsilon_{q}}
[n_{B}(\epsilon_{q})+n_{F}(E_{\nu})]  \notag \\
& +2E_{M}\left\{ \frac{\partial}{\partial E_{M}}\mathrm{p.V.}\sum _{\mathbf{%
q },\nu}\frac{\left\vert \left\langle \nu\left\vert H_{L}^{(+)}\right\vert 
\mathbf{q}\right\rangle \right\vert ^{2}+\left\vert \left\langle
\nu\left\vert H_{R}^{(+)}\right\vert \mathbf{q}\right\rangle \right\vert ^{2}%
}{E_{\nu}-\epsilon_{q}+E_{M}}[n_{B}(\epsilon_{q})+n_{F}(E_{\nu})]\right\}
_{E_{M}=0},  \label{SigmaT2Re}
\end{align}
while the dominant contribution to the imaginary part of $%
\Sigma_{TM}^{(2)}(E_{M}+i0)$ and, therefore, to the life-time $\tau_{M}$,
comes from the short-range correlations: 
\begin{equation}
\tau_{M}^{-1}=-\mathrm{\text{Im}}\Sigma_{TM}(E_{M}+i0)\approx2\pi \sum_{%
\mathbf{q},\nu}(\left\vert \left\langle \nu\left\vert H_{L}
^{(+)}\right\vert \mathbf{q}\right\rangle \right\vert ^{2}+\left\vert
\left\langle \nu\left\vert H_{R}^{(+)}\right\vert \mathbf{q}\right\rangle
\right\vert ^{2})[n_{B}(\epsilon_{q})+n_{F}(E_{\nu})]\delta(E_{\nu}
-\epsilon_{q}),  \label{tauM}
\end{equation}
where we have neglected terms which are exponentially small in the system
size $L$.

It follows from Eqs. (\ref{SigmaT1}) and (\ref{SigmaT2Re}) that the
correction to the energy $\delta_{T}E_{M}$ and, therefore, the energy
itself, remains exponentially small with the system size $L$, even at finite
temperatures $T\ll\Delta_{m}$. The leading temperature correction to the
zero-temperature result (\ref{deltaEMT0}) comes from the low-energy bosonic
excitations with $\epsilon_{q}\lesssim T\ll E_{\nu}\leq\Delta_{m}$ [the
number of fermionic excitations $n_{F}(E_{\nu})$ is exponentially small at
such temperatures, $n_{F}(E_{\nu})\lesssim\exp(-\Delta_{m}/T)\ll1$, and can
be neglected]. These bosonic excitations are phonons with $%
q\ll\xi_{BEC}^{-1} $, for which, as it follows from Eqs. (\ref{H3c2}), (\ref%
{HLplus}), and (\ref{HRplus}), one has 
\begin{equation}
\left\langle \nu\left\vert H_{L(R)}^{(+)}\right\vert \mathbf{q}\right\rangle
=-\left\langle \mathbf{q}\nu\left\vert H_{L(R)}^{(+)}\right\vert
0\right\rangle ,  \label{matrix elements relation}
\end{equation}
and the leading temperature correction $\delta_{T}E_{M}$ reads 
\begin{align*}
\delta_{T}E_{M} & \approx4\sum_{\mathbf{q},\nu}\left[ \frac{2\text{Re}
(\left\langle 0\left\vert H_{R}^{(+)}{}^{\dag}\right\vert \mathbf{q}
\nu\right\rangle \left\langle \mathbf{q}\nu\left\vert H_{L}^{(+)}\right\vert
0\right\rangle )}{E_{\nu}}-E_{M}\frac{\left\vert \left\langle \mathbf{q}
\nu\left\vert H_{L}^{(+)}\right\vert 0\right\rangle \right\vert
^{2}+\left\vert \left\langle \mathbf{q}\nu\left\vert H_{R}^{(+)}\right\vert
0\right\rangle \right\vert ^{2}}{E_{\nu}^{2}}\right] n_{B}(\varepsilon_{q})
\\
& \equiv\delta_{T}E_{M}^{(1)}+\delta_{T}E_{M}^{(2)},
\end{align*}
where, similar to the $T=0$ case, we introduce the two contributions $%
\delta_{T}E_{M}^{(1)}$ and $\delta_{T}E_{M}^{(2)}$ which contain long-range
and short-range correlations, respectively.

The calculation of the dominant (for large $L$) term $\delta _{T}E_{M}^{(1)}$
can be performed in the same way as for the case of zero temperature. With
the expressions (\ref{melementHL}) and (\ref{melementHR}) for the matrix
elements, the energy correction $\delta _{T}E_{M}^{(1)}$ reads 
\begin{align}
\delta _{T}E_{M}^{(1)}& \approx 8K_{0}^{2}\int \frac{d\mathbf{q}}{(2\pi
)^{3} }n_{B}(\varepsilon _{q})\int_{-\pi /a}^{\pi /a}\frac{adk}{2\pi }\frac{
f_{q}^{2}(u_{k}+v_{k})^{2}\sin ^{2}ka}{E_{k}}%
\sum_{j_{1},j_{2}}f_{Lj_{1}}f_{Rj_{2}}e^{i(q_{x}+k)(j_{1}-j_{2})a}  \notag \\
& =8K_{0}^{2}\sum_{j_{1},j_{2}}f_{Lj_{1}}f_{Rj_{2}}\int \frac{d\mathbf{q}}{%
(2\pi )^{3}}f_{q}^{2}n_{B}(\varepsilon
_{q})e^{iq_{x}(j_{1}-j_{2})a}\int_{-\pi /a}^{\pi /a}\frac{adk}{2\pi }\frac{%
(u_{k}+v_{k})^{2}\sin ^{2}ka}{E_{k}}e^{ik(j_{1}-j_{2})a}  \notag \\
&
=8K_{0}^{2}%
\sum_{j_{1},j_{2}}f_{Lj_{1}}f_{Rj_{2}}C_{B}(j_{1}-j_{2})h(j_{1}-j_{2}).
\label{deltaTE1}
\end{align}
We see that in this approximation, the correlation function of the pair of
excitations decouples into the product of bosonic 
\begin{equation}
C_{B}(j_{1}-j_{2})=\int \frac{d\mathbf{q}}{(2\pi )^{3}}f_{q}^{2}n_{B}(%
\varepsilon _{q})e^{iq_{x}(j_{1}-j_{2})a}  \label{CB}
\end{equation}
and fermionic 
\begin{equation}
h(j_{1}-j_{2})=\int_{-\pi /a}^{\pi /a}\frac{adk}{2\pi }\frac{%
(u_{k}+v_{k})^{2}\sin ^{2}ka}{E_{k}}e^{ik(j_{1}-j_{2})a}  \label{h}
\end{equation}
correlation functions, respectively. The function $h(j_{1}-j_{2})$ is
calculated in Appendix \ref{AppendixB}, Eqs. (\ref{correlationh}) and (\ref%
{correlationhanswers}), and is nonzero only for $\left\vert
j_{1}-j_{2}\right\vert \sim l_{M}/a\sim 1$. The bosonic correlation function 
$C_{B}(j_{1}-j_{2})$ can be represented in the form 
\begin{align*}
C_{B}(j_{1}-j_{2})& =\frac{1}{4\pi ^{2}}\frac{mT}{\hbar ^{2}}\frac{1}{%
\left\vert j_{1}-j_{2}\right\vert a}\int_{0}^{\infty }dx\frac{\sin [\frac{%
T\left\vert j_{1}-j_{2}\right\vert a}{\hbar c}x]}{e^{x}-1} \\
& =\frac{\pi }{2}-\frac{\hbar c}{2T\left\vert j_{1}-j_{2}\right\vert a}+ 
\frac{\pi }{\exp (2\pi T\left\vert j_{1}-j_{2}\right\vert a/\hbar c)-1}.
\end{align*}
For $\left\vert j_{1}-j_{2}\right\vert \sim l_{M}/a\sim 1$, one has $%
T\left\vert j_{1}-j_{2}\right\vert a/\hbar c\sim \left\vert
j_{1}-j_{2}\right\vert (T/J)(Ja/\hbar c)\sim \left\vert
j_{1}-j_{2}\right\vert (T/\Delta_{m})\lambda \ll 1$, and for such values of $%
\left\vert j_{1}-j_{2}\right\vert $ the correlation function takes the
simple form 
\begin{equation*}
C_{B}(j_{1}-j_{2})\approx \frac{1}{24}\frac{mT^{2}}{\hbar^{3}c},
\end{equation*}
where the power is determined by the space volume of phonons ($\sim T^{3}$)
and by the square of the matrix elements ($\sim q^{-1}\rightarrow T^{-1}$).
(Note that this expression for the bosonic correlation function is valid for
distances $r\ll \hbar c/T\sim a(J/T)\lambda ^{-1}\gg a$.)

For $\delta _{T}E_{M}^{(1)}$ we now have 
\begin{align*}
\delta _{T}E_{M}^{(1)}& =\frac{1}{3}\frac{mT^{2}K_{0}^{2}}{\hbar ^{3}c}
\sum_{j_{1},j_{2}}f_{Lj_{1}}f_{Rj_{2}}h(j_{1}-j_{2}) \\
& \approx -\frac{1}{3}\frac{mT^{2}K_{0}^{2}}{\hbar ^{3}c}2\left\vert
A\right\vert ^{2}\sum_{j_{1},j_{2}}\rho ^{L+1+j_{1}-j_{2}}\cos
[(L+1+j_{1}-j_{2})\theta ]h(j_{1}-j_{2}) \\
& =-\frac{1}{3}\frac{mT^{2}K_{0}^{2}}{\hbar ^{3}c}2\left\vert A\right\vert
^{2}L\rho ^{L+1}\left\{ h(0)\cos [(L+1)\theta ]+\sum_{s>0}\rho ^{s}h(s)\cos
[(L+1+s)\theta ]+\rho ^{-1}h(-1)\cos (L\theta )\right\} ,
\end{align*}
where we use Eq. (\ref{fLfR}) and keep only the dominant term for large $L$.
It follows from the results of Appendix \ref{AppendixB} that the leading
contribution for small $\rho $ comes from the last term, and we finally get 
\begin{align}
\delta _{T}E_{M}&\approx \delta _{T}E_{M}^{(1)}\approx \frac{1}{3}\frac{
mT^{2}K_{0}^{2}}{\hbar ^{3}c}2\left\vert A\right\vert ^{2}\frac{1}{
4J(1+\alpha )}L\rho ^{L}\cos (L\theta )  \label{deltaEMT} \\
& =\frac{1}{6}\frac{mT^{2}K_{0}^{2}}{\hbar ^{3}c}\frac{\Delta (4J^{2}-\mu
^{2})}{J(4J^{2}-4\Delta ^{2}-\mu ^{2})}\frac{1}{J+\Delta }L\rho ^{L}\cos
(L\theta )  \notag \\
& \sim \frac{\Delta ^{2}}{E_{R}}\frac{1}{n_{M}a^{3}}\left( \frac{T}{J}
\right) ^{2}\lambda Le^{-La/l_{M}}\sim E_{M}\frac{\Delta }{E_{R}}\left( 
\frac{T}{J}\right) ^{2}\lambda L\sim \delta E_{M}\left( \frac{T}{J}\right)
^{2}\lambda .  \notag
\end{align}
We see that in the considered temperatures $T\ll \Delta _{m}\sim J$, the
correction to the energy $E_{M}$ of the Majorana mode due to thermal
fluctuations is much smaller than that due to quantum fluctuations, and can
be neglected.

The life-time $\tau _{M}$, Eq. (\ref{tauM}), is determined by the
correlations at the edges and, hence, does not depend on the length of the
wire $L$. On the other hand, the dependence of $\tau _{M}$ on temperature is
exponential, $\tau _{M}\sim \exp (\Delta _{m}/T)$, as a result of
exponentially small number of thermal excitations, both bosonic and
fermionic, with energies larger than the gap $\Delta _{m}$ in the wire, $%
n_{B}(\varepsilon _{q})$,$\,n_{F}(E_{\nu })\approx \exp (-\Delta _{m}/T)$
for $E_{\nu },\,\epsilon _{q}\gtrsim \Delta _{m}$ and $T\ll \Delta _{m}$.
Note that relevant bosonic excitations must also have energies larger than $%
\Delta _{m}$ because of the energy conservation condition in Eq. (\ref{tauM}%
). The reason for this is the conservation of the parity: The change in the
population of the mode $\alpha _{M}$ has to be accompanied by the change in
the population of one of the gapped mode $\alpha _{\nu }$. For this to
happen, one needs either a bosonic excitation with the energy larger than $%
\Delta _{m}$ which excites a gapped fermionic mode (terms $\alpha _{M}\alpha
_{\nu }^{\dag }b_{\mathbf{q}}$ or $\alpha _{M}^{\dag }\alpha _{\nu }^{\dag
}b_{\mathbf{q}}$ in the Hamiltonian), or a gapped fermionic excitation which
is annihilated with emission of a bosonic excitation ($\alpha _{M}\alpha
_{\nu }b_{\mathbf{q}}^{\dag }$ or $\alpha _{M}^{\dag }\alpha _{\nu }b_{%
\mathbf{q}}^{\dag }$ terms). In both cases, the probability to find such
excitation is of the order of $\exp (-\Delta _{m}/T)$.

We now calculate $\tau _{M}$ for temperatures $T\ll \Delta _{m}$. In this
case, the relevant bosonic excitations have energies $\varepsilon
_{q}=E_{\nu }\gtrsim \Delta _{m}\sim J\ll \hbar c\xi _{\text{BEC}}^{-1}$,
and are, therefore, phonons with wave vectors $q\sim J/\hbar c\sim \lambda
a^{-1}\ll a^{-1}$. This allows us to use Eqs. (\ref{matrix elements relation}%
), (\ref{melementHL}), and (\ref{melementHR}) for the matrix elements in Eq.
(\ref{tauM}) with the result (the contributions from the right and left
edges are identical) 
\begin{align*}
\tau _{M}^{-1}& =4\pi K_{0}^{2}\sum_{j_{1},j_{2}}f_{Lj_{1}}f_{Lj_{2}}\int 
\frac{d\mathbf{q}}{(2\pi )^{3}}f_{q}^{2}\int_{-\pi /a}^{\pi /a}\frac{adk}{%
2\pi }\sin ^{2}(ka)e^{i(q_{x}+k)a(j_{1}-j_{2})}[n_{B}(\epsilon
_{q})+n_{F}(E_{k})]\delta (E_{k}-\epsilon _{q}) \\
& \approx 8\pi K_{0}^{2}\sum_{j_{1},j_{2}}f_{Lj_{1}}f_{Lj_{2}}\int_{-\pi
/a}^{\pi /a}\frac{adk}{2\pi }\sin
^{2}(ka)e^{-E_{k}/T}e^{ika(j_{1}-j_{2})}\int \frac{d\mathbf{q}}{(2\pi )^{3}}%
f_{q}^{2}\delta (E_{k}-\epsilon _{q})e^{iq_{x}a(j_{1}-j_{2})},
\end{align*}%
where we take into account that $n_{F}(E_{k})\approx n_{B}(\epsilon
_{q})\approx \exp (-E_{k}/T)$ for $E_{k}=\epsilon _{q}\gtrsim \Delta _{m}\gg
T$. The result of the integration over $\mathbf{q}$ is 
\begin{equation*}
\int \frac{d\mathbf{q}}{(2\pi )^{3}}f_{q}^{2}\delta (E_{k}-\epsilon
_{q})e^{iq_{x}a(j_{1}-j_{2})}=\frac{m}{4\pi ^{2}\hbar ^{3}}\frac{\sin
[E_{k}a(j_{1}-j_{2})/\hbar c]}{a(j_{1}-j_{2})},
\end{equation*}%
and, if we take into account that $E_{k}a\left\vert j_{1}-j_{2}\right\vert
/\hbar c\sim \lambda \left\vert j_{1}-j_{2}\right\vert \ll 1$ for $%
\left\vert j_{1}-j_{2}\right\vert \lesssim l_{M}/a\sim 1$, the expression
for $\tau _{M}^{-1}$ can be written in the form 
\begin{equation*}
\tau _{M}^{-1}=\frac{2mK_{0}^{2}}{4\pi\hbar ^{4}c}%
\sum_{j_{1},j_{2}}f_{Lj_{1}}f_{Lj_{2}}\int_{-\pi /a}^{\pi /a}\frac{adk}{2\pi 
}\sin ^{2}(ka)E_{k}e^{-E_{k}/T}e^{ika(j_{1}-j_{2})}.
\end{equation*}%
We next perform the summation over $j_{1}$ and $j_{2}$: 
\begin{equation*}
\sum_{j_{1},j_{2}}f_{Lj_{1}}f_{Lj_{2}}e^{ika(j_{1}-j_{2})}=\left\vert
\sum_{j}f_{Lj}e^{ikaj}\right\vert ^{2}\approx \frac{1-\beta ^{2}}{%
(E_{k}/2J)^{2}},
\end{equation*}%
where we use Eq. (\ref{fLj}) for $f_{Lj}$, and obtain 
\begin{equation}
\tau _{M}^{-1}=\frac{8mJ^{2}K_{0}^{2}(1-\beta ^{2})}{\pi \hbar ^{4}c}%
\int_{-\pi /a}^{\pi /a}\frac{adk}{2\pi }\frac{\sin ^{2}(ka)}{E_{k}}%
e^{-E_{k}/T}.  \label{tauMintegral}
\end{equation}

The final integral over $k$ can be calculated analytically in two limiting
cases when the temperature $T$, being much smaller than the gap $\Delta_{m}$%
, is much smaller (i) or much larger (ii) than the \emph{band-width} of
fermionic excitations $\Delta E_{b}\approx2J(1-\alpha+\left\vert
\beta\right\vert )$ (the latter case can be realized when the band of
fermionic excitations is narrow, $\Delta E_{b}\ll J$, which happens for $%
1-\alpha^{2}\ll1$ and $\left\vert \beta\right\vert \ll$ $1-\alpha^{2}$).

In the first case, $T\ll \Delta E_{b}$, the main contribution comes from the
vicinities of two minima of $E_{k}$ at $k=\pm k_{F}=\pm a^{-1}\arccos
[-\beta /(1-\alpha ^{2})]$ inside the Brillouin zone $-\pi /a\leq k\leq \pi
/a$. Near these minima, $E_{k}$ can be approximated as 
\begin{equation*}
E_{k}\approx \Delta _{m}+J\alpha ^{-1}\frac{(1-\alpha ^{2})^{2}-\beta ^{2}}{ 
\sqrt{(1-\alpha ^{2}-\beta ^{2})(1-\alpha ^{2})}}\delta k_{\pm }^{2},
\end{equation*}
where $\delta k_{\pm }=k\mp k_{F}$, and, after extending the integration
over $\delta k_{\pm }$ to infinite limits, we obtain 
\begin{equation}
\tau _{M}^{-1}\approx \frac{4mJ^{2}K_{0}^{2}}{\pi \hbar ^{4}c\Delta _{m}}
e^{-\Delta _{m}/T}(1-\beta ^{2})\sqrt{\frac{2}{\pi }\frac{T\Delta _{m}}{%
J^{2} }\frac{(1-\alpha ^{2})^{2}-\beta ^{2}}{(1-\alpha ^{2})^{3}}}.
\label{tauMbroadband}
\end{equation}%
The life-time $\tau _{M}$ estimated from this expression, 
\begin{equation}
\tau _{M}\sim \frac{\hbar }{J}\frac{E_{R}}{J\lambda }\sqrt{\frac{J}{T}}
e^{\Delta _{m}/T}\gg \frac{\hbar }{J},  \label{tauMbroad}
\end{equation}
contains not only the exponential factor $\exp (\Delta _{m}/T)$ but also
large prefactor ($E_{R}/J\lambda )\sqrt{J/T}\gg 1$, altogether making $\tau
_{M}$ much larger than the characteristic time $\hbar /J$ in the wire.

In the second case, $\Delta E_{b}\ll T\ll \Delta _{m}$, we can set $%
E_{k}\approx \Delta _{m}$ in Eq. (\ref{tauMintegral}) and obtain 
\begin{equation}
\tau _{M}^{-1}\approx \frac{4mJ^{2}K_{0}^{2}}{\pi \hbar ^{4}c\Delta _{m}}%
e^{-\Delta _{m}/T}(1-\beta ^{2}).  \label{tauMnarrowband}
\end{equation}%
An estimate of the life-time $\tau _{M}$ in this case, 
\begin{equation}
\tau _{M}\sim \frac{\hbar }{J}\frac{E_{R}}{J\lambda }e^{\Delta _{m}/T}\gg 
\frac{\hbar }{J},  \label{tauMnarrow}
\end{equation}%
also shows exponential dependence on temperature with the large
temperature-independent prefactor $E_{R}/J\lambda \gg 1$, such that also in
this case $\tau _{M}$ is much larger than the characteristic time $\hbar /J$
in the wire.

The life-time $\tau _{M}$ provides an estimate for the thermalization time
of the mode $\alpha _{M}$ and, therefore, for the \textquotedblleft{%
relaxation}\textquotedblright time of Majorana correlations -- the time
during which the correlations evolve from their initial values to the
stationary ones. If, for example, we the mode $\alpha _{M}$ is unpopulated
initially (i.e., $-i\left\langle \gamma _{L}\gamma _{R}\right\rangle =1$),
than its occupation $n_{M}(t)=\left
\langle \alpha _{M}^{\dag }(t)\alpha
_{M}(t)\right\rangle $ and the related Majorana correlation $-i\left\langle
\gamma _{L}(t)\gamma _{R}(t)\right\rangle $ for times $t>\tau _{M}$ can be
estimated as 
\begin{equation}
1-2n_{M}(t)=-i\left\langle \gamma _{L}(t)\gamma _{R}(t)\right\rangle \sim
\exp [-2L\exp (-\Delta _{m}/T)].  \label{correlationdecay}
\end{equation}
This estimate is based on purely statistical arguments with an account of
the parity constraint ($L$ in the exponent corresponds to the number of the
gapped modes in the systems). Without this constraint, the mode $\alpha _{M}$
will be effectively at infinite temperature with $n_{M}(t)-1/2\sim \exp
(-E_{M}/T)\approx 0$ for any realistic temperature $T$. Eq. (\ref%
{correlationdecay}) shows that no correlations between Majorana fermions
survive at finite temperature in the thermodynamic limit $L\rightarrow
\infty $. On the other hand, in a mesoscopic system, the thermal degradation
of the initial correlations can still be sufficiently small, allowing
quantum operations with Majorana fermions for times $t>\tau_{M}$ with
acceptable fidelity.

\section{Concluding remarks}

\label{Conclusion}

Our results show the prospect for creation and manipulation of Majorana
fermions in ultra-cold system of atoms and molecules. For a Kitaev's
topological wire which can be realized by coupling fermionic atoms in an
optical lattice to a superfluid molecular reservoir, we have shown that the
coupling between Majorana edge states in the wire and the corresponding
splitting in the ground state degeneracy decay exponentially with the length
of the wire. This results also holds at finite temperatures lower than the
gap $\Delta_{m}$ of the bulk fermionic excitations in the wire. With the
possibility of having the localization length of the Majorana edge states to
be of the order of few lattice spacings, this ensures that already
relatively short wires with $L\gtrsim 10$ are sufficient for creation of
well-separated Majorana edge states, their detection as \textquotedblleft{%
zero-energy}\textquotedblright edge states via, for example, spectroscopic
measurements \cite{LiangJiang,Nascimbene,MajAtom3}, and demonstration their
non-Abelian character via braiding \cite{MajAtom5,MajAtom6}.

Thermal fluctuations however result in the decay of the correlations between
the Majorana edge states on a time scale $\tau _{M}$ to the values which
decreases exponentially with the length of the wire $L$. This limits quantum
operations with Majorana fermions to times less than $\tau _{M}$. Note,
however, that under the rather general conditions of our implementation
scheme, see Appendix \ref{AppendixA}, one has $E_{R}/J\lambda \gtrsim 10^{3}$
and, already for $\Delta _{m}/T=3$, the life-time $\tau _{M}$ estimated from
Eq. (\ref{tauMnarrow}) is five orders of magnitude larger than $\hbar
/\Delta _{m}$.This is sufficient for the implementation of the brading
protocol and simple quantum computation algorithms, see Refs. \cite{MajAtom5}
and \cite{MajAtom6}, based on adiabatic manipulations of Majorana edge
states in atomic wires.

To perform quantum operations during longer times, $t>\tau _{M}$, one can
consider systems of mesoscopic wires with the length which is chosen to
obtain the highest fidelity in a given experimental setup. This optimal
length is a result of the competition between Eq. (\ref{correlationdecay})
which favours smaller $L$ in order to minimize the destructive effects of
thermal fluctuations on Majorana correlations, and Eqs. (\ref{EMexact1}), (%
\ref{deltaEMT0}), and (\ref{deltaEMT}) which suggest larger $L$ to minimize
the energy of the Majorana mode $E_{M}$ which determines the splitting of
the ground state. $E_{M}$ therefore sets the low bound on the speed of
adiabatic manipulations with Majorana states and, hence, on their error. As
an illustration of what one could expect, we consider the wire of the length 
$L=10$ with the localization length of the Majorana edge states $l_{M}=3a$.
From Eq. (\ref{correlationdecay}) we then find that thermal fluctuations
reduce the Majorana correlations to $\approx 70\%$ of their values when $%
\Delta _{m}/T=4$, and to $\approx 90\%$ when $\Delta _{m}/T=5$. At the same
time, Eq. (\ref{deltaEMT0}) gives $E_{M}\sim 10^{-2}\Delta _{m}$, such that
we can find the speed $t_{A}^{-1}$, $\hbar /\Delta _{m}\ll t_{A}\ll \hbar
/E_{M}$, at which operations with Majorana fermions are adiabatic with
respect to the gap $\Delta _{m}$ and diabatic with respect to the splitting $%
E_{M}$. The latter allows us to consider the ground-state manifold as being
degenerate during the operations -- the condition when non-Abelian
statistics of Majorana fermions determines the result of operations with
them. Based on the above estimates we can conclude that adiabatic quantum
manipulations with Majorana fermions in systems of ultracold atoms and
molecules are not unrealistic.

\section{Acknowledgement}

We would like to thank P. Zoller for raising our interest to the problem and
for many stimulating discussions during the work. We also acknowledge useful
discussions with M. Dalmonte, S. Diehl, C. Kraus, A. Kamenev, S. Nascimb\`{e}%
ne, H. Pichler, T. Ramos, E. Rico, and C. Salomon. This project was
supported by the ERC Synergy Grant UQUAM and the SFB FoQuS (FWF Project No.
F4016-N23). Y. H. acknowledges the support from the Institut f\"{u}r
Quanteninformation GmbH.

\appendix

\section{Microscopic Model}

\label{AppendixA}

Here we describe a realization of the Kitaev Hamiltonian using fermionic
atoms in an optical lattice coupled to a superfluid reservoir through Raman
lasers. We shall first illustrate our microscopic model for a setup in which
the reservoir is a molecular BEC, and derive the effective Hamiltonian (\ref%
{Htotal0}) in the main text. Later, we will extend to more general cases
where the superfluid reservoir consists of fermion pairs in the BEC-BCS
crossover regime.

\subsection{Setup and microscopic Hamiltonian}

We consider fermionic atoms in three internal states, labeled as $\mid
\uparrow\rangle$, $\mid\downarrow\rangle$ and $|3\rangle$, having energies $%
\varepsilon_{\uparrow}$, $\varepsilon_{\downarrow}$, and $\varepsilon_{3}$,
respectively. Atoms in the state $|3\rangle$ can be trapped in a strongly
anisotropic optical lattice where tunneling is only allowed in one
direction, leading to the realization of a quasi-$1D$ fermionic quantum gas
(wire). Atoms in the internal states $\mid\uparrow\rangle$ and $%
\mid\downarrow\rangle$ can form a Feshbach molecule. The molecules are
cooled to form a molecular BEC at sufficiently low temperature, which acts
as a reservoir for pairs of atoms in the lattice.

For the atoms in the wire, the corresponding field operator $\hat{\chi}_{3}(%
\mathbf{r})$ can be expanded on the basis of Wannier functions as 
\begin{align}
\hat{\chi}_{3}(\mathbf{r})=\sum_{j} w(\mathbf{r}-\mathbf{r}_{j})\hat{a}_{j},
\label{Chi3expand}
\end{align}
where $\hat{a}_{j}$ is the annihilation operator for an atom at the lattice
site $\mathbf{r}_{j}=ja \mathbf{e}_{x}+y_{0} \mathbf{e}_{y}+z_{0}\mathbf{e}%
_{z}$ with $a$ being the spatial period in the $x$-direction, and we assume
a Gaussian form for the Wannier function (in the lowest band tight binding
approximation) 
\begin{align}
w(\mathbf{r})=\frac{1}{\pi^{\frac{3}{4}}\sigma_{x}^{\frac{1}{2}%
}\sigma_{\perp }}e^{-x^{2}/2\sigma_{x}^{2}-(y^{2}+z^{2})/2\sigma_{%
\perp}^{2}},  \label{Wannier}
\end{align}
with $\sigma_{x}$ and $\sigma_{\perp}$ being the extension of the Wannier
function $w(\mathbf{r})$ in the $x$- and transverse directions,
respectively, which satisfy the condition $\sigma_{\perp}\ll\sigma_{x}\ll a$%
. The Hamiltonian for atoms hopping freely in the wire therefore reads 
\begin{equation}
H_{\mathrm{L}}=\sum_{j}[-J_{0}(\hat{a}_{j}^{\dag}\hat{a}_{j+1}+\hat{a}
_{j+1}^{\dagger}\hat{a}_{j})-\varepsilon_{3}^{\prime}\hat{a}_{j}^{\dag} \hat{%
a} _{j}],  \label{HL1}
\end{equation}
where $\epsilon_{3}^{\prime}=\epsilon_{3}-\varepsilon_{\mathrm{lat}}$ is the
chemical potential of a bare atom trapped in each well in the lattice and,
and as usual, we limit ourselves to the nearest-neighbor hopping $J_{0}$.

For the atoms in the internal state $|\sigma\rangle$ in the bulk reservoir
(with volume $V$), the corresponding field operator $\hat{\chi}_{\sigma}( 
\mathbf{r})$ can be written in terms of 'plane waves' as 
\begin{equation}
\hat{\chi}_{\sigma}(\mathbf{r})=\frac{1}{\sqrt{V}}\sum_{\mathbf{p}}\hat{c}_{ 
\mathbf{p}\sigma}e^{i\mathbf{p}\cdot\mathbf{r}},  \label{Chisigma}
\end{equation}
where $\hat{c}_{\mathbf{p}\sigma}$ is the annihilation operator for an atom
in the internal state $|\sigma\rangle$ with momentum $\mathbf{p}$. Two atoms
in the internal states $\mid\uparrow\rangle$ and $\mid\downarrow\rangle$,
respectively, can form a Feshbach molecule. A Feshbach molecule of a size $%
a_{s}$ (or the scattering length between $\mid\uparrow\rangle$ and $%
\mid\downarrow\rangle$\textbf{\ }atoms) has an energy $\epsilon_{\text{mol}%
}=\epsilon_{\uparrow}+\epsilon_{\downarrow}-E_{b}$ ($E_{b}=%
\hbar^{2}/ma_{s}^{2}$ is the binding energy). The corresponding molecular
field operator $\hat{\phi}^{\dag}(\mathbf{r})$ is expressed as $\hat{\phi}%
^{\dag }(\mathbf{r})=\frac{1}{\sqrt{V}}\sum_{\mathbf{q}}e^{-i\mathbf{q}\cdot%
\mathbf{r}}\hat{b}_{\mathbf{q}}^{\dag}$, where the molecular operator $\hat{b%
}_{\mathbf{q}}^{\dag}$ can be written in terms of the atomic operator $\hat{c%
}_{\mathbf{p}\sigma}^{\dag}$ as 
\begin{equation}
\hat{b}_{\mathbf{q}}^{\dag}=\sum_{\mathbf{k}}\varphi_{\mathbf{k}}\hat {c}_{%
\mathbf{q}/2+\mathbf{k},\uparrow}^{\dag}\hat{c}_{\mathbf{q}/2-\mathbf{k}%
,\downarrow}^{\dag},  \label{bq}
\end{equation}
with $\varphi_{k}$ being the molecular wave function (in the momentum space) 
\begin{equation*}
\varphi_{\mathbf{k}}=\left( \frac{8\pi}{a_{s}}\right) ^{1/2}\frac{1}{%
k^{2}+1/a_{s}^{2}}.
\end{equation*}
When the molecules are sufficiently cooled to form a molecular condensate,
the corresponding Hamiltonian reads, (for simplicity we assume that
molecules do not feel the optical lattice potential) 
\begin{equation}
{H}_{\mathrm{BEC}}=\int d\mathbf{r}\hat{\phi}^{\dag}\left( -\frac{\hbar^{2}}{%
2m}\nabla^{2}-\mu_{\text{M}}+\frac{g_{\text{M}}}{2}\hat{\phi}^{\dag}\hat{\phi%
}\right) \hat{\phi},  \notag
\end{equation}
where $m=2m_{a}$ is the mass of the molecule, $g_{\text{M}}=4\pi\hbar^{2}a_{%
\text{M}}/m$ is the coupling constant with $a_{\text{M}}\approx0.6a_{s}$ 
\cite{PetrovSolomonSchlyapnikov} being the molecule-molecule scattering
length, and $\mu_{\text{M}}$ is the chemical potential of molecules in the
condensate. Hereafter, we will assume weak interaction regime $n_{\text{M}%
}a_{\text{M}}^{3}<1$, where $n_{\text{M}}$ is the density of molecules.

The coupling between the atoms in the wire and the molecules in the
reservoir is introduced via a set of Raman transitions between the atomic
internal state $|3\rangle$ and the states $|\sigma\rangle$, described by the
Hamiltonian (after the rotating-wave approximation) 
\begin{equation}
H_{\mathrm{R}}=\sum_{\sigma=\uparrow,\downarrow}\int d\mathbf{r}\Omega
_{\sigma}[e^{(i\mathbf{k}_{\sigma}\mathbf{r}-i\omega_{\sigma}t)}\hat{\chi }_{%
\mathrm{L}}^{\dag}(\mathbf{r})\hat{\chi}_{\sigma}(\mathbf{r})+\text{h.c.}],
\label{HR1}
\end{equation}
where $\Omega_{\sigma}$ is the Rabi frequency, while $\omega_{\sigma}$ and $%
\mathbf{k}_{\sigma}$ are the frequency and momentum of the Raman laser,
respectively. A crucial condition in Eq. (\ref{HR1}) is to have $\mathbf{k}%
_{\uparrow}\neq\mathbf{k}_{\downarrow}$ for the reasons that will soon
become clear. By using Eqs. (\ref{Chi3expand}) and (\ref{Chisigma}), we
rewrite Hamiltonian (\ref{HR1}) as 
\begin{align}
H_{\text{R}}=\frac{1}{\sqrt{V}}\sum_{\mathbf{p},j,\sigma}[\Omega_{\sigma
}e^{i(\mathbf{p}+\mathbf{k}_{\sigma})\cdot\mathbf{r}_{j}-i\omega_{\sigma}t}
M^{*}_{\mathbf{p}+\mathbf{k}_{\sigma}}\hat{a}^{\dag}_{j} \hat{c}_{\mathbf{p}
\sigma}+\text{h.c.}],  \label{HRmoment}
\end{align}
with 
\begin{align}
M_{\mathbf{p}+\mathbf{k}_{\sigma}}=\int d\mathbf{r}_{1} w(\mathbf{r}%
_{1})e^{-i(\mathbf{p}+\mathbf{k}_{\sigma})\cdot\mathbf{r}_{1}},  \label{M}
\end{align}
being the Fourier transformation of the Wannier function $w(\mathbf{r})$.

Overall, the total Hamiltonian for an atomic wire coupled to a molecular
reservoir via Raman beams can be written as 
\begin{equation}
{H}={H}_{\mathrm{L}}+{H}_{\mathrm{BEC}}+{H}_{\mathrm{int}}+{H}_{\mathrm{R}},
\label{Htotal}
\end{equation}
where the Hamiltonian $H_{\mathrm{int}}$ describes the short-range
interaction between atoms in the lattice and molecules in the BEC, reading 
\begin{align}
{H}_{\mathrm{int}} & =g_{a\text{M}}\int d\mathbf{r}\hat{\chi}_{\mathrm{L}%
}^{\dag }( \mathbf{r})\hat{\chi}_{\mathrm{L}}(\mathbf{r})\hat{\phi}^{\dag}(%
\mathbf{r}) \hat{\phi}(\mathbf{r})  \notag \\
& \approx\sum_{j}\int d\mathbf{r}g_{j}(\mathbf{r})\hat{a}_{j}^{\dag}\hat {a}
_{j}\hat{\phi}^{\dag}(\mathbf{r})\hat{\phi}(\mathbf{r}),  \label{Hint1}
\end{align}
with $g_{a\text{M}}$ being the corresponding coupling constant (the
corresponding scattering length \ $a_{a\text{M}}\approx1.2a_{s}$, see \cite%
{PetrovSolomonSchlyapnikov,PetrovSolomonSchlyapnikov1}) and $g_{j}(\mathbf{r}
)=g_{a\text{M}}w(\mathbf{r}-\mathbf{r}_{j})^{2}$. As we shall show below,
the crucial ingredient in the Hamiltonian (\ref{Htotal0}) consists in the
Raman transitions between the atomic internal states ($H_{\mathrm{R}}$),
which provide a mechanism to inducing the $p$-wave pairing term in the wire
out of the $s$-wave superfluid reservoir.

\subsection{Raman-induced conversion of molecules into pairs of atoms}

Now, we will show in detail the realization of the conversion of a molecule
in the reservoir to a pair of atoms in the lattice described by the
Hamiltonian 
\begin{equation}
H_{\mathrm{conv}}=\sum_{j,j^{\prime}}\int d\mathbf{r[}K_{jj^{\prime}}( 
\mathbf{r})\hat{a}_{j}^{\dag}\hat{a}_{j^{\prime}}^{\dag}\hat{\phi }(\mathbf{%
r })+\mathrm{H.c.}]  \label{Hpair}
\end{equation}
from the setup described by Eq. (\ref{Htotal}). The physics behind the pair
transfer via Raman processes can be described as follows (see Fig. \ref%
{Fig:2}). 
\begin{figure}[tb]
\includegraphics[width=0.3\textwidth]{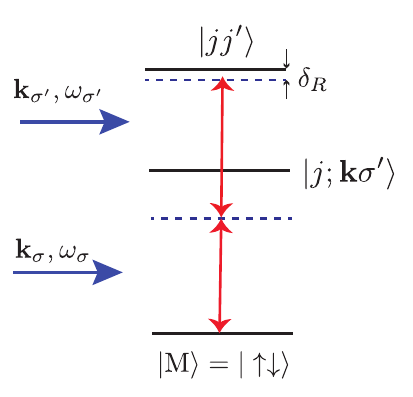}
\caption{ (Color online) A schematic illustration of the mechanism
converting a molecule from the condensate into a pair of atoms in the
optical lattice via two successive off-resonant Raman transitions. The first
Raman transition changes the internal state of a constituent atom in the
molecule ($|\text{M}\rangle=|\uparrow\downarrow\rangle$), from $%
\mid\downarrow(\uparrow)\rangle$ to $|3\rangle$. As a result, the molecule
is broken into one atom trapped in the lattice site $j$ and one unpaired $%
\mid\uparrow(\downarrow)\rangle$ atom with momenta $\mathbf{k}$. This
unpaired atom is transferred into the lattice after the second Raman
transition, which changes its internal state from $\mid\uparrow(\downarrow)
\rangle$ to $|3\rangle$. The overall process of transferring a molecule $| 
\text{M}\rangle$ in the reservoir into a pair of atoms in the lattice $%
|jj^{\prime}\rangle$ via absorbing two Raman photons is nearly resonant,
with a small two-photon detunning $\protect\delta_{R}$ determined by the
resonant condition in Eq. (\protect\ref{Condition1}). }
\label{Fig:2}
\end{figure}
The action of $H_{\text{R}}$ on a molecule, according to Eq. (\ref{bq}),
flips the internal state of one of the constituent atom from $%
|\sigma\rangle\rightarrow|3\rangle$, thereby generating processes where a
molecule breaks into an atom in the internal state $|3\rangle$ and an atom
in the internal state $|\sigma\rangle$, in particular, the process where the
generated $|3\rangle$ atom is trapped in the lattice. The Hamiltonian
describing the transfer of a molecule into an atom in the wire and a
unpaired atom in the internal state $|\sigma\rangle$ moving in the reservoir
(and vice versa) reads 
\begin{equation}
\!\!\!\!H_{\text{R}}^{\text{M}}=\!\!\frac{1}{\sqrt{V}}\!\!\sum_{\mathbf{k}
,j}\!\Big[\Omega_{\uparrow}\!e^{-i\omega_{\uparrow}t}e^{i(\frac{\mathbf{q }}{
2}\!+\!\mathbf{k}\!+\!\mathbf{k}_{\uparrow})\cdot\mathbf{r}_{j}}M_{ \frac{ 
\mathbf{q}}{2}\!+\!\mathbf{k}\!+\!\mathbf{k}_{\uparrow}}^{\ast}\varphi_{ 
\mathbf{k}}\hat{a}_{j}^{\dag}\hat{c}_{\frac{\mathbf{q}}{2}\!-\! \mathbf{k}
\downarrow}^{\dag}\hat{b}_{\mathbf{q}}\!-\!\Omega_{\downarrow
}e^{-i\omega_{\downarrow}t}e^{i(\frac{\mathbf{q}}{2}-\mathbf{k}+\mathbf{k}
_{\downarrow})\cdot\mathbf{r}_{j}}M_{\frac{\mathbf{q}}{2}\!-\!\mathbf{k}
\!+\!\mathbf{k}_{\downarrow}}^{\ast}\varphi_{\mathbf{k}}\hat{a}_{j}^{\dag} 
\hat{c}_{\frac{\mathbf{q}}{2}\!+\!\mathbf{k}\uparrow}^{\dag}\hat{b}_{ 
\mathbf{q}}\!+\!\text{h.c.}\Big].  \label{HRMol}
\end{equation}

Then, in the second Raman process, the unpaired $|\sigma\rangle$ atom in the
reservoir can be further transferred into the internal state $|3\rangle$ and
trapped in the lattice. Overall, after two successive Raman processes, a
transfer of a molecule in the reservoir into a pair of atoms in the wire is
achieved, corresponding to $\hat{b}^{\dag}\rightarrow\hat{a}_{j}^{\dag} \hat{
c}_{\mathbf{p}\sigma}^{\dag}\rightarrow\hat{a}_{j}^{\dag}\hat{a}
_{j^{\prime}}^{\dag}$, and vice versa.

Let us state the main conditions under which the two continuous Raman
processes lead to a \textit{resonant} transfer of a molecule from the BEC
into \emph{a pair} of atoms in the optical lattice (and vice versa), but
keeping the transfer of a single atom from the reservoir to the lattice
off-resonant. To this end, let us first briefly summarize the hierarchy of
relevant energy levels. A Feshbach molecule with a size $a_{s}$ in the BEC
has an energy $\epsilon_{\text{mol}}+\epsilon_{\text{MM}}$, where $\epsilon_{%
\text{MM}}=g_{\text{M}}n_{\text{M}}$ describes the interaction between
molecules in the BEC ($g_{\text{a\textrm{M}}}=3\pi\hbar^{2}a_{\text{a\textrm{%
M}}}/m$ with $a_{\text{M}}\approx0.6a_{s}$ and $m$ being the mass of an
atom). On the other hand, the average energy of a pair of atoms in a wire
can be written as $2(\epsilon _{3}^{\prime}-\frac{1}{2}\delta_{\text{R}%
}+\epsilon_{\text{aM}})$, where $\delta_{\text{R}}$ is the two-photon
detuning (see Fig. \ref{Fig:2}) and $\epsilon_{\text{aM}}=g_{\text{aM}}n_{%
\text{M}}$ is the mean-field interaction between an atom in the wire and
surrounding molecules. (For simplicity, we have assumed that the
atom-molecule interaction is independent of the internal state of an atom,
and thereby consider $g_{\text{aM}}=3\pi\hbar^{2}a_{\text{aM}}/m$ with $a_{%
\text{aM}}\approx1.2a_{s}$ being the atom-molecule scattering length.) As a
result, a nearly resonant transfer between a molecule in the BEC and a pair
of atoms in the wire is achieved when the two Raman photons provide an
energy satisfying the energy conservation reading 
\begin{equation}
\hbar\omega_{\uparrow}+\hbar\omega_{\downarrow}=2(\epsilon_{3}-\frac {
\delta_{\text{R}}}{2}+\epsilon_{\text{aM}})-(\epsilon_{\text{ mol}
}+\epsilon_{\text{MM}}),  \label{Condition1}
\end{equation}
where $\delta_{\text{R}}$ is a small detuning associated with the two-photon
Raman processes. In terms of $\delta_{\sigma}=\hbar\omega_{\sigma}+
\epsilon_{\sigma}-(\epsilon_{3}^{\prime}-\frac{1}{2}\delta_{ \text{R}})$
defined in the main text and assuming $\delta_{\uparrow}\approx\delta
_{\downarrow}$, the resonance condition in Eq. (\ref{Condition1}) can be
recast as $\delta_{\sigma}=\delta_{0}$ with 
\begin{equation}
\delta_{0}=\epsilon_{\text{aM}}+\frac{1}{2}E_{b}-\frac{1}{2}\epsilon_{ \text{%
MM}}.  \label{Condition2}
\end{equation}
Meanwhile, note that the energy cost for breaking a molecule into an atom in
the wire and an atom moving in the reservoir is 
\begin{align}
\Delta E_{\sigma} & =[\epsilon_{3}^{\prime}+(\epsilon_{\sigma}+\epsilon_{%
\mathbf{p}}^{0})+2\epsilon_{\text{aM}}]-[{\epsilon}_{\text{mol}}+\epsilon_{%
\text{MM}}-\hbar\omega_{\sigma}]  \notag \\
& =\epsilon_{\mathbf{p}}^{0}+\delta_{0},  \label{DE2}
\end{align}
where $\epsilon_{\mathbf{p}}^{0}$ is the kinetic energy of an unpaired atom
in the reservoir. Under the resonance condition in Eq. (\ref{Condition2}),
it is obvious that $\Delta E_{\sigma}\neq0$, and therefore, the state in
which an atom is generated in the wire and an atom remains unpaired in the
BEC is energetically prohibited, and serves as an intermediate state for the
ultimate realization of pair transfer.

Now, we are readily to derive the amplitude $K_{jj^{\prime}}(\mathbf{r})$ in
Eq. (\ref{Hpair}) for converting a molecule in the reservoir (labeled by the
state $|\text{M}\rangle$ ) into a pair of atoms at site $j$ and $j^{\prime}$
in the wire (labeled by the state $|jj^{\prime}\rangle$). By
straightforwardly applying the second-order perturbation theory, together
with Eqs. (\ref{Condition2}) and (\ref{DE2}), we obtain 
\begin{align}
K_{jj^{\prime}}(\mathbf{q})=-\sum_{k,\sigma}\frac{\langle jj^{\prime }|H_{%
\text{R}}|\mathbf{k}\sigma;j^{\prime}\rangle\langle\mathbf{k}%
\sigma;j^{\prime}|H^{\text{M}}_{\text{R}}|\text{M}\rangle}{\epsilon _{%
\mathbf{k}}^{0}+\delta_{0}}.  \label{Kq}
\end{align}
Substituting Eqs. (\ref{HRmoment}) and (\ref{HRMol}) into Eq. (\ref{Kq}), we
find 
\begin{align}
K_{jj^{\prime}}(\mathbf{q})=-\frac{16i}{V} \Omega\sin\left( \frac {\mathbf{k}%
_{d}\mathbf{r}_{jj^{\prime}}}{2}\right) e^{i(\mathbf{q}+\mathbf{k}_{c})\cdot%
\mathbf{R}_{jj^{\prime}}}M_{c}^{*}\left( \mathbf{q}+\mathbf{k}_{c}\right)
\times\frac{1}{a_{s}^{2}}\sum_{\mathbf{k}}\frac{\varphi_{k} e^{-\tilde{k}%
_{x}^{2}\sigma_{x}^{2}-\tilde{k}_{\perp}^{2}\sigma_{\perp}^{2}} e^{i\mathbf{k%
}\cdot\mathbf{r}_{jj^{\prime}}}}{k^{2}+1/l_{0}^{2}}.  \label{Kq1}
\end{align}
with 
\begin{align}
l^{2}_{0}=\frac{\hbar^{2}}{2m\delta_{0}}.
\end{align}
In Eq. (\ref{Kq1}), $\Omega=\Omega_{\uparrow}\Omega_{\downarrow}/E_{b}$ is
the effective Rabi frequency for pair transfer, $\mathbf{k} _{d}=\mathbf{k}%
_{\uparrow}-\mathbf{k}_{\downarrow}$, $\mathbf{k}_{c}=\mathbf{k}_{\uparrow }+%
\mathbf{k}_{\downarrow}$, $\mathbf{R}_{jj^{\prime}}=(\mathbf{r}_{j}+\mathbf{r%
}_{j^{\prime}})/2=\frac{(j+j^{\prime})a}{2}\mathbf{e}_{x}+y_{0}\mathbf{e}%
_{y}+z_{0}\mathbf{e}_{z}$, and $\mathbf{r}_{jj^{\prime}}=\mathbf{r}_{j}-%
\mathbf{r}_{j^{\prime}}=(j-j^{\prime})a\mathbf{e}_{x}$, $M_{c}^{*}(\mathbf{q}%
)=\exp[q_{x}^{2}\sigma_{x}^{2}/4-(q_{y}^{2}+q_{z}^{2})\sigma_{\perp}^{2}/4]$%
, $\tilde{\mathbf{k}}=\mathbf{k}+\mathbf{k}_{d}/2$ and $\tilde{k}%
_{\perp}^{2}=\tilde{k}_{y}^{2}+\tilde{k}_{z}^{2}$. Note that for a molecule
of a size $a_{s}\sim a$, the dominant contribution to the sum in Eq. (\ref%
{Kq1}) comes from $k\sim1/a_{s}$, and therefore under the condition $%
\sigma_{\perp}\ll\sigma_{x}\ll a$ imposed previously, one has $k\sigma
_{x(\perp)}\ll1$. Also taking into account $\mathbf{k}_{d}\sim1/a$, we can
thus simplify Eq. (\ref{Kq1}) by approximating $\exp[-\sigma_{x(\perp)}^{2}%
\tilde{k}_{x(\perp)}^{2}]\approx1$. Consequently, after transforming back to
the real space using $K_{jj^{\prime}}(\mathbf{r})=\int K_{jj^{\prime}}(%
\mathbf{q})e^{-i\mathbf{q}\cdot\mathbf{r}}d\mathbf{q}$, we obtain the
amplitude $K_{jj^{\prime}}(\mathbf{r})$ in the Hamiltonian (\ref{Hpair}) as 
\begin{align}
K_{jj^{\prime}}(\mathbf{r}) =i\Omega F(\mathbf{r}-\mathbf{R}%
_{jj^{\prime}})e^{i\mathbf{k}_{c}\cdot\mathbf{R}_{jj^{\prime}}}\sin\left( 
\frac {\mathbf{k}_{d}\mathbf{r}_{jj^{\prime}}}{2}\right) \frac{2a_{s}}{|%
\mathbf{r}_{jj^{\prime}}|}\frac{e^{-|\mathbf{r}_{jj^{\prime}}|/l_{0}}-e^{-|%
\mathbf{r}_{jj^{\prime}}|/a_{s}}}{1-a_{s}^{2}/l_{0}^{2}}.
\end{align}
with $F(\mathbf{r})=8\sqrt{2/\pi a_{s}^{3}}e^{-x^{2}/%
\sigma_{x}^{2}-(y^{2}+z^{2})/\sigma_{\perp}^{2}}$. In the weak interaction
regime $[n_{\text{M}} a_{s}^{3}\ll1]$ under consideration, 
\begin{align}
a_{s}/l_{0}\approx1+3\pi n_{\text{M}} a_{s}^{3}.  \label{Expand}
\end{align}
Thus to the leading order of $n_{\text{M}}a_{s}^{3}$, we obtain 
\begin{equation}
K_{jj^{\prime}}(\mathbf{r})=i\Omega\sin(\frac{\mathbf{k}_{d}\mathbf{r}%
_{jj^{\prime}}}{2})F(\mathbf{r}-\mathbf{R}_{jj^{\prime}})e^{i\mathbf{k}%
_{c}\cdot\mathbf{R}_{jj^{\prime}}-|r_{jj^{\prime}}|/a_{s}},  \label{Kpair}
\end{equation}
where $\Omega=\Omega_{\uparrow}\Omega_{\downarrow}/E_{b}$ is the effective
Rabi frequency for pair transfer, $F\left( \mathbf{r}\right) =8\sqrt{2/\pi
a_{s}^{3}}\exp[-x^{2}/\sigma_{x}^{2}-(y^{2}+z^{2})/\sigma_{\perp}^{2}]$, $%
\mathbf{k}_{d}=\mathbf{k}_{\uparrow}-\mathbf{k}_{\downarrow}$, $\mathbf{k}%
_{c}=\mathbf{k}_{\uparrow}+\mathbf{k}_{\downarrow}$, $\mathbf{R}%
_{jj^{\prime}}=(\mathbf{r}_{j}+\mathbf{r}_{j^{\prime}})/2=\frac{%
(j+j^{\prime})a}{2}\mathbf{e}_{x}+y_{0}\mathbf{e}_{y}+z_{0}\mathbf{e}_{z}$,
and $\mathbf{r}_{jj^{\prime}}=\mathbf{r}_{j}-\mathbf{r}_{j^{\prime}}=(j-j^{%
\prime })a\mathbf{e}_{x}$.

Equation (\ref{Kpair}) shows that, in order to engineer a $p$-wave pairing,
the condition $\mathbf{k}_{\uparrow}\neq\mathbf{k}_{\downarrow}$ must be
fulfilled, such that the amplitude $K_{jj^{\prime}}$ is antisymmetric, $%
K_{jj^{\prime}}=-K_{j^{\prime}j}$. In addition, $K_{jj^{\prime}}$ is in
general complex: $K_{jj^{\prime}}=|K_{jj^{\prime}}|e^{i\theta_{jj^{\prime}}}$
with $\theta_{jj^{\prime}}=\frac{\pi}{2}+\mathbf{k}_{c}\mathbf{R}%
_{jj^{\prime }}$. We can, however engineer a homogeneous phase $%
\theta_{jj^{\prime}}$ along the $x$-direction (direction of the lattice) by
choosing $\mathbf{k}_{c}$, $\,$say, along the $y$-axis, $\mathbf{k}_{c}=k_{c}%
\mathbf{e}_{y}$, such that $\theta_{jj^{\prime}}=\frac{\pi}{2}+k_{c}y_{0}$
depends only on the wire position in $y$-direction. Taking into account the
exponential fall-off $K_{jj^{\prime}}\sim e^{-|r_{jj^{\prime}}|/a_{s}}$ and $%
a_{s}\sim a$, we will consider $K_{jj^{\prime}}$ to be nonzero only for the
nearest-neighbor sites $\left\vert j-j^{\prime}\right\vert =1$ with $%
K_{j,j+1}=K_{j}$.

\subsection{Raman-induced hopping}

\begin{figure}[tb]
\includegraphics[width=0.45\textwidth]{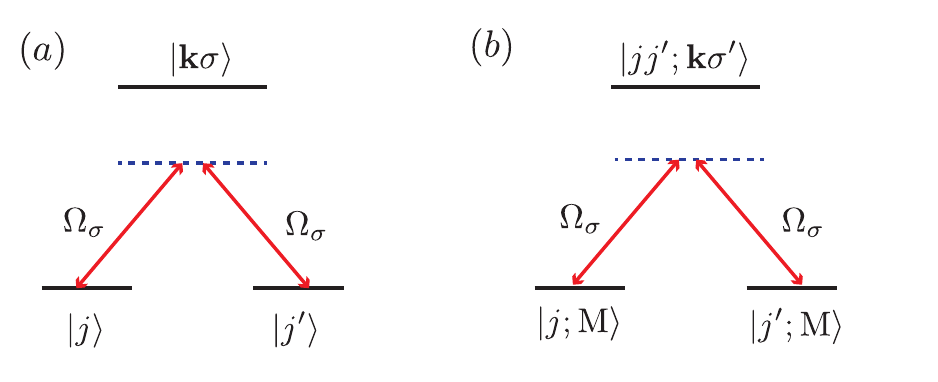}
\caption{ (Color online) Two Raman-induced processes contributing to the
correction in the hopping amplitude $\protect\delta J_{jj^{\prime}}$. In the
process (a), an optically trapped atom at the lattice site $j$ hops to the
lattice site $j^{\prime}$ via an intermediate single-atom process, in which
the atom changes its internal state to $|\protect\sigma\rangle$ and
untrapped from the lattice (and vice versa) under the Raman drive. (b)
describes a molecule-mediated hopping, where the intermediate process
involves breaking of a molecule $|\text{M}\rangle$ into an atom on the
lattice site and an unpaired $|\protect\sigma^{\prime}\rangle$ atom with
momenta $\mathbf{k}$, and vice versa.}
\label{Fig:3}
\end{figure}

Apart from inducing the pair transfer, the Raman processes also contribute
to the correction to the hopping term in Eq. (\ref{HL1}) via the
reservoir-mediated intermediated processes, corresponding to a Hamiltonian 
\begin{equation}
H_{\mathrm{J}}=\sum_{j,j^{\prime}}\delta J_{jj^{\prime}}\hat{a}_{j}^{\dag}%
\hat{a}_{j^{\prime}}+\text{h.c.}.  \notag
\end{equation}
As will be seen below, there are two processes (labeled as process $a$ and
process $b$, respectively) that contribute to $\delta J_{jj^{\prime}}$ (see
Fig. \ref{Fig:3}): 
\begin{align}
\delta J_{jj^{\prime}}=\delta J_{jj^{\prime}}^{a}+\delta J_{jj^{\prime}}^{b},
\end{align}
where the process $a$ involves only single-atom states, while the process $b$
also involves molecules in the reservoir. In what follows, we derive the
hopping amplitude $\delta J_{jj^{\prime}}^{a(b)}$ in detail, respectively.

(i) In the process $a$ (see Fig. \ref{Fig:3}(a)), 
an atom in the wire, say, at the lattice site $\mathbf{r}_{j^{\prime}}$
labeled as $|j^{\prime}\rangle$, when acted under the Hamiltonian $H_{\text{R%
}}$, flips its internal state from $|3\rangle$ to $|\sigma\rangle$ and
transfers into a unpaired atom moving in the BEC, labeled as $|\mathbf{k}%
\sigma\rangle$. Such single-atom transfer costs an energy 
\begin{equation*}
\Delta E^{(a)}=\epsilon_{\sigma}+\epsilon_{\mathbf{k}}^{0}+\hbar\omega
_{\sigma}-\epsilon_{3}^{\prime}=\epsilon_{\mathbf{k}}^{0}+\delta_{0}\neq0,
\end{equation*}
and is thereby off-resonant. Then, via the second Raman transition $H_{\text{%
R}}$, the atom in the state $|\mathbf{k}\sigma\rangle$ can be transferred
back into an atom in the wire, but at position $\mathbf{r}_{j}$, labeled as $%
|j\rangle$. Overall, one realizes a process $\hat{a}_{j}^{\dag
}a_{j^{\prime}}$ (and vice versa) with the second-order hopping amplitude
given by 
\begin{equation}
\delta J_{jj^{\prime}}^{(a)}=-\sum_{\sigma}\sum_{\mathbf{k}}\frac{\langle
j|H_{\text{R}}|\mathbf{k}\sigma\rangle\langle\mathbf{k}\sigma|H_{\text{R} }|{%
j^{\prime}}\rangle}{\epsilon_{k}^{0}+\delta_{0}}.  \label{Ja0}
\end{equation}
The matrix element in Eq. (\ref{Ja0}) can be straightforwardly evaluated
with Eq. (\ref{HRmoment}), and after some calculation, we obtain 
\begin{align}
\delta J_{jj^{\prime}}^{(a)} =-\frac{16\pi^{3/2}\Omega_{J}}{a_{s}^{2}}\frac{%
\sigma_{x}\sigma_{\perp}^{2}}{V} \sum_{\mathbf{k}}\frac{e^{-\sigma _{x}^{2}%
\tilde{k}_{x}^{2}-\sigma_{\perp}^{2}\tilde{k}_{\perp}^{2}+i(\mathbf{k}+%
\mathbf{k}_{c})\cdot\mathbf{r}_{jj^{\prime}}}}{k^{2}+1/l_{0}^{2}},
\label{Ja1}
\end{align}
with 
\begin{equation*}
\Omega_{J}=\frac{\Omega_{\uparrow}^{2}e^{i\mathbf{k}_{d}\mathbf{r}%
_{jj^{\prime}}/2}+\Omega_{\downarrow}^{2}e^{-i\mathbf{k}_{d}\mathbf{r}%
_{jj^{\prime}}/2}}{E_{b}}.
\end{equation*}
Having in mind $\tilde{k}_{x(\perp)}\sigma_{x(\perp)}\ll1$ under the
condition $\sigma_{\perp}\ll\sigma_{x}\ll a$, we evaluate Eq. (\ref{Ja1} )
as 
\begin{equation}
\delta J_{jj^{\prime}}^{(a)}=-\Omega_{J}\frac{4\pi^{1/2}\sigma_{x}\sigma_{%
\perp}^{2}}{a_{s}^{3}}\frac{e^{-|\mathbf{r}_{jj^{\prime}}|/l_{0}}}{| \mathbf{%
r}_{jj^{\prime}}|/a_{s}}e^{i\frac{\mathbf{k}_{c}}{2}\cdot\mathbf{r}
_{jj^{\prime}}}.  \label{Ja2}
\end{equation}
For weak interaction ($n_{\text{M}}a_{s}^{3}\ll1$), we submit the expansion (%
\ref{Expand}) into Eq. (\ref{Ja2}), and obtain in the first order in $n_{%
\text{M}}a_{s}^{3}$ (we set $\mathbf{k}_{c}=y_{0}\mathbf{e}_{y}$ such that $%
\mathbf{k}_{c}\cdot\mathbf{r}_{jj^{\prime}}=0$ as in the main text) 
\begin{align}
\delta J_{jj^{\prime}}^{(a)} & =-\Omega_{J}\frac{4\pi^{1/2}\sigma_{x}
\sigma_{\perp}^{2}}{a_{s}^{3}}\frac{e^{-|\mathbf{r}_{jj^{\prime}}|/a_{s}}}{| 
\mathbf{r}_{jj^{\prime}}|/a_{s}}  \notag \\
& \times(1-\frac{|\mathbf{r}_{jj^{\prime}}|}{a_{s}}3\pi n_{M}a_{s}^{3}).
\label{Jafinal}
\end{align}
It follows from Eq. (\ref{Jafinal}) that $|\mathbf{r}_{jj^{\prime}}|\sim
a_{s}$ because of the exponential decay $\exp(-|\mathbf{r}
_{jj^{\prime}}|/a_{s})$, and as a result, the contribution $\frac{|\mathbf{r}
_{jj^{\prime}}|}{a_{s}}n_{\text{M}}a_{s}^{3}\ll1$.

(ii) The process $b$ (see Fig. \ref{Fig:3} (b)) involves simultaneously an
atom at lattice site $\mathbf{r}_{j^{\prime}}$ and a molecule in the BEC,
labeled as $|j^{\prime};\text{M}\rangle$. The action of $H_{\text{R}}^{\text{%
M}}$ on the state $|j^{\prime};\text{M}\rangle$ leads to an intermediate
state where two atoms are in the wire and one unpaired atom moves in the
BEC, labeled as $|jj^{\prime};\mathbf{p}\sigma\rangle$, with an energy cost
given by 
\begin{align}
\Delta E^{(b)} & =[2\epsilon_{3}^{\prime}+\epsilon_{\sigma}+\epsilon _{%
\mathbf{p}}^{0}+2\epsilon_{\text{aM}}]-[\epsilon_{3}^{\prime}+\epsilon _{%
\text{mol}}+\epsilon_{\text{MM}}+\hbar\omega_{\sigma^{\prime}}]  \notag \\
& =\epsilon_{\mathbf{p}}^{0}+\delta_{0}.  \label{DE31}
\end{align}
Then, the action of $H_{\text{R}}^{\text{M}}$ on the intermediate state $%
|jj^{\prime};\mathbf{p}\sigma\rangle$ generates a process where a molecule
is created in the BEC and an atom remains at the lattice site $\mathbf{r}%
_{j} $ in the wire, labeled as $|j;\text{M}\rangle$. The overall amplitude
between the initial state $|j^{\prime};\text{M}\rangle$ and the final state $%
|j;\text{M}\rangle$ is given by 
\begin{equation}
\delta J_{jj^{\prime}}^{(b)}=\sum_{\mathbf{k}}\frac{\langle j;{\text{M}}|H^{%
\text{M}}_{\text{R}}|jj^{\prime};\mathbf{p}\sigma\rangle\langle jj^{\prime};%
\mathbf{p}\sigma|H^{\text{M}}_{\text{R}}|j^{\prime};{\text{M}}\rangle}{%
\epsilon_{k}^{0}+\delta_{0}}.  \label{Jb0}
\end{equation}
It follows from Eq. (\ref{HRMol}) that the matrix element of $H^{\text{M}}_{%
\text{R}}$ between the intermediate state $|jj^{\prime};\mathbf{p}%
\sigma\rangle$ and the state $|j^{\prime};{\text{M}}\rangle$ is derived as 
\begin{align}
\langle\mathbf{p}\uparrow;jj^{\prime}|H^{\text{M}}_{\text{R}}|j^{\prime };%
\text{M}\rangle & =-\frac{\Omega_{\downarrow}}{\sqrt{V}}M^{*}_{\mathbf{p}+%
\mathbf{k}_{\downarrow}} e^{i(\mathbf{p}+\mathbf{k}_{\downarrow })\cdot%
\mathbf{r}_{j^{\prime}}}\varphi_{\mathbf{k}}\sqrt{n_{\text{M}}}\delta _{%
\mathbf{p},\frac{\mathbf{q}}{2}-\mathbf{k}},  \notag \\
\langle\mathbf{p}\downarrow;jj^{\prime}|H^{\text{M}}_{\text{R}}|j^{\prime };%
\text{M}\rangle & =\frac{\Omega_{\uparrow}}{V}M^{*}_{\mathbf{p}+\mathbf{k}%
_{\uparrow}} e^{i(\mathbf{p}+\mathbf{k}_{\downarrow})\cdot \mathbf{r}%
_{j^{\prime}}}\varphi_{\mathbf{k}}\sqrt{n_{\text{M}}}\delta_{\mathbf{p},%
\frac{\mathbf{q}}{2}+\mathbf{k}},  \label{Jmatrixb}
\end{align}
where $n_{\text{M}}$ is the condensate density of molecular BEC.
Substituting Eq. (\ref{Jmatrixb}) into Eq. (\ref{Jb0}), and after
straightforward calculation, we obtain (to the first order of $n_{\text{M}%
}a_{s}^{3}\ll1$), 
\begin{align}
\delta J_{jj^{\prime}}^{(b)}\approx\Omega_{J}\left( \pi n_{\text{M}}
a_{s}^{3}\right) \frac{4\sqrt{\pi}\sigma_{x}\sigma_{\perp}^{2}}{a_{s}^{3}}
e^{-\frac{|\mathbf{r}_{jj^{\prime}}|}{a_{s}}} \left( 1+\frac{|\mathbf{r}
_{jj^{\prime}}|}{a_{s}}\right) .  \label{Jb1}
\end{align}

Consequently, combination of Eqs. (\ref{Ja2}) and (\ref{Jb1}) yields (in the
limit $n_{\text{M}}a_{s}^{3}\ll1$) 
\begin{align}
\delta J_{jj^{\prime}} =(4\sqrt{\pi})(\sigma_{x}\sigma_{%
\perp}^{2}/a_{s}^{3})e^{i\mathbf{k}_{c}\mathbf{r}_{jj^{\prime}}/2}\Omega_{J} 
\frac {e^{-|\mathbf{r}_{jj^{\prime}}|/a_{s}}}{|\mathbf{r}_{jj^{%
\prime}}|/a_{s}} \left[ 1-\pi n_{\text{M}}a_{s}^{3}\frac{|\mathbf{r}%
_{jj^{\prime}}|}{a_{s}} \left( 4+\frac{|\mathbf{r}_{jj^{\prime}}|}{a_{s}}%
\right) \right] .  \label{deltaJ2}
\end{align}
Note that by tuning $\mathbf{k}_{c}=k_{c}\mathbf{e}_{y}$, the phase factor $%
\exp(i\mathbf{k}_{c}\mathbf{r}_{jj^{\prime}}/2)$ in Eq. (\ref{deltaJ2})
vanishes, and $\delta J_{jj^{\prime}}$ can be made real by choosing $%
\Omega_{1}=\Omega_{2}$. Similar to the pair transfer amplitude, $\delta
J_{jj^{\prime}}$ also decays exponentially with increasing $\left\vert
j-j^{\prime}\right\vert$, and therefore, we will take into account only the
nearest-neighbor contribution $\delta J_{jj+1}$. As a result, the
nearest-neighbour hopping amplitude $J_{0}$ in Eq. (\ref{HL}) will be
renormalized to 
\begin{equation}
J=J_{0}+\delta J_{j,j+1}.  \label{Jfinal}
\end{equation}

Collecting above results, it is clear that after elimination of the Raman
processes, we arrive at the effective Hamiltonian ({\ref{Htotal0}}) in the
main text for the setup. There, the renormalized chemical potential for a
fermionic atom in the wire is given by $\mu_{0}=\epsilon_{3}^{\prime}
-\delta_{\text{R}}/2$.

\subsection{Reservoir in the regime of BEC-BCS crossover}

In above derivations, we note that when $n_{\text{M}}a_{s}^{3}$ approaches
unity, $n_{\text{M}}a_{s}^{3}<1$, the intermediate processes involving
molecules in the BEC plays increasingly important role compared to
single-particle process, and previous expansions in terms of $n_{M}a_{s}^{3}$
are no longer valid. In order to evaluate the Raman-induced pairing
amplitude $K_{j,j^{\prime}}(\mathbf{r})$ and hopping amplitude $\delta
J_{j,j^{\prime}}$ in this case, we use the theory of BCS-BEC crossover \cite%
{Ketterle08}, which corresponds to considering a reservoir in the molecular
side of the BCS-BEC crossover regime. \newline

We begin with writing the particle operator $\hat{c}_{\mathbf{k}\sigma}$
introduced in Eq. (\ref{Chisigma}) in terms of the Bogoliubov quasi-particle
operators $\hat{\gamma}_{\mathbf{k}\sigma}$: 
\begin{align}
\hat{\gamma}_{\mathbf{k}\uparrow} & =u_{\mathbf{k}}\hat{c}_{\mathbf{k}
\uparrow}-\upsilon_{\mathbf{k}}\hat{c}_{-\mathbf{k}\downarrow}^{\dag }, 
\notag \\
\hat{\gamma}_{-\mathbf{k}\downarrow}^{\dag} & =u_{\mathbf{k}}\hat {c}_{- 
\mathbf{k}\downarrow}^{\dag}+\upsilon_{\mathbf{k}}\hat{c}_{\mathbf{k}
\uparrow}^{\dag},  \label{Gamma}
\end{align}
where $u_{\mathbf{k}}$ and $\upsilon_{\mathbf{k}}$ are the standard wave
functions of the Bogoliubov quasi-particles, and $E_{k}$ is the
corresponding excitation energy given by 
\begin{equation*}
E_{k}=\sqrt{\Delta_{b}^{2}+(\epsilon_{k}^{0}-\mu_{b})^{2}},
\end{equation*}
where $\epsilon_{\mathbf{k}}^{0}$ is the kinetic energy of a free atom,
while $\mu_{b}$ and $\Delta_{b}$ are the chemical potential and the gap of
the superconducting reservoir, respectively. In the BCS-BEC crossover
regime, both $\mu_{b}$ and $\Delta_{b}$ are self-consistently determined
from the gap equation and the number equation (see Ref. \cite{Ketterle08}
for expressions and the derivations). While subsequent derivations apply to
the whole crossover regime, for our purpose, here we will limit ourselves to
the molecular side of the crossover.

Substituting Eqs. (\ref{Gamma}) into Eqs. (\ref{HRmoment}) and (\ref{HRMol}%
), and using, as before, the second-order perturbation theory, we obtain the
paring amplitude 
\begin{align}
K_{jj^{\prime}}\left( \mathbf{r}\right) =-16i\Omega\sin\left( \frac{\mathbf{k%
}_{d}\cdot\mathbf{r}_{jj^{\prime}}}{2}\right) F\left( \mathbf{r}\!-\mathbf{R}%
_{jj^{\prime}}\right) e^{i\mathbf{k}_{c}\cdot\mathbf{R}_{jj^{\prime}}}\frac{%
E_{b}}{2V}\sum_{k}\frac{\Delta_{\text{b}}}{E_{k}^{2}}e^{i\mathbf{k}\cdot%
\mathbf{r}_{jj^{\prime}}},  \label{Kinter}
\end{align}
and the hopping amplitude 
\begin{equation}
\delta J_{jj^{\prime}}=\Omega_{J}\left( 8\pi^{3/2}\sigma_{x}\sigma_{\perp
}^{2}\right) \frac{E_{b}}{V}\sum_{\mathbf{k}}\frac{\epsilon_{ \mathbf{k}%
}^{0}-\mu_{b}}{E_{k}^{2}}e^{i\mathbf{k}\cdot\mathbf{r} _{jj^{\prime}}}.
\label{Jinter}
\end{equation}
After performing the summations in $\mathbf{k}$ in Eqs. (\ref{Kinter}) and ( %
\ref{Jinter}), respectively, we arrive at 
\begin{align}
K_{jj^{\prime}}\left( \mathbf{r}\right) & =i\frac{4\Omega}{\pi}\sin\left( 
\frac{\mathbf{k}_{d}\mathbf{r}_{jj^{\prime}}}{2}\right) \frac{M\left( \frac{|%
\mathbf{r}_{jj^{\prime}}|}{a_{s}}\right) }{\sqrt{ n_{\text{M}}a_{s}^{3}}}
F\left( \mathbf{r}-\mathbf{R}_{jj^{\prime}}\right) e^{i\mathbf{k} _{c}\cdot%
\mathbf{R}_{jj^{\prime}}},  \label{Kinter1} \\
\delta J_{jj^{\prime}} & =4\pi^{1/2}\Omega_{\text{J}}\frac{%
\sigma_{x}\sigma_{\perp}^{2}}{a_{s}^{3}}Q\left( \frac{|\mathbf{r}%
_{jj^{\prime}}| }{a_{s}}\right) ,  \label{Jinter1}
\end{align}
where we have introduced the functions 
\begin{align}
& M(d)=\frac{e^{-d\sqrt{\rho}\sin\frac{\theta}{2}}}{d}\sin\left[ d\sqrt{\rho}%
\cos\frac{\theta}{2}\right] ,  \notag \\
& Q(d)=\frac{e^{-d\sqrt{\rho}\sin\frac{\theta}{2}}}{d}\cos\left[ d\sqrt{ \rho%
}\cos\frac{\theta}{2}\right] ,  \notag
\end{align}
with 
\begin{align}
& \rho=\frac{\sqrt{\mid\mu_{b}\mid^{2}+\mid{\Delta}_{b}\mid^{2}}}{ E_{F}}; 
\notag \\
& \sin^{2}\left( \frac{\theta}{2}\right) =\frac{1}{2}\left[ 1-\frac {\mu_{b}%
}{\sqrt{\mid\mu_{b}\mid^{2}+\mid{\Delta}_{b}\mid^{2}}}\right] ;  \notag \\
& \cos^{2}\left( \frac{\theta}{2}\right) =\frac{1}{2}\left[ 1+\frac {\mu_{b}%
}{\sqrt{\mid\mu_{b}\mid^{2}+\mid{\Delta}_{b}\mid^{2}}}\right] .  \notag
\end{align}
Here, $E_{F}={\hbar^{2}(6\pi^{2}n_{\text{M}})^{2/3}}/{2m}$ is the Fermi
energy of the reservoir.

\subsection{Optimal conditions for Majorana edge states}

\begin{figure}[tb]
\includegraphics[width=0.35\textwidth]{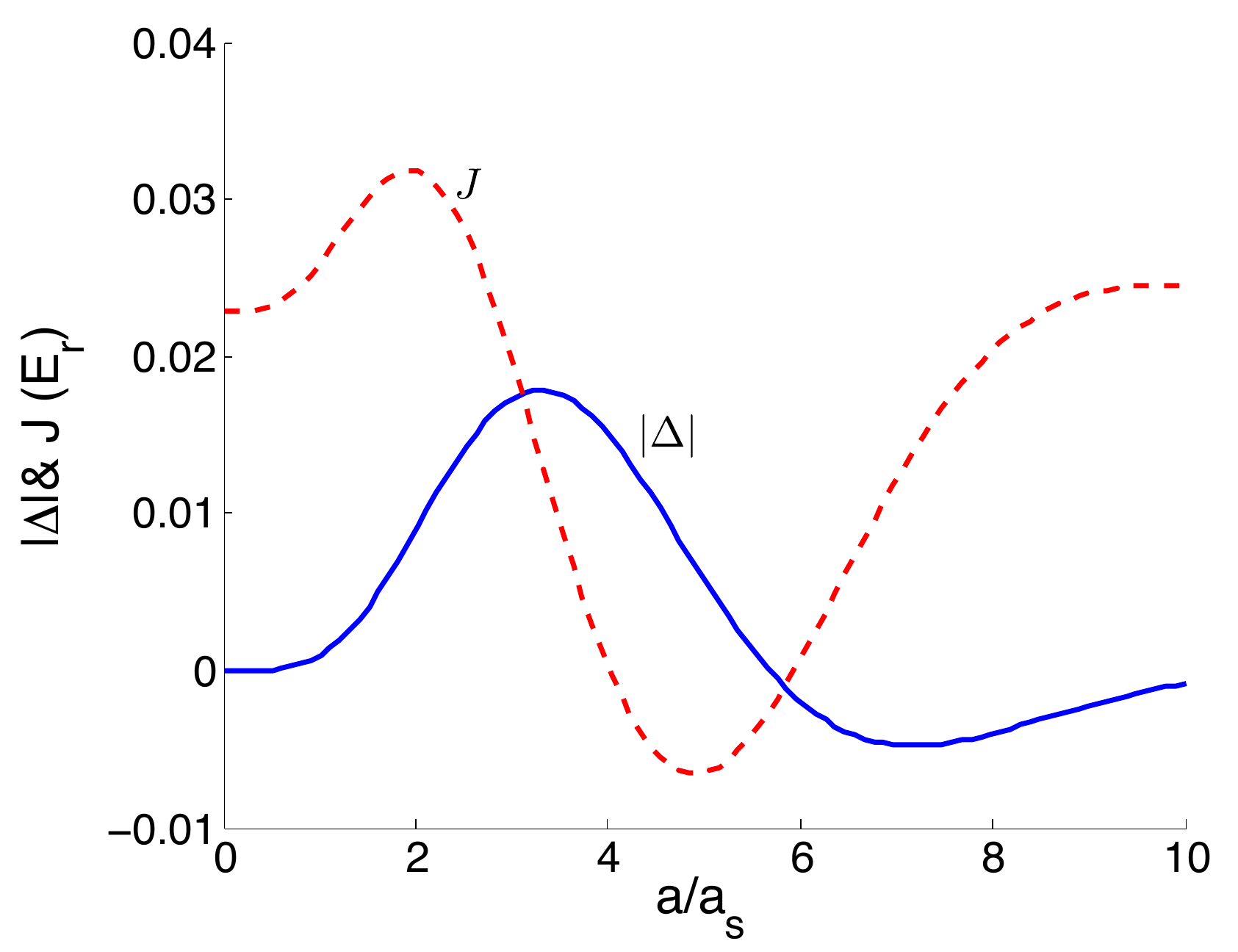}
\caption{ (Color online) The pairing amplitude $|\Delta|$ (solid line) and
the hopping parameter $J$ (dashed line) in the Kitaev Hamiltonian in the
units of the recoil energy as a function of the ratio $a/a_{s}$ between the
lattice constant $a$ and the molecular size $a_{s}$, based on Eqs. (\protect
\ref{Delta}) and (\protect\ref{Kinter1}) for $|\Delta|$ and Eqs. (\protect
\ref{Jfinal}) and (\protect\ref{Jinter1}) for $J$. Other parameters are $%
\protect\sigma_{x}\protect\sigma_{\perp}^{2}/a^{3}=0.03$, and $n_{\text{M}%
}a_{s}^{3}=0.01$. The maximal value for the pairing amplitude $%
|\Delta|\sim0.02E_{r}$ occurs for $a/a_{s}\sim3$. }
\label{Fig:4}
\end{figure}

\begin{figure}[tb]
\includegraphics[width=0.35\textwidth]{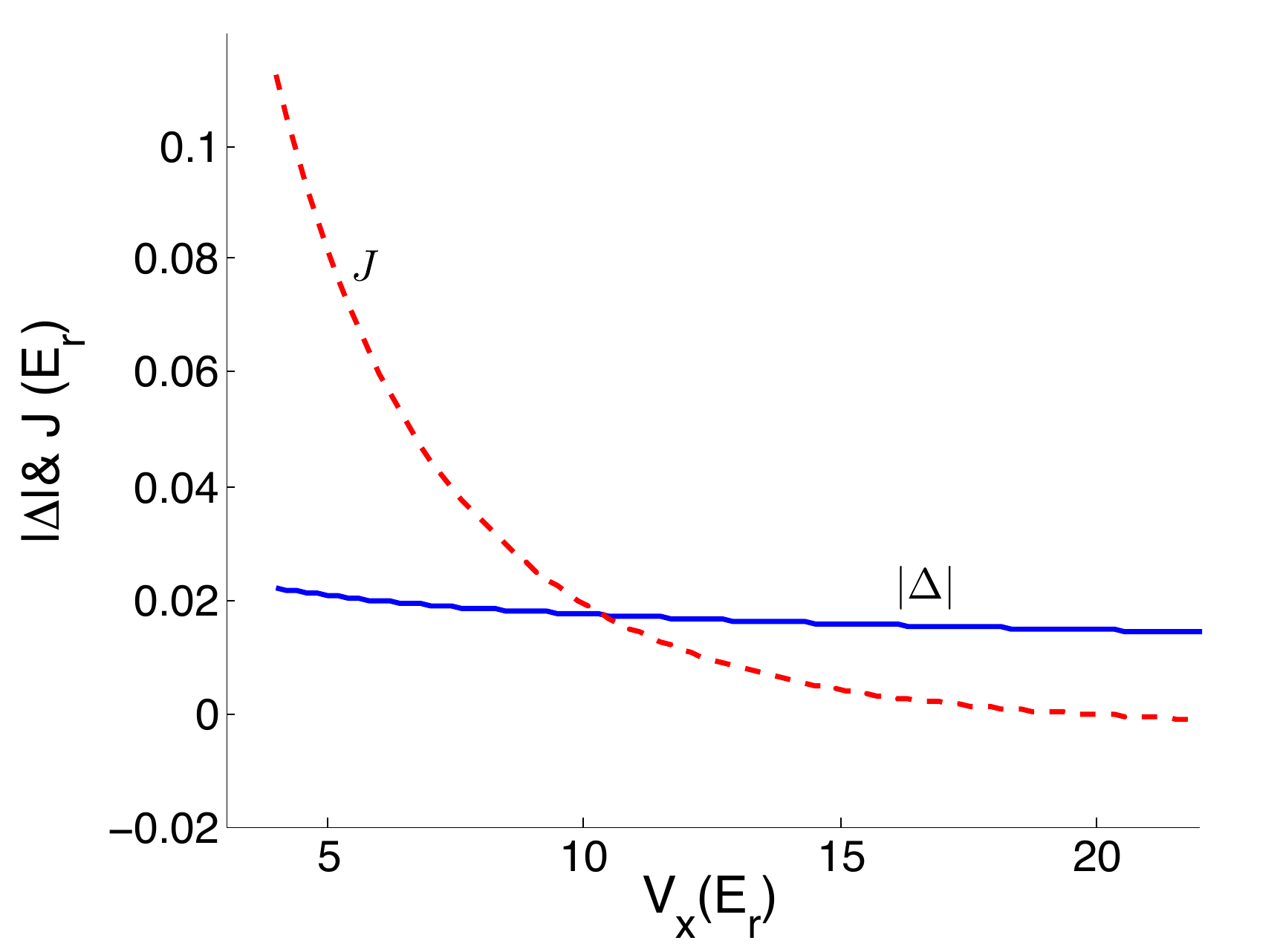}
\caption{ (Color online) The pairing amplitude $|\Delta|$ (solid line) and
the hopping amplitude $J$ (dashed line) in the Kitaev Hamiltonian in units
of the recoil energy as a function of the lattice depth $V_{x}$ for $%
a_{s}/a=1/3$ and $n_{\text{M}}a_{s}^{3}=0.01$. The optimal condition $%
|\Delta| \approx J$ for the localization of the Majorana modes is achieved
for $V_{x}\sim10E_{r}$.}
\label{Fig:5}
\end{figure}

We now look for the optimal conditions, under which (1) the overlap between
the two Majorana edge modes are minimized, i.e. the Majoranas modes ares
strongly localized at the edges; (2) the gap in the bulk spectrum is as
large as possible. This can be achieved by tuning $J\sim|\Delta|$ and $%
\mu_{f}\sim0$ in Eq. (\ref{HK}) in the main text (here we drop the subscript
in $\Delta_{\phi_{0}}$ for clarity), corresponding to the realization of a
nearly ideal Kitaev chain. The chemical potential $\mu_{f}\approx0$ can be
realized via a fine control of the two-photon detuning $\delta_{R}$, as
described earlier. On the other hand, the hopping amplitude $J$ [see Eqs. ( %
\ref{Jfinal}) and (\ref{Jinter1})] and pairing amplitude $|\Delta|$ [see
Eqs. (\ref{Delta}) and (\ref{Kinter1}) ] depend on characteristic parameters
for the reservoir (e.g. molecular size $a_{s}$, density $n_{M}$) and for the
wire (e.g. lattice depth $V_{x}$, lattice constant $a$). In order to find
the optimal ratio $a/a_{s}$ between the lattice constant $a$ and the
molecule size $a_{s}$, we scan $|\Delta|$ and $J$ as a function of $a/a_{s}$
while fixing other parameters in Eqs. (\ref{Kinter1}) and (\ref{Jinter1}),
as illustrated in Fig. \ref{Fig:4}. There, for typical parameters $%
\sigma_{x}\sigma_{\perp}^{2}/a^{3}=0.03$ and $n_{\text{M}}a_{s}^{3}=0.01$,
we find a maximum gap arising at $a_{s}\sim a/3$. Then, we fix the molecular
size at $a_{s}=a/3$, and scan $|\Delta|$ and $J$ as a function of the
lattice depth $V_{x}$, respectively, as shown in Fig. \ref{Fig:5}. We see
that the condition $J\sim|\Delta|$ can be achieved for $V_{x}\sim10E_{r}$
with $E_{r}=\hbar^{2}/2m\lambda^{2}$ denoting the recoil energy, which is
well in reach in current experiment facilities.

\section{Majorana edge states in a finite Kitaev wire}

\label{AppendixB}

We present in this Appendix a detailed derivation of the analytical
expressions for the wave function and eigenenergy of the Majorana edge
states in a finite Kitaev chain of $L$ sites with open boundary conditions,
described by the Hamiltonian 
\begin{equation*}
H_{K}=\sum_{j=1}^{L-1}[-J\hat{a}_{j}^{\dag }\hat{a}_{j+1}+\Delta \hat{a}_{j}%
\hat{a}_{j+1}+\text{h.c.}]-\sum_{j=1}^{L}\mu \hat{a}_{j}^{\dag }\hat{a}_{j}.
\end{equation*}%
Without loss of generality, we consider the hopping amplitude $J$ and the
gap parameter $\Delta $ as real and positive. Our starting point is the
Bogoliubov-de Gennes equations for the Bogoliubov amplitudes $u_{j,n}$ and $%
v_{j,n}$ at sites $j=1,...,L$, 
\begin{align}
-J(u_{j+1,n}+u_{j-1,n})-\mu u_{j,n}+\Delta (\upsilon _{j-1,n}-\upsilon
_{j+1,n})& =E_{n}u_{j,n},  \notag \\
-J(\upsilon _{j+1,n}+\upsilon _{j-1,n})-\mu \upsilon _{j,n}+\Delta \left(
u_{j-1,n}-u_{j+1,n}\right) & =-E_{n}\upsilon _{j,n},  \label{uv}
\end{align}%
supplemented with the open boundary conditions 
\begin{equation}
u_{0,n}=\upsilon _{0,n}=u_{L+1,n}=\upsilon _{L+1,n}=0.  \label{Boundary}
\end{equation}%
Here, the definition of $u_{j,n}$ and $v_{j,n}$ has been formally extended
to the sites $j=0$ and $j=L+1$. Next, we will look for the edge states $%
(u_{j,M},\upsilon _{j,M})$ with the energy $E_{M}$ that satisfy the BdG
equations (\ref{uv}) under the boundary condition in Eq. (\ref{Boundary}),
in the regime $|\mu |<2J$.

To this end, let us introduce new functions 
\begin{equation}
f_{\pm ,j}=u_{jM}\pm \upsilon _{jM}.  \label{fpm}
\end{equation}%
In terms of $f_{\pm ,j}$, the BdG equations (\ref{uv}) is transformed into
(for $j=1,...,L$) 
\begin{align}
-J\left( f_{+,j+1}\!+\!f_{+,j-1}\right) -\mu f_{+,j}+\Delta
(f_{+,j-1}-f_{+,j+1})& =E_{M}f_{-,j},  \notag \\
-J\left( f_{-,j+1}\!+\!f_{-,j-1}\right) -\mu f_{-,j}-\Delta
(f_{-,j-1}-f_{-,j+1})& =E_{M}f_{+,j},  \label{bdgfE}
\end{align}%
which is supplemented with the corresponding open boundary conditions at $%
j=0 $ and $j=L+1$ 
\begin{equation}
f_{\pm ,0}=0,\hspace{6mm}f_{\pm ,L+1}=0.  \label{fpmBoundary}
\end{equation}

Equations (\ref{bdgfE}) can be solved by the following ansatz 
\begin{align}
f_{+,j}=\alpha z^{j}, \hspace{6mm} f_{-,j}=\beta z^{j}.  \label{fansatz}
\end{align}
Substitution of Eqs. (\ref{fansatz}) into Eqs. (\ref{bdgfE}) yields two
coupled equations 
\begin{align}
F_{1}(z)\alpha+E_{M} \beta=0, \hspace{2mm}E_{M} \alpha+F_{2}(z)\beta=0,
\label{bdgfE_2}
\end{align}
with 
\begin{align}
\!\!\!\!F_{1}(z) & =(J+\Delta)z\!+\mu\!+(J-\Delta)/z,  \label{F1} \\
\!\!\!\!F_{2}(z) & =(J-\Delta)z\!+\mu\!+(J+\Delta)/z\!=\!F_{1}(1/z).
\label{F2}
\end{align}
From the condition for the existence of nonzero solutions $(\alpha,\beta)$
to Eq. (\ref{bdgfE_2}), we immediately obtain 
\begin{align}
F_{1}(z)F_{2}(z)=E_{M}^{2}.  \label{Eigen}
\end{align}

\subsection{The case of $L\rightarrow\infty$}

First, we consider the limiting case $L\rightarrow \infty $, for which $%
E_{M}=0$ is exact. Equation (\ref{bdgfE_2}) can immediately be decoupled
into two equations 
\begin{align}
F_{1}(z)& =0,  \label{F1_0} \\
F_{2}(z)& =0,  \label{F2_0}
\end{align}%
which can be easily solved. Denoting the solutions to Eq. (\ref{F1_0}) as $%
z_{1},z_{2}$ and that to Eq. (\ref{F2_0}) as $z_{3},z_{4}$, we find 
\begin{equation*}
z_{1}=x_{+},\hspace{3mm}z_{2}=x_{-},\hspace{3mm}z_{3}=x_{+}^{-1},\hspace{3mm}%
z_{4}=x_{-}^{-1},
\end{equation*}%
with 
\begin{equation}
x_{\pm }=\frac{-\mu \pm \sqrt{\mu ^{2}-4(J^{2}-\Delta ^{2})}}{2(J+\Delta )}.
\label{xpm}
\end{equation}%
In the topological phase of the chain when $|\mu |<2J$, it follows from Eq. (%
\ref{xpm}) that $|x_{\pm }|<1$. As a result, the solutions in Eq. (\ref{F1_0}%
), $z_{1,2}^{j}=x_{\pm }^{j}$, decay exponentially with increasing $j$;
whereas, the solutions in Eq. (\ref{F2_0}), $z_{3,4}^{j}=x_{\pm }^{-j}$,
decay exponentially with decreasing $j$. Therefore, we see that, for a
Kitaev wire with $L\rightarrow \infty $, there exists an exact solution of
BdG Eq. (\ref{bdgfE}) corresponding $E_{M}=0$ with $f_{+,j}\sim
(x_{+}^{j}-x_{-}^{j})$ and $f_{-.j}\sim (x_{+}^{L+1-j}-x_{-}^{L+1-j})$ that
fulfill the boundary condition of Eq. (\ref{fpmBoundary}). \newline

\subsection{The case of finite $L$}

Now, we turn to the case when $L$ is finite but large, in which $E_{M}$ is
nonzero but exponentially small. Since Eq. (\ref{Eigen}) cannot be decoupled
for $E_{M}\neq0$, the corresponding four solutions become $E_{M}$-dependent.
Let us label these solutions as $z_{i}(E_{M})$ (for $i=1,2,3,4$), so that in
the limit $E_{M}\rightarrow0$ they approaches $z_{i}$ in an infinite wire,
i. e. $z_{i}(E_{M}\rightarrow0)=z_{i}$. Notice that, as $F_{1}(z)=F_{2}(1/z)$
from Eq. (\ref{F2}), we have the relation $z_{3}(E_{M})=1/z_{1}(E_{M})$ and $%
z_{4}(E_{M})=1/z_{2}(E_{M})$ between the pair of solutions $z_{1,2}(E_{M})$
and $z_{3,4}(E_{M})$. The exact expressions for $z_{i}(E_{M})$ can be found,
by casting Eq. (\ref{Eigen}) into a quadratic equation $(J^{2}-\Delta
^{2})y^{2}+4\mu Jy+(4\Delta^{2}+\mu^{2}-E_{M}^{2})=0$ for $y=z+z^{-1}$.
However, they are very lengthy and will not be presented here.

Corresponding to each $z_{i}(E_{M})$, Equation (\ref{bdgfE_2}) allows us to
derive the ratio between $\alpha^{(i)}$ and $\beta^{(i)}$. Specifically, for
the pair of solutions $z_{1,2}(E_{M})$, by noting $F_{1}\left( x_{\pm
}\right) =0$ but $F_{2}\left( x_{\pm}\right) \neq0$, we use Eq. (\ref{Eigen}%
) to obtain $F_{1}\left[ z_{1,2}(E_{M})\right] =E_{M}^{2}/F_{2}\left[
z_{1,2}(E_{M})\right] $, which is substituted into Eq. (\ref{bdgfE_2}) to
give (for $i=1,2$) 
\begin{align}
{\beta}^{(i)}=-\frac{E_{M}}{F_{2}\left[ z_{i}(E_{M})\right] }\alpha^{(i)}.
\label{ratio1}
\end{align}
On the other hand, for the pair of solutions $%
z_{3,4}(E_{M})=1/z_{1,2}(E_{M}) $, we recall $F_{2}\left( 1/x_{\pm}\right)
=0 $ but $F_{1}\left( 1/x_{\pm }\right) =F_{2}(x_{\pm})\neq0$, and thus
substitute $F_{2}\left[ z_{3,4}(E_{M})\right] =E_{M}^{2}/F_{1}\left[
z_{3,4}(E_{M})\right] $ into Eq. (\ref{bdgfE_2}) to obtain (for $i=3,4$) 
\begin{align}
{\alpha}^{(i)}=-\frac{E_{M}}{F_{1}\left[ z_{i}(E_{M})\right] }{\beta}^{(i)}.
\label{ratio2}
\end{align}

Now, we are readily to find the general solutions to Eq. (\ref{bdgfE}) with
Eqs. (\ref{fansatz}), (\ref{ratio1}) and (\ref{ratio2}). Keeping in mind
that $z_{3}(E_{M})=1/z_{1}(E_{M})$ and $z_{4}(E_{M})=1/z_{2}(E_{M})$, we can
express the general solutions of Eq. (\ref{bdgfE}) as 
\begin{align}
f_{+,j}& =\alpha ^{(1)}z_{1}^{j}(E_{M})+\alpha ^{(2)}z_{2}^{j}(E_{M})+\tilde{%
\alpha}^{(3)}z_{1}^{L+1-j}(E_{M})+\tilde{\alpha}^{(4)}z_{2}^{L+1-j}(E_{M}), 
\notag \\
f_{-,j}& =\beta ^{(1)}z_{1}^{j}(E_{M})+\beta ^{(2)}z_{2}^{j}(E_{M})+\tilde{%
\beta}^{(3)}z_{1}^{L+1-j}(E_{M})+\tilde{\beta}^{(4)}z_{2}^{L+1-j}(E_{M}),
\label{f}
\end{align}%
with $\tilde{\alpha}^{(3)}={\alpha }^{(3)}/z_{1}^{L+1}$, $\tilde{\alpha}%
^{(4)}={\alpha }^{(4)}/z_{2}^{L+1}$, $\tilde{\beta}^{(3)}={\beta }%
^{(3)}/z_{1}^{L+1}$, and $\tilde{\beta}^{(4)}={\beta }^{(4)}/z_{2}^{L+1}$.
In the limit $E_{M}\rightarrow 0$, Equations (\ref{f}) naturally approaches
the corresponding expressions in a $L\rightarrow \infty $ chain. After
imposing the open boundary conditions in Eq. (\ref{Boundary}), we obtain the
following equations

\begin{align}
-\frac{E_{M}}{F_{2}\left[ z_{1}(E_{M})\right] }\alpha^{(1)}-\frac{E_{M}}{%
F_{2}\left[ z_{2}(E_{M})\right] }\alpha^{(2)} +z_{1}^{L+1}(E_{M}) \tilde{%
\beta}^{(3)}+z_{2}^{L+1}(E_{M})\tilde{\beta}^{(4)} & =0,  \label{Eqcoef2} \\
z_{1}^{L+1}(E_{M})\alpha^{(1)}+z_{2}^{L+1}(E_{M})\alpha^{(2)}-\frac{E_{M}}{%
F_{1}\left[ z_{3}(E_{M}\right] }\tilde{\beta}^{(3)}-\frac{E_{M}}{F_{1}\left[
z_{4}(E_{M}\right] }\tilde{\beta}^{(4)} & =0,  \label{Eqcoef3} \\
\alpha^{(1)}+\alpha^{(2)}-\frac{E_{M}z_{1}^{L+1}(E_{M})}{F_{1}\left[
z_{3}(E_{M})\right] }\tilde{\beta}^{(3)}-\frac{E_{M}z_{2}^{L+1}(E_{M})}{F_{1}%
\left[ z_{4}(E_{M})\right] }\tilde{\beta}^{(4)} & =0,  \label{Eqcoef1} \\
-\frac{E_{M}z_{1}^{L+1}(E_{M})}{F_{2}\left[ z_{1}(E_{M})\right] }%
\alpha^{(1)}-\frac{E_{M}z_{2}^{L+1}(E_{M})}{F_{2}\left[ z_{2}(E_{M}\right] }%
\alpha^{(2)} +\tilde{\beta}^{(3)}+\tilde{\beta}^{(4)} & =0.  \label{Eqcoef4}
\end{align}

The resolutions of Eqs. (\ref{Eqcoef2})-(\ref{Eqcoef4}) and the exact
determination of $E_{M}$ are possible but very complicated. For our purpose,
it suffices to noting the exponentially smallness of $E_{M}$ and thus
seeking approximate solutions in the linear order of $E_{M}$. Keeping in
mind $x_{\pm }^{L+1}\approx e^{-La/l_{M}}\sim E_{M}$, we ignore terms $\sim
E_{M}^{2}$ and beyond, such that Eqs. (\ref{Eqcoef2})-(\ref{Eqcoef4}) reduce
to 
\begin{align}
-\frac{E_{M}}{s_{+}}\alpha ^{(1)}-\frac{E_{M}}{s_{-}}\alpha ^{(2)}+\tilde{%
\beta}^{(3)}x_{+}^{L+1}+\tilde{\beta}^{(4)}x_{-}^{L+1}& =0,  \notag \\
\alpha ^{(1)}x_{+}^{L+1}\!+\!\alpha ^{(2)}x_{-}^{L+1}\!-\!\frac{E_{M}}{s_{+}}%
\tilde{\beta}^{(3)}\!-\!\frac{E_{M}}{s_{-}}\tilde{\beta}^{(4)}& =0,  \notag
\\
\alpha ^{(1)}\!+\!\alpha ^{(2)}=0,\tilde{\beta}^{(3)}+\tilde{\beta}^{(4)}&
=0,  \label{Eqcoeff}
\end{align}%
where we have introduced $s_{\pm }=F_{2}\left[ x_{\pm }\right] =F_{1}\left[
x_{\pm }^{-1}\right] $ given by 
\begin{equation*}
s_{\pm }=\frac{2\Delta \left( \mu \Delta \pm J\sqrt{\mu ^{2}-4(J^{2}-\Delta
^{2})}\right) }{\Delta ^{2}-J^{2}}.
\end{equation*}%
Consequently by solving Eq. (\ref{Eqcoeff}), we can obtain the eigenenergy 
\begin{equation}
E_{M}=\Big|\frac{\Delta (4J^{2}-\mu ^{2})(x_{+}^{L+1}-x_{-}^{L+1})}{J(\Delta
+J)(x_{+}-x_{-})}\Big|,  \label{EMfinal}
\end{equation}%
and the corresponding eigenfunctions 
\begin{align}
f_{+,j}& =A\Big[x_{+}^{j}-x_{-}^{j}-\frac{E_{M}}{s_{+}}x_{+}^{L+1-j}+\frac{%
E_{M}}{s_{-}}x_{-}^{L+1-j}\Big],  \notag \\
f_{-,j}& =A\Big[x_{+}^{L+1-j}\!-\!x_{-}^{L+1-j}\!-\!\frac{E_{M}}{s_{+}}%
x_{+}^{j}+\frac{E_{M}}{s_{-}}x_{-}^{j}\Big].  \label{fREfinal}
\end{align}%
We emphasize that $f_{\pm ,j}$ in Eq. (\ref{fREfinal}) fulfills the open
boundary condition in Eq. (\ref{Boundary}) approximately (to the order of $%
\sim E_{M}^{2}$). As is manifest from Eq. (\ref{fREfinal}), for large but
finite $L$, $f_{+,j}$ is localized near the left edge but involves small
admixture (at the order $\sim E_{M}$) from $x_{\pm }^{L+1-j}$ which decays
from the right edge, while $f_{-,j}$ is localized near the right end with
small admixtures ($\sim E_{M}$) from $x_{\pm }^{j}$ that decays from the
left. The coefficient $A$ in Eq. (\ref{fREfinal}) can be determined from the
renormalization condition $\sum_{j}|u_{jM}|^{2}+|\upsilon _{jM}|^{2}=1$
(ignoring terms $\sim E_{M}^{2}$). In this way, we obtain 
\begin{equation}
A=\sqrt{\frac{\Delta (4J^{2}-\mu ^{2})}{J(\mu ^{2}+4\Delta ^{2}-4J^{2})}},
\label{A1}
\end{equation}%
for $\mu ^{2}-4(J^{2}-\Delta ^{2})>0$, when $x_{\pm }$ and $s_{\pm }$ are
real; and 
\begin{equation*}
A=-i\sqrt{\frac{\Delta (4J^{2}-\mu ^{2})}{J(4J^{2}-4\Delta ^{2}-\mu ^{2})}},
\end{equation*}%
for $4(J^{2}-\Delta ^{2})-\mu ^{2}>0$, when $x_{+}=x_{-}^{\star }=(-\mu +i%
\sqrt{4J^{2}-4\Delta ^{2}-\mu ^{2}})/2(\Delta +J)$ and $s_{+}^{\ast }=s_{-}$%
. Consequently, in both regimes, the resulting $f_{\pm ,j}$ are real. Having
found $f_{\pm ,j}$, we can obtain the expressions for $u_{j,M},\upsilon
_{j,M}$ ( in the linear order of $E_{M}$) from Eq. (\ref{fpm}). The results
are 
\begin{eqnarray}
u_{jM} &=&\frac{A}{2}\Big[(1-\frac{E_{M}}{s_{+}}%
)(x_{+}^{j}+x_{+}^{L+1-j})-(1-\frac{E_{M}}{s_{-}})(x_{-}^{j}+x_{-}^{L+1-j})%
\Big],  \notag \\
\upsilon _{jM} &=&\frac{A}{2}\Big[(1-\frac{E_{M}}{s_{+}}%
)(x_{+}^{j}-x_{+}^{L+1-j})-(1-\frac{E_{M}}{s_{-}})(x_{-}^{j}-x_{-}^{L+1-j})%
\Big].  \label{uvfinal}
\end{eqnarray}

Now, the Majorana wave functions $f_{L/R,j}$ can be readily derived using
Eq. (\ref{uvfinal}) according to the main text. Since $f_{\pm,j}$ can always
be made real, we have $f_{L,j}=f_{+,j}$ and $f_{R,j}=f_{-,j}$. Let us
illustrate our results in the considered regime $\mu^{2}-4(J^{2}-%
\Delta^{2})<0$, in which it is more convenient to write $x_{\pm}=\rho e^{\pm
i\theta}$ with $\rho =\sqrt{(J-\Delta)/(J+\Delta)}$ and $\theta=\arccos[%
-\mu/2\sqrt{J^{2} -\Delta^{2}}]$. The Majorana wave function $f_{j,L/R}$ can
then be written (in the leading order $E_{M}$) as 
\begin{align}
f_{Lj} & =2\left\vert A\right\vert \rho^{j}\sin(j\theta),  \notag \\
f_{Rj} & =2\left\vert A\right\vert \rho^{L-j+1}\sin[(L-j+1)\theta];
\label{fLRphase}
\end{align}
and the energy, Eq. (\ref{EMfinal}), as 
\begin{align}
E_{M} & =\Delta\rho^{L}\frac{4J^{2}-\mu^{2}}{J(\Delta+J)}\Big|\frac {\sin[%
(L+1)\theta]}{\sin\theta}\Big|  \notag \\
& =\Delta e^{-La/l_{M}}\frac{4J^{2}-\mu^{2}}{J(\Delta+J)}\Big|\frac {\sin[%
(L+1)\theta]}{\sin\theta}\Big|.  \label{EMcom}
\end{align}
We thus clearly see that the energy of the edge mode decays exponentially
with $L$, and the localization length of the Majorana wave functions near
the edges is 
\begin{equation*}
l_{M}=\frac{a}{\ln\rho^{-1}}=\frac{a}{\ln(\sqrt{(J+\Delta)/(J-\Delta})}.
\end{equation*}

As an example, consider the case of $\mu =0$ and $J\neq \Delta $, when Eq. (%
\ref{EMcom}) indicates $E_{M}=0$ when $L$ is odd. In fact, $E_{M}=0$ is an
exact result for $\mu =0$ and odd $L$, which can be most easily seen by
expressing the Kitaev Hamiltonian in the Majorana basis \cite{Kitaev}. In
this basis, the Hamiltonian matrix $(2L\times 2L)$ For $\mu =0$ can be
brought into a a block diagonal form $H_{\mathrm{K}}=H_{1}\oplus H_{2}$, in
which $H_{1}$ matrix couples Majorana operators $(c_{4n+1},c_{4n+4})$ and $%
H_{2}$ matrix couples Majorana operators $(c_{4n+2},c_{4n+3})$,
respectively, for $n=0,1,2..$. Both $H_{1}$ and $H_{2}$ matrices are
antisymmetric and are of dimension $L$, such that we can immediately infer
the existence of the zero energy-eigenvalue when $L$ is odd.

\subsection{Bulk correlations}

Let us also calculate some correlations functions in the bulk of the wire,
which are determined by the gapped modes. In the thermodynamic limit $%
L\rightarrow\infty$, the bulk gapped modes can be characterized by their
quasi-momentum $\hbar k$ from the Brillouin zone (BZ), $k\in\lbrack-\pi
/a,\pi/a]$, with the corresponding energy $E_{k}=\sqrt{(2J\cos ka+\mu
)^{2}+4\Delta^{2}\sin^{2}ka}$. In this case, Eq. (18) takes the form 
\begin{equation}
a_{j}^{\prime}=\frac{1}{\sqrt{L}}\sum_{k\in\mathrm{BZ}}(u_{k}\alpha
_{k}e^{ikaj}+v_{k}^{\ast}\alpha_{k}^{\dag}e^{-ikaj}),
\label{BogoliubovFermi}
\end{equation}
where 
\begin{equation}
u_{k}=\sqrt{\frac{E_{k}+\xi_{k}}{2E_{k}}},\;v_{k}=i\frac{2\Delta\sin ka} {%
E_{k}+\xi_{k}}u_{k}=\frac{2i\Delta\sin ka}{\sqrt{2E_{k}(E_{k}+\xi_{k})}}
\label{bulk u and v}
\end{equation}
satisfy the condition $u_{k}^{2}+\left\vert v_{k}\right\vert ^{2}=1$ and $%
\xi_{k}=-2J\cos ka-\mu$.

\begin{figure}[th]
\includegraphics[width=0.3\textwidth]{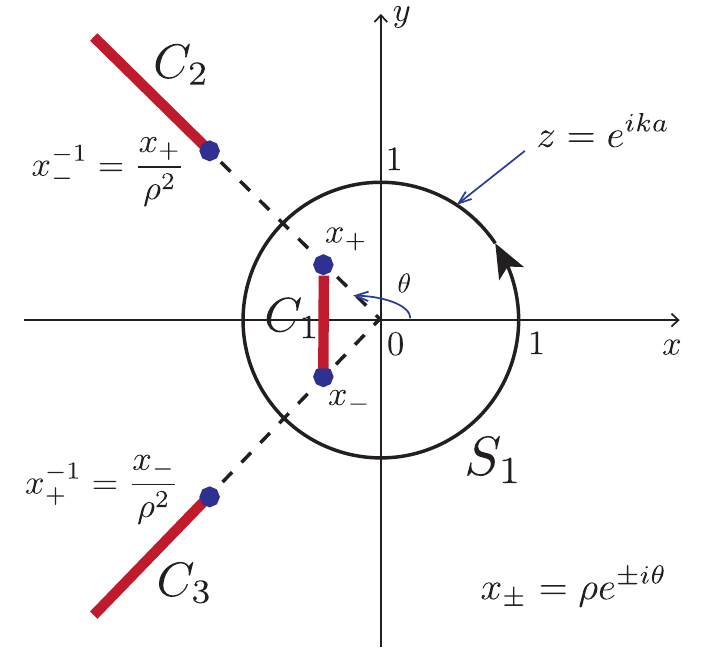}
\caption{Contour of the integration $\mathrm{S}_{1}$ in the complex $z$%
-plane, four brancing points $x_{\pm }$ and $x_{\pm }^{-1}$, and three cuts $%
C_{1}$, $C_{2}$, and $C_{3}$ defining the branch of the function $\protect%
\sqrt{F_{1}(z)F_{2}(z)}$.}
\label{Pole}
\end{figure}

We start with the correlation function $f(r)\equiv \left\langle
a_{j+r}^{\prime }a_{j}^{\prime }\right\rangle =-f(-r)$ which can be written
as 
\begin{equation}
f(r)=\left\langle a_{j+r}^{\prime }a_{j}^{\prime }\right\rangle =\frac{1}{L}%
\sum_{k\in \mathrm{BZ}}u_{k}v_{k}^{\ast }e^{ikar}=-\int_{-\pi }^{\pi }\frac{d%
\widetilde{k}}{2\pi }\frac{2i\Delta \sin \widetilde{k}}{2E_{\widetilde{k}}}%
e^{i\widetilde{k}r},  \notag
\end{equation}%
where $\widetilde{k}=ka$. After introducing the complex variable $z=\exp (i%
\widetilde{k})$, the expression for $f(r)$ can be rewritten as a contour
integral over the unit circle \textrm{S}$_{1}$ in the complex $z$-plane (see
Fig. \ref{Pole}): 
\begin{equation}
f(r)=-\frac{\Delta }{2\pi }\oint_{\mathrm{S}_{1}}\frac{dz}{iz}\frac{%
(z-z^{-1})}{2\sqrt{F_{1}(z)F_{2}(z)}}z^{r}  \notag
\end{equation}%
with $F_{1}(z)$ and $F_{2}(z)$ being defined in Eqs. (\ref{F1}) and (\ref{F2}%
), respectively. Note that the integrand in the above expression has four
branching points $x_{+}$, $x_{-}$, $x_{+}^{-1}=x_{-}/\rho ^{2}$, and $%
x_{-}^{-1}=x_{+}/\rho ^{2}$ [zeros of $F_{1}(z)$ and $F_{2}(z)$], and the
branch of this multivalued function is specified by making three cuts 
\textrm{C}$_{1}$, \textrm{C}$_{2}$, and \textrm{C}$_{3}$ in the complex
plane, see Fig. \ref{Pole}. After simple manipulations, the integral can be
rewritten in the form 
\begin{equation}
f(r)=-\frac{1}{2\pi }\frac{\alpha }{2\sqrt{1-\alpha ^{2}}}\oint_{\mathrm{S}%
_{1}}\frac{dz}{iz}\frac{(z^{2}-1)}{\sqrt{%
(z-x_{+})(z-x_{-})(z-x_{+}^{-1})(z-x_{-}^{-1})}}z^{r},  \notag
\end{equation}%
where $\alpha =\Delta /J<1$. Without loss of generality, we consider $r>0$
and deform \textrm{S}$_{1}$ to the contour around the cut \textrm{C}$_{1}$
which connects the points $x_{+}$ and $x_{-}$. To simplify the calculations
we consider the case $1-\alpha ^{2}\ll 1$, when $\rho =\left\vert x_{\pm
}\right\vert \ll 1$ and $\left\vert z\right\vert \ll 1$ for $z\in $\textrm{C}%
$_{1}$, and, using the approximate expression for $%
(z-x_{+}^{-1})(z-x_{-}^{-1})\approx (x_{+}x_{-})^{-1}=\rho ^{-2}$ for $z\in $%
\textrm{C}$_{1}$, simplify the integral to the form 
\begin{equation}
f(r)=\frac{1}{2\pi }\frac{\alpha }{2(1+\alpha )}\oint_{\mathrm{C}_{1}}\frac{%
dz}{iz}\frac{1}{\sqrt{(z-x_{+})(z-x_{-})}}z^{r}.  \notag
\end{equation}%
After writing 
\begin{equation}
z(y)=\frac{1}{2}(x_{+}+x_{-})+\frac{y}{2}(x_{+}-x_{-})=\rho (\cos \theta
+iy\sin \theta )\in \mathrm{C}_{1},  \notag
\end{equation}%
where $y\in \lbrack -1,1]$, we find (the value of the function $\sqrt{%
(z-x_{+})(z-x_{-})}$ is chosen to be positive on the right side of the cut) 
\begin{eqnarray}
f(r) &=&\frac{1}{2\pi }\frac{\alpha }{1+\alpha }\rho ^{r-1}\int_{-1}^{1}%
\frac{dy}{\sqrt{1-y^{2}}}(\cos \theta +iy\sin \theta )^{r-1}
\label{fcorrelation} \\
&=&\frac{1}{2}\frac{\alpha }{1+\alpha }\rho ^{r-1}\mathrm{P}_{r-1}(\cos
\theta ),  \notag
\end{eqnarray}%
where $\mathrm{P}_{n}(x)$ is the Legendre polynomial of degree $n$. The
asymptotics of $f(r)$ for large $r,$ 
\begin{equation}
f(r\gg 1)\approx \frac{1}{2}\frac{\alpha }{1+\alpha }\sqrt{\frac{2}{\pi
r\sin \theta }}\rho ^{r-1}\cos [(r-1/2)\theta -\pi /4],
\label{fcorrelationasympt}
\end{equation}%
shows exponential decay with the characteristic length $\xi
_{BCS}=l_{M}=-a\ln \rho $.

The correlation function 
\begin{equation}
g(r)=\left\langle a_{j+r}^{\prime \dag }a_{j}^{\prime }\right\rangle =g(-r)=%
\frac{1}{L}\sum_{k\in \mathrm{BZ}}\left\vert v_{k}\right\vert
^{2}e^{ikar}=\int_{-\pi }^{\pi }\frac{d\widetilde{k}}{2\pi }\frac{1}{2}%
\left( 1+\frac{\xi _{\widetilde{k}}}{E_{\widetilde{k}}}\right) e^{i%
\widetilde{k}r}
\end{equation}%
can be calculated in the same way (we again consider the case $1-\alpha
^{2}\ll 1$, that is $\rho \ll 1$): For $r\geq 1$ we obtain 
\begin{equation}
g(r)=\frac{1}{2}\frac{1}{1+\alpha }\rho ^{r-1}\mathrm{P}_{r-1}(\cos \theta )
\label{gcorrelation}
\end{equation}%
or asymptotically 
\begin{equation}
g(r\gg 1)\approx \frac{1}{2}\frac{1}{1+\alpha }\sqrt{\frac{2}{\pi r}}\left( 
\frac{1-\alpha ^{2}}{1-\alpha ^{2}-\beta ^{2}}\right) ^{1/4}\rho ^{r-1}\cos
[(r-1/2)\theta -\pi /4].  \label{gcorrelationasympt}
\end{equation}

Finally, we calculate the correlation function 
\begin{equation}
h(r)=\frac{1}{L}\sum_{k\in \mathrm{BZ}}\frac{(u_{k}+v_{k})^{2}\sin ^{2}ka}{%
E_{k}}e^{ikar}  \label{correlationh}
\end{equation}%
which appears in the temperature-dependent correction to the energy of the
Majorana mode, Eq. (\ref{deltaTE1}). After using the expressions for $u_{k}$
and $v_{k}$ from Eq. (\ref{bulk u and v}) and dimensional variable $%
\widetilde{k}=ka$, the expression for $h(r)$ reads 
\begin{equation*}
h(r)=\int_{-\pi }^{\pi }\frac{d\widetilde{k}}{2\pi }\frac{\xi _{\widetilde{k}%
}+2i\Delta \sin \widetilde{k}}{E_{\widetilde{k}}^{2}}\sin ^{2}\widetilde{k}%
\;e^{i\widetilde{k}r}=\int_{-\pi }^{\pi }\frac{d\widetilde{k}}{2\pi }\frac{%
\sin ^{2}\widetilde{k}}{\xi _{\widetilde{k}}-2i\Delta \sin \widetilde{k}}e^{i%
\widetilde{k}r},
\end{equation*}%
where we use the identity $E_{\widetilde{k}}^{2}=(\xi _{\widetilde{k}%
}+2i\Delta \sin \widetilde{k})(\xi _{\widetilde{k}}-2i\Delta \sin \widetilde{%
k})$. The last integral can be transformed into the contour integral in the
complex $z$-plane as 
\begin{equation*}
h(r)=\frac{1}{2\pi }\oint_{\mathrm{S}_{1}}\frac{dz}{iz}\frac{(z-z^{-1})^{2}}{%
4F_{1}(z)}z^{r}=\frac{1}{4J(1+\alpha )}\oint_{\mathrm{S}_{1}}\frac{dz}{2\pi i%
}\frac{1}{z^{2}}\frac{(1-z^{2})^{2}}{(z-x_{+})(z-x_{-})}z^{r},
\end{equation*}%
and then calculated by deforming the contour of integration and using the
Cauchy theorem: For $r<-1$, we deform the contour to infinity with zero
result for the integral; for $r\geq -1$, the contour is deformed to zero and
the result of the integration is given by the contribution of the poles at $%
z=x_{+}$, $z=x_{-}$, and $z=0$ (for $r=0$, $\pm 1$). After calculating the
corresponding residues we obtain 
\begin{eqnarray}
h(r) &=&\frac{1}{4J}\frac{%
(1-x_{+}^{2})^{2}x_{+}^{r-2}-(1-x_{-}^{2})^{2}x_{-}^{r-2})}{2i\sqrt{1-\alpha
^{2}-\beta ^{2}}}  \notag \\
&=&\frac{1}{4J(1+\alpha )}\frac{\rho ^{r-3}}{\sin \theta }\left\{ \sin
[(r-2)\theta ]-2\rho ^{2}\sin (r\theta )+\rho ^{4}\sin [(r+2)\theta
]\right\} ,\hspace{2mm}r>1  \notag \\
h(1) &=&-\frac{1}{4J(1+\alpha )}\left[ 1+2\frac{1+\alpha -2\beta ^{2}}{%
(1+\alpha )^{2}}\right] =\frac{1}{4J(1+\alpha )}[-2+\rho ^{2}(4\cos
^{2}\theta -1)],  \notag \\
h(0) &=&-\frac{\beta }{2J(1+\alpha )^{2}}=\frac{1}{4J(1+\alpha )}2\rho \cos
\theta ,  \notag \\
h(-1) &=&\frac{1}{4J(1+\alpha )},  \notag \\
h(r) &=&0,\hspace{2mm}r<-1,  \label{correlationhanswers}
\end{eqnarray}%
where we $\alpha =\Delta /J$ and $\beta =\mu /2J$. Note that this
correlation function behaves differently for positive and negative $r$ due
to the specific analytical structure of the integrand. The replacement of $%
(u_{k}+v_{k})^{2}$ with $(u_{k}-v_{k})^{2}$ in the $h(r)$ results in the
reflected ($r\rightarrow -r$) behavior.

The above expressions for the correlation functions show that the bulk
correlation length $\xi_{BCS}$ is identical with the localization length of
the Majorana edge states $l_{M}$. Mathematically it follows from the fact
that both lengths originates from the zeros $x_{+}$ and $x_{-}$ of $E_{k}$
[or $F_{1}(z)F_{2}(z)$] in the complex plane of $z=\exp(ika)$.


\begin{thebibliography}{99}
\bibitem{Wilczek2009} {F. Wilczek, Nat. Phys. \textbf{5}, 614 (2009).} 


\bibitem{AnyonReview0} {C. Nayak, Steven H. Simon, A. Stern, M. Freedman,
and S. Das Sarma, Rev. Mod. Phys. \textbf{80}, 1083 (2008).} 

\bibitem{AnyonReview1} {A. Stern, Ann. Phys. \textbf{323}, 204 (2008).} 

\bibitem{AnyonReview2} {A. Stern, Nature (London) \textbf{464}, 187 (2010).} 


\bibitem{TQC1} {A. Kitaev, Ann. Phys. (Amsterdam) \textbf{303}, 2 (2003).} 

\bibitem{TQC2} {M. H. Freedman, A. Kitaev, J. Larsen, and Z. Wang, Bull. Am.
Math. Soc. 40, \textbf{31} (2003).} 

\bibitem{TQC3} {S. Das Sarma, M. Freedman, and C. Nayak, Phys. Rev. Lett. 
\textbf{94}, 166802 (2005).} 

\bibitem{TQC4} {J. K. Pachos, \textit{Introduction to Topological Quantum
Computation} (Cambridge University Press, Cambridge, England, 2012).} 


\bibitem{PfQHE} {G. Moore and N. Read, Nucl. Phys. \textbf{B360 }, 362
(1991).} 

\bibitem{Read2000} {N. Read and D. Green, Phys. Rev. B \textbf{61}, 10267
(2000).} 

\bibitem{Anyon1} {D. A. Ivanov, Phys. Rev. Lett. \textbf{86}, 268 (2001).} 

\bibitem{Kitaev} {A. Kitaev, Phys. Usp. \textbf{44}, 131 (2001).} 

\bibitem{Cheng} {M. Cheng and H. H. Tu, Phys. Rev. B \textbf{84}, 094503
(2011).} 


\bibitem{Majreview1} {J. Alicea, Rep. Prog. Phys. \textbf{75}, 076501 (2012)}

\bibitem{Majreview2} {C. W. J. Beenakker, Annu. Rev. Condens. Matter Phys. 
\textbf{4}, 113 (2013)}

\bibitem{Majreview3} {T. D. Stanescu and S. Tewari, J. Phys.: Condens.
Matter \textbf{25}, 233201 (2013).} 

\bibitem{MajCMP0} {L. Fu and C. L. Kane, Phys. Rev. Lett. \textbf{100},
096407 (2008); Phys. Rev. B \textbf{79}, 161408(R) (2009).} 

\bibitem{MajCMP1} {J. D. Sau, R. M. Lutchyn, S. Tewari, and S. Das Sarma,
Phys. Rev. Lett. \textbf{104}, 040502 (2010).} 

\bibitem{MajCMP2} {R. M. Lutchyn, J. D. Sau, and S. Das Sarma, Phys. Rev.
Lett. \textbf{105}, 077001 (2010).} 

\bibitem{MajCMP12} {J. Alicea, Phys. Rev. B \textbf{81}, 125318 (2010).} 

\bibitem{MajCMP3} {Y. Oreg, G. Refael, and F. von Oppen, Phys. Rev. Lett. 
\textbf{105}, 177002 (2010).} 



\bibitem{LiangJiang} {L. Jiang, T. Kitagawa, J. Alicea, A. R. Akhmerov, D.
Pekker, G. Refael, J. I. Cirac, E. Demler, M. D. Lukin, and P. Zoller, Phys.
Rev. Lett. \textbf{106}, 220402 (2011).} 

\bibitem{Nascimbene} {S. Nascimbene, J. Phys. B: At. Mol. Opt. Phys \textbf{%
46}, 134005 (2013).} 


\bibitem{MajAtom2} {S. Diehl, E. Rico, M. A. Baranov, and P. Zoller, Nat.
Phys. \textbf{7}, 971 (2011).} 

\bibitem{MajAtom3} {C. V. Kraus, S Diehl, P. Zoller, and M. A. Baranov, New
J. Phys. \textbf{14}, 113036 (2012).} 


\bibitem{MajAtom10} {B. Sundar and E. J. Mueller, Phys. Rev. A \textbf{88},
063632 (2013).} 

\bibitem{MajAtom11} {A. B\"uhler, N. Lang, C.V. Kraus, G. M\"oller, S. D.
Huber, and H.P. B\"uchler, Nat. Commun. \textbf{5}, 4504 (2014)} 


\bibitem{MajEx1} {V. Mourik, K. Zuo, S. M. Frolov, S. R. Plissard, E. P. A.
M Bakkers, and L.P. Kouwenhoven, Science \textbf{336}, 1003 (2012).} 

\bibitem{MajEx2} {M. T. Deng, C. L. Yu, G. Y. Huang, M. Larson, P. Caroff,
and H. Q. Xu, Nano Lett. \textbf{12}, 6414 (2012).} 

\bibitem{MajEx30} {L. P. Rokhinson, X. Liu, and J. K. Furdyna, Nat. Phys. 
\textbf{8}, 795 (2012). } 

\bibitem{MajEx3} {A. Das, Y. Ronen, Y. Most, Y. Oreg, M. Heiblum, and H.
Shtrikman, Nat. Phys. \textbf{8}, 887 (2012). } 

\bibitem{Churchill2013} {H. O. H. Churchill, V. Fatemi, K. Grove-Rasmussen,
M. T. Deng, P. Caroff, H. Q. Xu, and C. M. Marcus, {Phys. Rev.
B} \textbf{87}, 241401(R) (2013). }

\bibitem{Finck2013} {A. D. K. Finck, D. J. Van Harlingen, P. K. Mohseni, K.
Jung, and X. Li,  {Phys. Rev. Lett}. \textbf{110}, 126406 (2013). }

\bibitem{MajEx4} {S. Nadj-Perge, I. K. Drozdov, J. Li, H. Chen, Sangjun
Jeon, Jungpil Seo, A. H. MacDonald, B. A.Bernevig, A. Yazdani, Science 
\textbf{346}, 602 (2014).} 



\bibitem{Dloss} {A.A. Zyuzin, D. Rainis, J. Klinovaja, and D. Loss, Phys.
Rev. Lett. \textbf{111}, 056802 (2013).} 

\bibitem{BlochDalibardZwerger} {I. Bloch, J. Dalibard, and W. Zwerger, Rev.
Mod. Phys. \textbf{80}, 885 (2008).} 

\bibitem{PetrovSolomonSchlyapnikov} {D. S. Petrov, C. Salomon, and G. V.
Shlyapnikov, Phys. Rev. Lett. \textbf{93}, 090404 (2004).} 

\bibitem{PetrovSolomonSchlyapnikov1} {D. S. Petrov, Phys. Rev. A \textbf{67}%
, 010703 (2003).} 

\bibitem{Fetter} {A. L. Fetter and J. D. Walecka, \textit{Quantum Theory of
Many-Particle Systems}, (McGraw-Hill Book Company, New York, 1971).} 

\bibitem{MajAtom5} {C. V. Kraus, P. Zoller, and M. A. Baranov, Phys. Rev.
Lett. \textbf{111}, 203001 (2013).} 

\bibitem{MajAtom6} {C. Laflamme, M. A. Baranov, P. Zoller, and C. V. Kraus
Phys. Rev. A \textbf{89}, 022319 (2014).} 


\bibitem{Ketterle08} {W. Ketterle and M. W. Zwierlein, Rivista del Nuovo
Cimento \textbf{31}, 247-422 (2008).} 
\end{thebibliography}
\end{document}